%% file: main.tex
\documentclass[a4paper,10pt]{article}

\usepackage[pdftex]{color,graphicx,hyperref}
\usepackage{amsmath,amssymb,amsthm}
\usepackage[margin=1in]{geometry}
\usepackage{setspace}
\usepackage{booktabs}
\usepackage[font=scriptsize,labelfont=bf,labelsep=period]{caption}
\usepackage{subcaption}
\usepackage[T1]{fontenc}
\usepackage[affil-it]{authblk}
\graphicspath{ {./figures/} }

\usepackage{placeins} 
\usepackage{rotating} 
\usepackage{enumitem}
\usepackage{xr-hyper} 

\usepackage[backend=biber,style=nature,url=false]{biblatex}
\newcommand{\be}{\begin{equation}}
\newcommand{\ee}{\end{equation}}

\newcommand{\eref}[1]{Eq.~(\ref{#1})}
\newcommand{\esref}[2]{Eqs.~(\ref{#1})-(\ref{#2})}
\newcommand{\sref}[1]{Section~\ref{#1}}
\newcommand{\fref}[1]{Fig.~\ref{#1}}

\newcommand{\srefsi}[1]{SI Section~{\ref{#1}}}
\newcommand{\frefsi}[1]{SI Fig.~S{\ref{#1}}}
\newcommand{\trefsi}[1]{SI Table~S\ref{#1}}

\graphicspath{ {./figures/} }

\title{Universal time scales linking structure and dynamics in temporal networks}

\author[1,*]{Giulia de Meijere}
\author[2,3]{Márton Karsai}
\author[1,4,*]{Gerardo Iñiguez}

\affil[1]{\small{Tampere Complexity Lab, Data Science Research Centre, Tampere University, FI-33720 Tampere, Finland}}
\affil[2]{\small{Department of Network and Data Science, Central European University, AT-1100 Vienna, Austria}}
\affil[3]{\small{HUN-REN R\'enyi Institute of Mathematics, HU-1053 Budapest, Hungary}}
\affil[4]{\small{Centro de Ciencias de la Complejidad, Universidad Nacional Autonóma de México, 04510 Ciudad de México, Mexico}}
\affil[*]{\small{Corresponding author email: giulia.demeijere@tuni.fi, gerardo.iniguez@tuni.fi}}

\date{}

\begin{document}

\maketitle	

\begin{abstract}
Temporal networks underpin a wide range of social, technological, and biological phenomena by mapping the dynamics of complex systems onto inhomogeneous temporal interactions. While temporal network connectivity is of key interest, its evolution across time scales remains poorly understood. By analyzing various temporal network data of informational and societal origin, we find systematic evidence of an optimal time scale for coarse-graining interaction events. At this level of aggregation, networks are maximally dynamic in their local structure, while retaining system-wide connectivity. To understand the origins of this generic interplay of structure and dynamics, we explore a minimal temporal network model based on uncorrelated renewal processes, and show that intermittent yet globally connected activity arises solely due to heterogeneities in inter-event times and degrees, and no other system-specific details. All coarse-grained empirical networks we study show persistent patterns of cyclic node degree change yet stationary system-level degree distributions, a striking coexistence of microscopic self-regulation and macroscopic stability. Our results suggest a universal pattern in temporal networks that involves both time and structure, via the nontrivial interplay of aggregation and temporal inhomogeneity. This may have fundamental consequences for dynamical processes on networks and the balance between robustness and adaptability in complex systems.
\end{abstract}

\section*{Introduction}
\label{sec:intro}

Networks are key to our understanding of complex systems dynamics, highlighting how large sets of components interacting locally can self-organize into structures and behaviors at higher scales \cite{newman_networks_2018}. These patterns of interactions, often sparse, heterogeneous, and multilayered, underpin a variety of natural and human-made phenomena, as they facilitate the spread of pathogens, influence and information in society \cite{pastor-satorras_epidemic_2015,starnini_opinion_2025} and regulate the resilience of networks, infrastructures, and ecosystems to global epidemics~\cite{pastor-satorras_epidemic_2015}, systemic risks~\cite{acemoglu_systemic_2015}, and cascading failures \cite{gao_universal_2016}. A fundamental feature of network structure is to evolve in time \cite{holme_temporal_2012,holme_modern_2015,li_fundamental_2017}. While slowly-varying interactions might be well captured by quenched static graphs where ties are assumed to be permanent, most complex systems exhibit structural dynamics across multiple time scales. By considering instantaneous or short-duration events via timestamped edges (such as brief chemical reactions, proximity contacts, or digital communication), temporal networks generalize static structural measures \cite{pan_path_2011,scholtes_higher-order_2016} to uncover the role of temporal inhomogeneities in processes like social contagion and network control \cite{barabasi_origin_2005,unicomb_dynamics_2021}. Dynamical networks, in turn, account for links of longer but still finite duration, such as ongoing social relationships, exploring how tie concurrency shapes epidemic spreading \cite{holme_network_2003} and group formation \cite{palla_quantifying_2007,asikainen_cumulative_2020}. These two paradigms of changeable network structure, temporal and dynamical, are joined by efforts to extract sustained but time-varying interactions from high-resolution spatio-temporal data \cite{psorakis_inferring_2012,gelardi_detecting_2019}, typically via aggregation over time. While straightforward to implement in practice, coarse-graining complicates temporal network analysis due to a broad distribution of potentially relevant time scales that obscures patterns within time intervals \cite{peixoto_modelling_2017} due to their length and placement \cite{sulo_meaningful_2010,krings_effects_2012,kivela_estimating_2015}. This in turn leads to biases in the estimation of network structure \cite{clauset_persistence_2012} and in the characterization of dynamical processes over evolving networks \cite{ribeiro_quantifying_2013,unicomb_dynamics_2021}.

The detection of relevant time scales in temporal networks of social, biological, and technological systems remains an active challenge \cite{holme_epidemiologically_2013,darst_detection_2016,gelardi_temporal_2021,andres_detecting_2024}, with applications to change point detection \cite{peel_detecting_2015,masuda_detecting_2019}, anomaly and structural break dynamics \cite{gelardi_detecting_2019,wilson_modeling_2019}, time series forecasting \cite{sulo_meaningful_2010}, and transport monitoring and control \cite{gautreau_microdynamics_2009,tang_dynamic_2014}. At its core, the detection of time scales aims to separate meaningful structural change from random fluctuations \cite{rosvall_mapping_2010,peixoto_modelling_2017}, in order to relate patterns of interactions to mechanisms operating at specific temporal resolutions \cite{saramaki_seconds_2015,holme_fundamental_2023,holme_map_2023}. At short time scales, temporal networks display correlated, bursty trains of events \cite{karsai_universal_2012} and temporal motifs \cite{kovanen_temporal_2013} that are characterized by broad inter-event time distributions \cite{barabasi_origin_2005}, link weight heterogeneity \cite{barrat_architecture_2004}, degree-weight correlations \cite{onnela_structure_2007}, and circadian rhythms of edge and node activity \cite{jo_circadian_2012,aledavood_daily_2015}. When coarse-grained into longer time periods, these patterns smoothen into dynamical network observations with tie growth/decay, persistent patterns at the node level \cite{iniguez_universal_2023,saramaki_persistence_2014}, and mesoscopic community structure \cite{mucha_community_2010,traag_significant_2013,peixoto_modelling_2017,bovet_flow_2022}, ultimately leading to static networks with robust connectivity driven by low-activity ties \cite{miritello_limited_2013}. Overall, the multiscale nature of networked complex systems under temporal coarse-graining suggests a fundamental trade-off between dynamic interactions and global cohesion: aggregation over short periods implies short-lived, asynchronous ties that hinder connectivity, while coarser aggregation regains network structure by averaging out changes at faster time scales.

Here, we find ample evidence for the existence of two relevant time scales in the evolution of networked data under temporal aggregation: one driving structural changes locally, and another determining long-range connectivity for the system as a whole. We gather many datasets on social and informational networks, amounting to tens of thousands of individuals and millions of offline/online interactions in time spans from hours to years, and analyze them via a minimal framework of temporal coarse-graining that continuously interpolates between temporal, dynamical, and static representations of these systems as aggregation increases. At intermediate levels of aggregation, all networks display a specific time scale at which edge dynamics is maximal, and a second time scale where a largest connected component appears due to dynamic percolation. A simple model of temporal networks based on renewal processes \cite{unicomb_dynamics_2021} shows that both time scales are mainly driven by heterogeneity in inter-event times (burstiness) and arrival times (openness), even in the absence of other correlations. For the vast majority of systems, the characteristic time scale of local dynamicity is larger than that of global connectivity. This indicates that the aggregated dynamical networks constructed from repeated social interactions can retain extensive variability at the micro level (including circadian patterns of node degree change), while retaining network cohesion and stationarity of structural features at the macro level.

\section*{Results}
\label{sec:results}

\begin{figure}[t]
    \centering
    \includegraphics[width=0.9\textwidth]{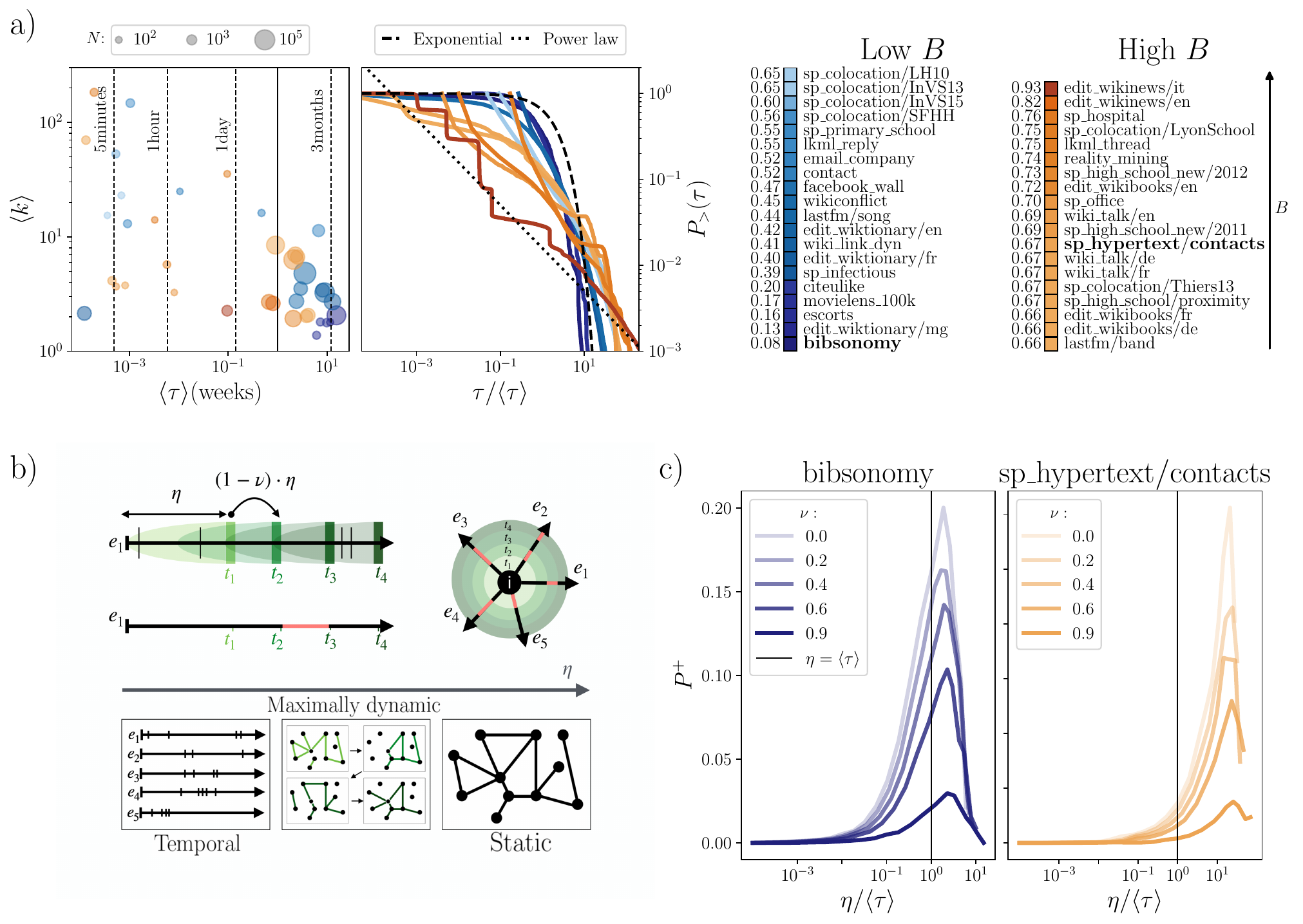} 
    \caption{\textbf{Temporal coarse-graining leads to maximally dynamical networks}.
    \textbf{a)} (left) Correlation between average inter-event time (IET) $\langle \tau \rangle$ and average static degree $\langle k \rangle$ in the fully aggregated network, for all temporal networks studied. Systems vary greatly in their size $N$ (size of dots), connectivity, and time scales of interaction, from minutes to months. (middle) Complementary cumulative distribution (CCDF) $P_>(\tau)$ as a function of rescaled IET $\tau / \langle \tau \rangle$, for 13 selected datasets (all networks in \frefsi{fig:IETs}). IET distributions vary across systems from exponential (dashed line, rate 0.45) to Zipf-like power-law (dotted line, exponent -0.5), signaling increasing bursty behavior at the system level (see \sref{SI:analytical} for functional forms). (right) Dots and lines are colored by burstiness $B$~\cite{goh_burstiness_2008}, shown for all networks as an ordered list.
    \textbf{b)} Framework of temporal network coarse-graining explored in this study. Events in the temporal network are aggregated within rolling time windows of size $\eta$ and fractional overlap $\nu \in [0, 1]$, such that the elapsed time between window endpoints $t_n$ and $t_{n+1}$ is $\omega = (1 - \nu) \eta$. Undirected and binary edges $e_i$ in the resulting dynamical network are active if a window has at least one event, and inactive otherwise. The level of aggregation $\eta$ interpolates between regimes of temporal ($\eta \to 0$) and static ($\eta \to \infty$) networks, with an intermediate regime where the dynamical network is maximally dynamic.
    \textbf{c)} Activation probability $P^+$ as a function of rescaled aggregation window $\eta / \langle \tau \rangle$, for varying overlap $\nu$ in selected datasets with low (bibsonomy) and high (sp\_hypertext/contacts) levels of burstiness. The maximum in $P^+$ signals a time scale $\eta^*$ of maximal dynamicity where edges persist within windows but also appear and disappear frequently. Overlap $\nu$ doesn't change the optimal time scale $\eta^*$, while temporal inhomogeneities (encoded by $B$) drive $\eta^*$ from its homogeneous baseline $\eta^* \sim \langle \tau \rangle$ (solid vertical line) to larger values ($P^+$ and its standard deviation for all systems in \frefsi{fig:Pplus_eta} and \frefsi{fig:Pplus_eta_std}).}
    \label{fig1}
\end{figure}

We start our study by analyzing 39 datasets of recurring, time-stamped interactions of online activity, physical proximity, and information transfer including emails, social media posts, movie ratings, webpage edits, online purchases, tag assignments, face-to-face interactions, and hyperlinks. The observation periods range from 8.5 hours to 16.3 years (\fref{fig1}). Unipartite and bipartite undirected temporal networks include, among others, 24.7M common song listenings of 179k users and 49k bands in the music website last.fm \cite{celma_long_2010}, 4.5M edits in 179k pages and 16.6M messages between 110k editors in 4 languages of Wikipedia, and 8.5M co-locations among 1.3k people in schools, hospitals, workplaces, and conferences in France \cite{genois_can_2018}. We consider events as instantaneous, filter out edges with less than 10 events and keep only datasets with more than 100 active edges (see Supplementary Information [SI] Section \ref{SI:datasets} for details on data collection). Structural properties vary widely across datasets, with network size $N$ roughly going from $10^2$ to $10^5$ nodes, average static degree $\langle k \rangle$ (aggregated over the entire observation period) ranging from $10^0$ to $10^2$ connections, and average inter-event time $\langle \tau \rangle$ (IET, the time between consecutive events in an edge) spanning from minutes to months (\fref{fig1}a, left panel). The IET distribution $P(\tau)$, the probability that a randomly chosen IET has value $\tau$, shows diverse levels of inhomogeneous temporal patterns \cite{goh_burstiness_2008,karsai_universal_2012}, as demonstrated in \fref{fig1}a (middle panel). Measured by the burstiness index~\cite{goh_burstiness_2008}, defined as $B = (r-1)/(r+1)$ with $r = \sigma_{\tau} / \langle \tau \rangle$ the coefficient of variation of $P(\tau)$ \cite{kim_measuring_2016}, the shape of the IET distribution varies across datasets from almost exponential ($B = 0.08$) to broad-tailed ($B = 0.93$), as listed in \fref{fig1}a (right panel). For a given system, the functional shape of the IET distribution is robust to the choice of observation period $T$ (\frefsi{fig:IETs_period}). This agrees with studies that relate temporal inhomogeneities to decision mechanisms rather than periodic variability \cite{jo_circadian_2012,gandica_stationarity_2017}, suggesting stationarity of network features at the system level \cite{miller_constant_2019,gonzalez-casado_evidence_2025} (see \srefsi{SI:basic_stats} for details on $B$, $T$, and other basic statistics of datasets).

\subsection*{Aggregation in temporal networks leads to maximal dynamicity}

The aggregation of timestamped interaction events into dynamical networks where edges have duration and weight has been approached in many different ways \cite{holme_epidemiologically_2013}. A common route is to slice the observation period into single or consecutive windows at periodic times and aggregate contact events within \cite{gelardi_detecting_2019,krings_effects_2012}, or use parametrized sliding windows with either an infinitesimal \cite{unicomb_dynamics_2021} or finite \cite{kossinets_origins_2009,andres_detecting_2024} slide step. Edges might be considered as ongoing if events happen outside the window \cite{holme_network_2003,miritello_limited_2013}. Edge weight is typically proportional to the number or total duration of contacts within the time window \cite{gelardi_temporal_2021} or assumed to decay exponentially with time \cite{holme_epidemiologically_2013}. Window length might also be variable and adjusted to meaningful time scales \cite{sulo_meaningful_2010,darst_detection_2016} or change points \cite{peel_detecting_2015} in the dynamics. Despite its simplicity, aggregation via time windows naturally implements a notion of memory in temporal networks \cite{scholtes_causality-driven_2014,williams_shape_2022}, highlighting how bursty contact patterns lead to heterogeneous degree and weight distributions in dynamic networks \cite{vestergaard_how_2014}. Here we take a minimal approach combining these ideas and consider time windows of length $\eta$ and fractional overlap $\nu \in [0, 1]$, such that the part of a window not contained in the previous one has length $\omega = \eta (1 - \nu)$ (for a schematic explanation see \fref{fig1}b). Edges in the aggregated dynamical network are undirected and binary; they exist (or are active) if a window has at least one event, and disappear (become inactive) otherwise. Thus, the aggregation parameter $\eta$ interpolates between a truly temporal network where windows of vanishing length ($\eta \to 0$) contain at most one event per active edge, and a static network formed by active edges between any pair of nodes who have ever interacted ($\eta \to \infty$).

Intermediate values of the aggregation window $\eta$ reveal a time scale at which the emergent dynamical networks are maximally dynamic. This is calculated by defining the activation probability $P^+$ that an inactive edge in window $n$ becomes active in the next window $n + 1$, averaged over all edges in the empirical temporal network for given values of $\eta$ and $\nu$. $P^+$ is analogous to an edge transition probability between states with zero or more events in temporal network models \cite{unicomb_dynamics_2021}, for finite windows and overlap (see Materials and Methods [MM]). When varying $\eta$, all datasets show an optimal aggregation time scale $\eta^*$ with a maximum in $P^+$, which we refer to as \textit{maximal dynamicity}. This is demonstrated in \fref{fig1}c for two sample datasets with low and high levels of burstiness (see $P^+$ for all systems ranked by $B$ in \frefsi{fig:Pplus_eta} and \frefsi{fig:Pplus_eta_std}). Intermediate aggregation consistently leads to networks where edges might endure for long times but also appear and disappear frequently. In contrast, both temporal ($\eta \to 0$) and static ($\eta \to \infty$) network limits have low $P^+$, since on average edges remain inactive or active across windows, respectively. Increasing the overlap $\nu$ retains the optimal time scale $\eta^*$ while reducing the rate of edge activity change (as in \fref{fig1}c). Maximal dynamicity increases with inhomogeneities in the underlying temporal network; memoryless point processes with low $B$ imply $\eta^* \sim \langle \tau \rangle$, while bursty signals with high $B$ have a larger $\eta^* > \langle \tau \rangle$ that might even exceed the entire observation period. The deactivation probability $P^-$, the probability that an active edge becomes inactive across consecutive windows, shows the same qualitative behavior as $P^+$. Indeed, the removal of first (residual) IETs from all edges gives $P^+ = P^-$ for any aggregation $\eta$, implying time-reversal symmetry in edge dynamics \cite{unicomb_dynamics_2021} (see $P^{\pm}$ for all datasets in \srefsi{SI:Pplus}).

\subsection*{Coexistence of local dynamicity and global connectivity}

\begin{figure}[t]
    \centering
    \includegraphics[width=1.0\textwidth]{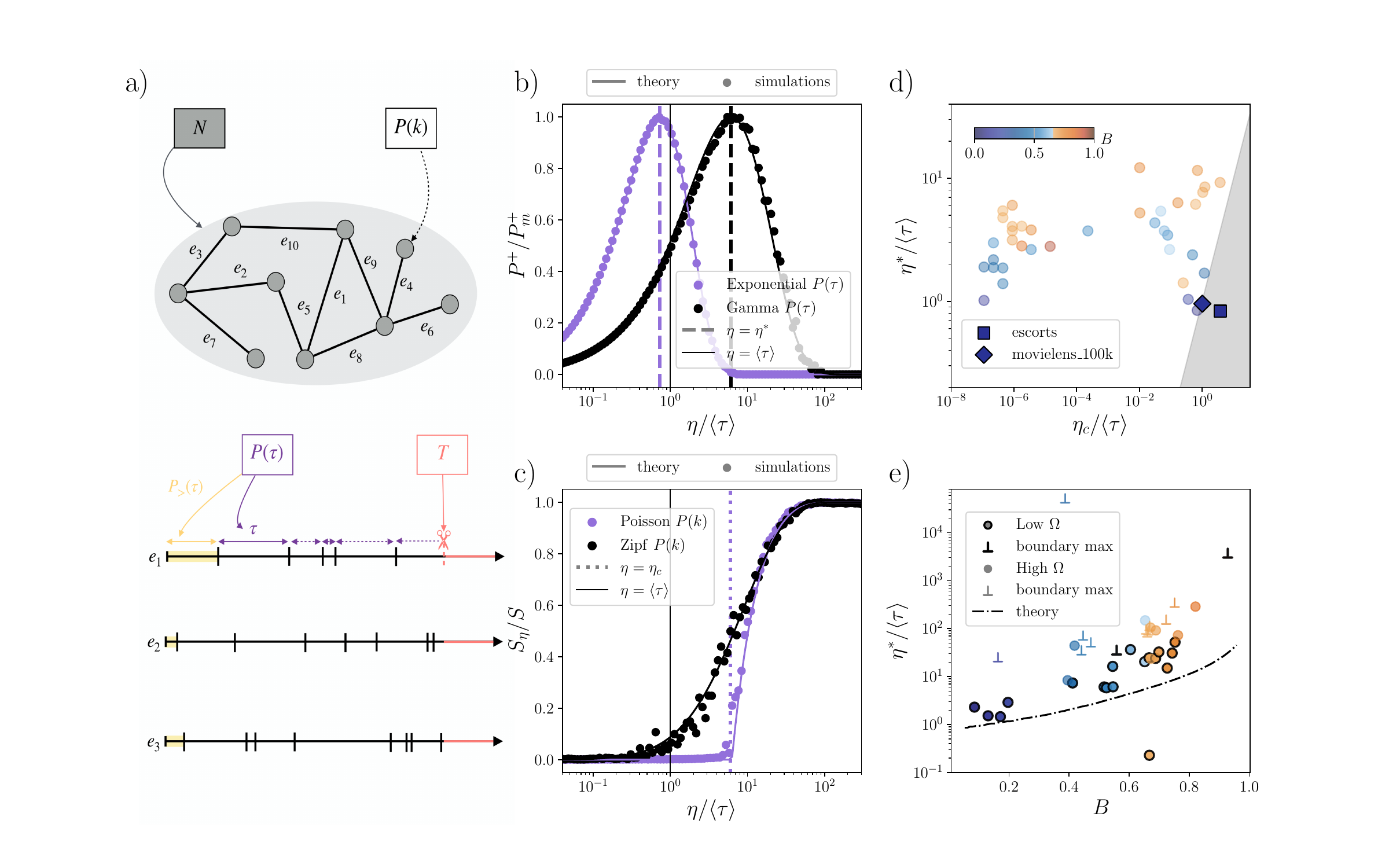} 
    \caption{\textbf{Local dynamicity and global connectivity coexist due to temporal and structural heterogeneity}. \textbf{a)} Schematic diagram of our temporal network model. During an observation period of length $T$, a renewal process with IET distribution $P(\tau)$ generates contact events over edges $e_i$ of a configuration-model network backbone of size $N$ and degree distribution $P(k)$. In this stationary dynamics, residual times follow the IET CCDF $P_>(\tau)$. \textbf{b)} Activation probability $P^+$ (relative to its maximum value $P^+_m$) as a function of rescaled aggregation scale $\eta / \langle \tau \rangle$ in two cases: homogeneous [exponential $P(\tau)$ with $\langle \tau \rangle = 2.5$], and heterogeneous [gamma $P(\tau)$ with $\alpha = 0.05$ and $\beta = 50$], both with the same average IET ($\langle \tau \rangle = \alpha \beta = 2.5$). As IET heterogeneity increases, the time scale $\eta^*$ of maximal dynamicity (dashed lines) grows beyond $\eta = \langle \tau \rangle$ (vertical line). \textbf{c)} Fractional size $S_{\eta}$ of dynamical LCC (relative to fractional size $S$ of static LCC) as a function of $\eta / \langle \tau \rangle$, for the same heterogeneous $P(\tau)$ as in (b), for homogeneous degrees [Poisson $P(k)$ with $\langle k \rangle = 1.95$] and heterogeneous degrees [Zipf $P(k)$ with $\gamma = 2.5$], both with the same $\langle k \rangle$ [for functional forms of $P(\tau)$ and $P(k)$ see \srefsi{SI_sec:functions}]. Degree heterogeneity pushes the dynamical percolation transition to a lower time scale $\eta_c$ (dotted lines; $\eta_c$ in the Zipf case goes to zero). Analytical calculations of $P^+$ and $S_{\eta}$ (continuous lines) agree with numerical simulations ($N = 5000$ and $\nu = 0.2$) and predict a coexistence of local dynamicity and global connectivity ($\eta_c < \eta^*$) for large enough heterogeneity. \textbf{d)} Location of empirical datasets in phase diagram $(\eta_c, \eta^*)$, rescaled by mean IET $\langle \tau \rangle$ and calculated with \esref{eq:opt_scale}{eq:perc_scale} by using the empirical $P(\tau)$ and $P(k)$. Most datasets are in the coexistence phase $\eta_c < \eta^*$ (circles in light area). Color follows $B$ (same colormap as in \fref{fig1}). \textbf{e)} Time scale $\eta^* / \langle \tau \rangle$ as a function of burstiness $B$ for empirical datasets, and analytical prediction assuming a gamma 
    $P(\tau)$ (dot-dashed line) [for Gamma fits of $P(\tau)$ see \srefsi{SI:Gammafit}], for $\nu = 0.6$. Dots indicate datasets with a non-boundary maximum in $P^+$ (i.e. a point estimate for $\eta^*$), inverted T-shaped markers indicate censored lower bounds for $\eta^*$, and color follows $B$ (same colormap as in \fref{fig1}). Contour of dots indicates the level of openness $\Omega$ (see \fref{fig3}b). Open datasets have systematically larger values of $\eta^*$ than closed systems.}
    \label{fig2}
\end{figure}

Generative models of temporal and dynamical networks are useful tools to emulate structural and temporal patterns of interactions in complex systems and find mechanistic rules driving their evolution \cite{holme_modern_2015,holme_map_2023}. Modeling efforts have so far been mostly disconnected; dynamical network models focus on the evolution of nodes and links via mechanisms like stochastic growth \cite{zhang_random_2017}, measurable constraints \cite{krivitsky_separable_2014}, and edge rewiring \cite{asikainen_cumulative_2020,berner_adaptive_2023,starnini_opinion_2025}. In contrast, temporal network models concentrate on generating time-stamped events via a variety of mechanisms including multiplicative growth \cite{gautreau_microdynamics_2009}, activity reinforcement \cite{perra_activity_2012}, random walks \cite{zhang_characterizing_2015}, memory \cite{vestergaard_how_2014}, and Hawkes processes \cite{holme_self-exciting_2013}. These two approaches (dynamics of ongoing links or instantaneous contacts) can be mixed in models of link activation via renewal processes over static underlying networks \cite{kivela_estimating_2015,hiraoka_modeling_2020,unicomb_dynamics_2021}. In order to understand the origin of the time scale of maximal dynamicity $\eta^*$ and find an explicit link between temporal and dynamical networks, we explore the role of aggregation in a related and recently introduced model of networked renewal processes \cite{unicomb_dynamics_2021} (for a schematic definition see \fref{fig2}). We take a static, unipartite, undirected, and unweighted configuration-model network \cite{newman_networks_2018} of size $N$ with an arbitrary degree distribution $P(k)$, lacking any degree-degree correlations or mesoscopic structure. We generate instantaneous and time-stamped contact events over each edge according to an independent renewal process with IET distribution $P(\tau)$, during an observation period of length $T$ (\fref{fig2}a). The shape of $P(\tau)$ controls the level of temporal inhomogeneities in the system, while inducing no event correlations (within or across edges) beyond burstiness $B$. Our temporal network model is minimal, analytically tractable \cite{unicomb_dynamics_2021}, and represents a closed system where the time of the first event (the residual waiting time \cite{kivela_estimating_2015}) follows the complementary cumulative IET distribution $P_>(\tau_R) = \int_{\tau_R}^{\infty} P(\tau) d\tau$ (see MM and \srefsi{SI:analytical}).

The model allows for aggregation into dynamical networks in a similar fashion as with the empirical data above. We accumulate renewal process events over windows of length $\eta$ and fractional overlap $\nu$, and consider edges as active if there is at least one event in a given window. Since events are independent, the probability that an edge has no events within a window is $E_0(\eta) = \frac{1}{\langle \tau \rangle} \int_{\eta}^{\infty} P_{>}(\tau) d\tau$, a monotonically decreasing function of $\eta$ (other values of $E_j$ for $j \geq 0$ in \srefsi{SI:analytical}). Assuming a memoryless Poissonian process, the activation probability can then be approximated as $P^+(\eta) = E_0(\eta)[1-E_0(\eta[1-\nu])]$ (the probability that a window does not have events, and the remainder of the next window has at least one). In the more general case of a non-Poissonian renewal process, this approximation introduces a bias that scales with burstiness and overlap (for details see \srefsi{SI_sec:independence}). Since the renewal process is symmetric under time reversal, we also have $P^+ = P^-$. For overlap $\nu = 0$, $P^+$ is a logistic function that maximizes at $E_0(\eta^*) = 1/2$, while for $\nu \geq 0$ it has a stationary point $\eta^*$ given implicitly by
\begin{equation}
\label{eq:opt_scale}
\frac{P_{>}(\eta^*)}{E_0(\eta^*)} = \frac{(1 - \nu) P_{>}(\eta^* [1 - \nu])}{1 - E_0(\eta^* [1 - \nu])}.
\end{equation}
Matching empirical observations, the model shows maximal dynamicity at the optimal aggregation time scale $\eta^*$ given by \eref{eq:opt_scale}, which increases monotonically with burstiness $B$ ($B\approx 0$ or $0.6$ for an exponential or Gamma IET distribution, \fref{fig2}b). For given $\langle \tau \rangle$, the homogeneous case of exponential IETs gives $\eta^* / \langle \tau \rangle \in [\ln 2, 1]$ as we vary $\nu$, while the heterogeneous case of gamma-distributed IETs systematically gives larger $\eta^*$ values driven by the dispersion index of $P(\tau)$ [see \srefsi{SI:analytical} and \srefsi{SI:Gammafit} for Gamma fits of $P(\tau)$]. In other words, temporal networks with bursty event activity require more coarse-graining to show the same degree of dynamic behavior at the aggregated edge level.

Our temporal network model has a static underlying configuration-model network structure with degree distribution $P(k)$, over which independent renewal processes generate instantaneous events across time. By aggregating events in windows of length $\eta$, each edge is independently and temporally active with probability $E_+(\eta) = 1 - E_0(\eta)$, which increases with $\eta$. Thus, we expect an active edge percolation phase transition at a critical window size $\eta_c$, a second optimal aggregation scale characterized by the emergence of a largest connected component (LCC) of active links \cite{newman_networks_2018,unicomb_dynamics_2021} (\fref{fig2}c). For $\eta < \eta_c$, active edges form dynamic but finite components with exponentially distributed sizes smaller than $N \to \infty$ (even when the underlying network itself has a giant component). For increasing $\eta > \eta_c$, a growing and dynamic active LCC appears, which scales with $N$ and has a varying composition of nodes and edges percolating throughout the network. Following \cite{newman_spread_2002}, the average size $\langle s \rangle$ of an active connected component (excluding the active LCC) is given by $\langle s \rangle = 1 + [E_+(\eta) g_0'(1)] / [1 - E_+(\eta) g_1'(1)]$, where $g_0(x)$ and $g_1(x)$ are the generating functions of $P(k)$ and its excess degree, respectively. The active edge percolation critical point $\eta_c$ is given implicitly by the condition at which $\langle s \rangle$ diverges, that is,
\begin{equation}
\label{eq:perc_scale}
\frac{1}{\langle \tau \rangle} \int_0^{\eta_c} P_{>}(\tau) d\tau = \frac{\langle k \rangle}{\langle k^2 \rangle - \langle k \rangle},
\end{equation}
which relates temporal inhomogeneities of events [coded by $P(\tau)$] with features of the static underlying network [the first two moments of $P(k)$] (see \srefsi{SI:analytical}). The right-hand side of \eref{eq:perc_scale} is essentially the Molloy-Reed criterion of classic network percolation and epidemic spreading \cite{newman_networks_2018,pastor-satorras_epidemic_2015}, and the left-hand side is valid for all IET distributions (as long as edge renewal processes are independent). As static degree heterogeneities become dominant ($\langle k^2 \rangle \rightarrow \infty$), the percolation time scale $\eta_c$ goes to zero and an active LCC exists for any level of temporal aggregation.

Overall, the model shows that aggregating temporal network data necessarily implies the existence of a time scale $\eta^*$ where aggregated dynamical networks are maximally dynamic, and a time scale $\eta_c$ where such networks are also globally connected due to dynamic percolation. While temporal inhomogeneities in $P(\tau)$ increase $\eta^*$, degree heterogeneities in $P(k)$ decrease $\eta_c$. Thus, we expect a wide parameter regime where local dynamicity and global connectivity coexist,
\begin{equation}
\label{eq:univ_ineq}
\eta_c \leq \eta^*.
\end{equation}
To verify this prediction, we measure $P(\tau)$ and $P(k)$ from empirical data and calculate $\eta^*$ and $\eta_c$ via \esref{eq:opt_scale}{eq:perc_scale} (see \fref{fig2}d). A vast majority of datasets (37 out of 39, $95\%$) shows coexistence of local dynamicity and global connectivity, while only two are not maximally dynamic when the network percolates, and even those two are close to the critical line $\eta_c = \eta^*$. To assess whether the empirical agreement with \eref{eq:univ_ineq} is disproportionately driven by repeated sampling of closely related datasets, we perform a leave-one-family-out analysis: each of twelve families (Wikipedia, co-location data \cite{genois_can_2018}, Linux mailing lists, Last.fm \cite{celma_long_2010}, and eight standalone datasets, see \srefsi{sec:list_data}) is excluded in turn, and the coexistence percentage recomputed on the remaining datasets. The percentage remains strikingly high across all twelve exclusions ($92\%-95\%$, with mean $94\%$). Overall, the generic condition in \eref{eq:univ_ineq} seems to be applicable to a variety of systems widely ranging in their structural and temporal properties, suggesting a universal pattern in temporal networks relating time and structure via coarse-graining \cite{holme_modern_2015}.

A simplifying assumption in our model is that renewal processes are stationary and time-reversal symmetric, such that the distribution of residual times $P(\tau_R)$ is given by the IET CCDF $P_> (\tau)$, and (de-)activation probabilities are equal, $P^+ = P^-$. In such an ideally closed system, the dynamicity time scale $\eta^*$ increases with burstiness $B$ (shown as a dashed line in \fref{fig2}e). In empirical data (symbols in \fref{fig2}e), estimated values of $\eta^*$ do increase with $B$ but are systematically larger than the theoretical prediction (assuming gamma-distributed IETs). We attribute this discrepancy to the openness of empirical temporal networks. Indeed, defining openness $\Omega$ as the distance between the residual waiting time distribution $P(\tau_R)$ and the IET CCDF $P_> (\tau)$ (see \srefsi{parSI:openness}), the most open systems (symbols with no contour in \fref{fig2}e) show the largest dynamicity time scales relative to their theoretical expectations. This result is consistent with previous analyses of biases in IETs due to a short observation period $T$ \cite{kivela_estimating_2015}. For very open systems, $T$ might not be long enough to capture the maximum of $P^+$ via estimation, making the time scale $\eta^* > T$ unobservable due to lack of data (i.e. only boundary maxima are observed, leading to censored lower bounds rather than point estimates).

\subsection*{Microscopic self-regulation and macroscopic stability}

\begin{figure}[t]
    \centering
    \includegraphics[width=0.95\textwidth]{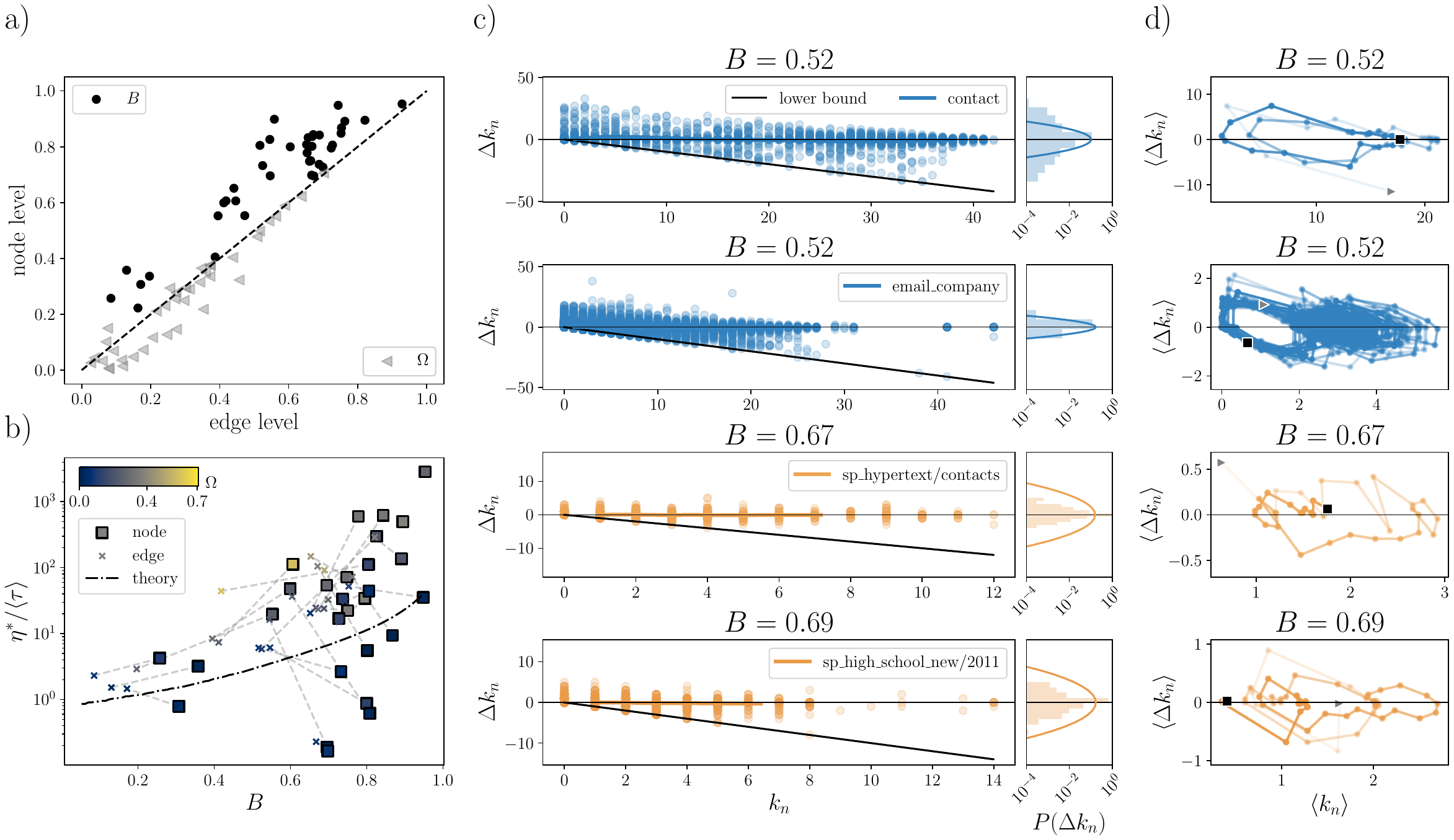} 
    \caption{\textbf{Periodic node degree changes in maximally dynamical networks}. \textbf{a)} Correlation between burstiness $B$ (circles) and openness $\Omega$ (triangles) values at the edge and node levels. With structural coarse-graining, burstiness increases and openness stays roughly constant or decreases. \textbf{b)} Rescaled time scale of maximal dynamicity $\eta^* / \langle \tau \rangle$ (only for datasets with point estimates for $\eta^*$, see \fref{fig2}e) as a function of burstiness $B$, calculated at both the edge (crosses) and node (squares) levels. Dots are colored by openness $\Omega$ and values for edge and node levels are connected by dashed lines. For most datasets, maximum dynamicity increases with $B$ and $\Omega$, regardless of level of structural coarse-graining, and is higher at the node level. \textbf{c)} Degree change $\Delta k_n = k_{n+1} - k_n$ between consecutive windows as a function of dynamic degree $k_n$, averaged over all corresponding nodes and windows, in selected datasets (all systems in \srefsi{sec:micro_regulation}). Inset: Distribution of degree change $P(\Delta k_n)$ (histograms), and theoretical prediction (lines) using \eref{eq:deg_change_binom} and the empirical $P(\tau)$ (see \srefsi{SI:analytical}). Under temporal and structural coarse-graining, nodes regulate their degree over time in cyclic patterns. \textbf{d)} Averages of degree change and dynamic degree, $\langle \Delta k_n \rangle$ and $\langle k_n \rangle$, computed over all nodes in a given pair of consecutive windows and plotted for increasing time for the same datasets as in (c) (all systems in \srefsi{SI:fig3c_supp}). Triangles/squares indicate the start/end of a dataset's trajectory. Results shown for overlap $\nu=0.6$ (b), $\nu = 0.2$ (c), and $\nu = 0.9$ (d) at the maximal dynamicity time scale $\eta^*$ in the node level. Dataset colors follow the burstiness colormap in \fref{fig1}. Periodic patterns in $( \Delta k_n, k_n)$ space have the largest amplitude in maximally dynamic networks, and gain granularity with $\nu$.}
    \label{fig3}
\end{figure}

The existence of a time scale of maximal dynamicity is also robust to structural coarse-graining of the temporal network (\fref{fig3}). To show this, we shift our perspective from edges to nodes and calculate the activation probability $P^+$ at the \textit{node level}, that is, the probability that a node has no active edges in window $n$ and has at least one event on window $n+1$, averaged over all nodes and consecutive time windows in the network for given $\eta$ and $\nu$. Increasing structural scale from edges to nodes potentially changes values of $P(\tau)$, $B$ and $\Omega$ for a given dataset, since IET $\tau$ is now the time between consecutive events concerning any edge of a node. Coarse-graining from edges to nodes tends to increase burstiness, in line with previous studies \cite{saramaki_seconds_2015,hiraoka_modeling_2020}, but might decrease openness slightly (as shown in \fref{fig3}a). When aggregating contact events over time at the nodel level, $P^+$ also has a maximum at a particular time scale $\eta^*$ that increases with node burstiness $B$ and openness $\Omega$ (see \fref{fig3}b). In other words, burstiness and openness continue to be strong predictors of maximal dynamicity under temporal aggregation, despite changes in $P(\tau)$ due to structural coarse-graining. In most systems, going from edge to node levels increases $\eta^* / \langle \tau \rangle$, meaning that networks become maximally dynamic in larger windows of event aggregation. 

Analyzing dynamical networks at the node level also allows us to explore persistence patterns in node neighborhoods \cite{saramaki_persistence_2014,iniguez_universal_2023}, particularly their size \cite{heydari_disentangling_2024}. We calculate the degree change of a node between consecutive windows $n$ and $n + 1$, $\Delta k_n = k_{n+1} - k_n$, averaged over all nodes with dynamic degree $k_n$ in some window and all such windows in the observation period (for results see \fref{fig3}c). Interestingly, all systems show signs of microscopic self-regulation at the node level: when degree $k_n$ is low in a window, nodes tend to gain edges in the next one ($\Delta k_n > 0$), and when degree $k_n$ is high, nodes tend to lose edges instead ($\Delta k_n < 0$). This persistent cyclic behavior is robust to the choice of aggregation parameters $\eta$ and $\nu$ and shows the largest magnitudes of $\Delta k_n$ at the time scale of maximal dynamicity $\eta^*$. Strikingly, some nodes populate the lower bound $\Delta k_n = -k_n$, implying a total loss of edges within consecutive aggregation windows. The distribution of degree changes $P(\Delta k_n)$ is symmetric around zero, with similar decays for both positive and negative values of $\Delta k_n$ (\fref{fig3}c, inset). This symmetry suggests a level of macroscopic, network-wide stability \cite{miller_constant_2019,gonzalez-casado_evidence_2025} that coexists with degree self-regulation; most nodes and pairs of consecutive windows show no degree change ($\Delta k_n = 0$), but the few that do cancel each other out and lead to a stable degree distribution (see \srefsi{SI:basic_stats}). According to the renewal process model of \eref{eq:opt_scale}, $P(\Delta k_n)$ has symmetric tails with exponential decay (see derivation in \srefsi{SI:analytical}). While a good fit for some networks, the model tends to predict a faster decay in $P(\Delta k_n)$, meaning that degree changes seen empirically are in general larger than theoretically expected. This discrepancy is likely related to the assumption of independence of renewal processes across edges in the model, since event sequences in empirical temporal networks are usually correlated \cite{karsai_universal_2012,gelardi_temporal_2021} [see $\Delta k_n$ and $P(\Delta k_n)$ for all datasets in \srefsi{sec:micro_regulation}].

In order to visualize such periodic regulation of node connectivity \cite{lahiri_mining_2008} due to temporal aggregation, we plot averages of degree change and dynamic degree, $\langle \Delta k_n \rangle$ and $\langle k_n \rangle$, computed over all nodes in a given pair of consecutive windows, with results shown in \fref{fig3}d for selected datasets (for all systems see \srefsi{SI:fig3c_supp}). As time advances, these averages form constrained, clockwise and roughly cyclic patterns in $(\Delta k_n, k_n)$ space that might move horizontally as well. As dynamical networks move across this space, the speed of their trajectory changes, with a systematic slow down close to the $\langle \Delta k_n \rangle = 0$ axis. Just like before, these periodic patterns show the largest changes at maximal dynamicity $\eta^*$, and gain granularity in degree change for large window overlap $\nu$. Some datasets show two-cycle trajectories (i.e. performing a cycle several times before moving horizontally onto another cycle, and going back), which we readily relate in \srefsi{SI:fig3c_supp} to weekday and weekend routines, in line with the extensive research on circadian rhythms in temporal networks \cite{jo_circadian_2012,aledavood_daily_2015}.

\section*{Discussion}
\label{sec:disc}

Temporal and dynamical networks are fundamental notions in the description of structural regularities in complex social systems. Temporal networks, often arising from pairwise interaction timeseries of information spreading or task solving, display correlated bursty behavior and circadian rhythms \cite{li_fundamental_2017,karsai_universal_2012,jo_circadian_2012}. Dynamical networks, in turn, originate from persistent yet slowly changing social ties and are driven by evolution mechanisms like homophily, closure, and contagion \cite{kossinets_origins_2009,asikainen_cumulative_2020}. Despite huge interest in these time-varying patterns of interactions and the associated vast literature \cite{holme_temporal_2012}, it has been difficult to find universal relations in temporal and dynamical networks that involve both time and structure \cite{holme_modern_2015}. This stands in contrast with the abundance of scaling laws in heavily aggregated complex systems data \cite{altmann_statistical_2024}, including Zipf-like frequency distributions in economics, linguistics and geology \cite{newman_power_2005}, power-law degree distributions in static networks \cite{newman_networks_2018}, gravitational laws in urban mobility \cite{barthelemy_structure_2016}, and stability-openness relations in rank dynamics \cite{iniguez_dynamics_2022}.

Here, we provide ample evidence for the existence of one such statistical relation between time and structure that becomes apparent in coarse-grained temporal network data. When aggregating interaction events into windows of variable length $\eta$ representing the time scale at which the system is studied, we find a time scale $\eta^*$ at which edges and nodes change their activity and connectivity patterns maximally, and a second time scale $\eta_c$ where dynamical percolation drives the network towards global connectivity. As coarse-graining increases, the vast majority of social and informational temporal networks studied reach percolation before maximal dynamicity ($\eta_c \leq \eta^*$), meaning they can keep high levels of local variability (including cyclic changes of node degree) while retaining system-wide cohesion and macroscopic stability. Via a simple model of temporal networks based on renewal processes, we show that this generic relation between time scales is solely driven by inhomogeneities in edge activity and node arrival times, and no other structural or temporal correlations. This suggests our results might be valid in other highly bursty and open networked systems (of biological, economic, and technological origin, for example) beyond those studied here.

Our results have intriguing consequences for the study of temporal networks and dynamical processes evolving on top of them, such as epidemic spreading, social contagion, and information transfer \cite{pastor-satorras_epidemic_2015,starnini_opinion_2025}. We expect the strongest coupling between underlying network structure and a spreading process to happen at the time scale of maximal dynamicity $\eta^*$, since smaller aggregation windows would reduce connectivity and larger ones would destroy the temporality of the network. Exploring existing datasets and models at the time scales unveiled by our framework may reveal additional effects that such strong coupling has on empirical spreading processes \cite{scholtes_causality-driven_2014,delvenne_diffusion_2015,unicomb_dynamics_2021}, such as the bias in diffusion dynamics due to aggregation reported for activity-driven networks \cite{ribeiro_quantifying_2013}. Recent results in mapping percolation and spreading processes might be useful here ~\cite{badie-modiri_directed_2022_1}. Similarly, exploring link prediction and network reconstruction problems \cite{lu_link_2011} close to maximal dynamicity can prove effective, since smaller or larger time scales might not detect links (and link turnover) that actually exist \cite{holme_network_2003,holme_epidemiologically_2013,kivela_estimating_2015}. While the simplicity of our coarse-graining framework highlights the general nature of the reported relationship between dynamicity and percolation time scales, our approach has limitations. Beyond our analysis, it would be fruitful to explore this phenomenon with the extensive range of tools to detect relevant time scales in temporal networks \cite{rosvall_mapping_2010,darst_detection_2016,gelardi_temporal_2021,andres_detecting_2024}. Our approach could be further applied to biological, technological, and other systems where temporal networks are relevant. Extensions of our temporal network model with arbitrary arrival times and link-link correlations might also explain the non-trivial dependency of $\eta^*$ with $B$, $\Omega$ and structural aggregation seen in \fref{fig2}e and \fref{fig3}b.

An interesting corollary of the generic interplay between structure and dynamics reported here is its interpretation in terms of the balance between robustness and adaptability in complex systems \cite{langton_computation_1990,hidalgo_cooperation_2016}. The small, fast, and numerous structural changes at the maximal dynamicity scale $\eta^*$ (providing adaptability) are offset by a temporal network that remains globally connected, giving robustness and preserving functionality. The fact that the inequality $\eta_c \leq \eta^*$ broadens with temporal inhomogeneities (as seen in \fref{fig2}b--c) further strengthens the notion that heterogeneity promotes critical-like behavior without fine-tuning \cite{sanchez-puig_heterogeneity_2023}, a signal of the evolutionary advantages of complex adaptive systems.

\section*{Material and Methods}

\small{\subsection*{Temporal network model}
\label{sec:methods}

We consider an undirected, unweighted static network of $N$ nodes as the underlying structure over which temporal interactions take place. The degree of a node (its number of neighbors in the static network) has discrete values $k = 0, 1, \ldots, N-1$ sampled from a given degree distribution $P(k)$. The static network might be the (undirected and unweighted) aggregate of an empirical temporal network over a long observation time $T$, or a synthetic, e.g., configuration-model network \cite{newman_networks_2018} with $P(k)$ measured from the data or having a parametrized functional form. Pairwise temporal interactions, or events, occur independently at random over each static edge via a renewal process \cite{whitt_approximating_1982,unicomb_dynamics_2021} with IET distribution $P(\tau)$. Time $t$ is continuous, events are instantaneous, and IETs $\tau > 0$ are uncorrelated, both along a single edge and across the $k$ edges surrounding a given node. We assume that the first IET ($\tau_R$) is distributed according to the complementary cumulative (e.g. residual or survival) distribution $P_{>}(\tau) = \int_{\tau}^{\infty} P(t)dt$, while every other IET $\tau_j$ with $j = 1, 2, \ldots$ follows $P(\tau)$.

We identify active edges in the dynamical network as those that have one or more events in a time window of size $\eta$. Thus, we need to compute the fraction $E_0$ of edges that have no event in the aggregation window, which is equal to the probability that the first IET $\tau_R$ is larger than the window size $\eta$, i.e.,
\begin{equation}
\label{eq:frac_0_edges1}
E_0(\eta) = \frac{1}{\langle \tau \rangle} \int_{\eta}^{\infty} P_{>}(\tau) d\tau.
\end{equation}
Now, consider two consecutive aggregation windows of size $\eta$ and fractional overlap $\nu \in [0, 1]$, such that the part of the second window not contained in the first has length $\omega = \eta (1 - \nu)$. As seen above, the probability of having no events in the first window (including the overlap) is $E_0(\eta)$, while the probability of having one or more events in the second window (without the overlap) is $1 - E_0(\omega)$. Thus, we can approximate the probability $P^+$ that an edge gets activated (i.e. goes from having no events to having at least one, across consecutive windows) as
\begin{equation}
\label{eq:act_prob1}
    P^{+}(\eta) = E_0(\eta) [1 - E_0(\eta [1 - \nu])],
\end{equation}
only exact for Poisson processes.
}

{\small\subsection*{Time scale of maximal dynamicity}

To find the optimal aggregation time scale $\eta^*$ at which the activation probability $P^{+}$ gets maximized, we take the derivative of Eq.~\ref{eq:act_prob1} with respect to $\eta$ (for constant but arbitrary $\nu$), evaluate at $\eta^*$ and equate to zero. Using Eq.~\ref{eq:frac_0_edges1}, we find that the optimal window length is given implicitly by
\begin{equation}
\label{eq:opt_scale1}
\frac{P_{>}(\eta^*)}{E_0(\eta^*)} = \frac{(1 - \nu) P_{>}(\eta^* [1 - \nu])}{1 - E_0(\eta^* [1 - \nu])},
\end{equation}
from which we might be able to explicitly derive $\eta^*$ for particular functional forms of the IET distribution $P(\tau)$. From Eq.~\ref{eq:opt_scale1}, the maximum activation probability $P^{\pm}_m = P^{\pm}(\eta^*)$ is
\begin{equation}
\label{eq:act_prob_max1}
    P^{\pm}_m = (1 - \nu) \frac{P_{>}(\eta^* [1 - \nu])}{P_{>}(\eta^*)} E_0^2(\eta^*),
\end{equation}
which reduces to $P^{\pm}_m = 1/4$ for $\nu = 0$, and has a limit $P^{\pm}_m \to 0$ for $\nu \to 1$, as expected.

We study archetypal examples of homogeneous and heterogeneous IET distributions. 
As a homogeneous IET distribution, we consider exponentially distributed IETs. In this case, we show in SI~\ref{SI:analytical} that $E_0(\eta) = P_>(\eta)$ is satisfied, implying that the fraction of inactive edges decays exponentially with the window length $\eta$ and yielding a maximally dynamic scale $\eta^* / \mu = \ln 2$ for $\nu = 0$ and $\eta^* / \mu \to 1$ for $\nu \to 1$.
As a heterogeneous IET distribution, we consider Gamma distributed IETs with parameters $\alpha, \beta$, mean $\mu = \alpha \beta$ and dispersion index $D =\sigma^2/\mu = \beta$. As heterogeneity increases in the temporal network (regulated by a growing dispersion $D$), the fraction of inactive edges:
\begin{equation}
    E_0(\eta) = P_>(\eta) + \frac{\eta}{\nu} (D P(\eta) - P_>(\eta)),
\end{equation}
can remain relatively large for increasing $\eta$, leading to an optimal window length $\eta^*$ larger than the homogeneous baseline at mean IET $\mu$.

For empirical data, we compute the optimal time scale of dynamic aggregation in two distinct ways: with a mixed analytical and empirical approach, or purely empirically. The `mixed' approach to computing the optimal time scale of dynamic aggregation (used in Fig.~\ref{fig2}d) relies on the exploration of a range of values of $\eta$ through a divide-and-conquer algorithm~\cite{cormen_introduction_2022} with the goal of minimizing the difference between the left-hand-side and right-hand-side of Eq.~\ref{eq:opt_scale1}, with a tolerance of $10^{-6}$, and using the empirical IET distribution of the temporal network considered for $P_{>}(\tau)$. We also use this mixed empirical and analytical method to find the analytical expectation of the time scale of global connectivity (used in Fig.~\ref{fig2}d) and the distribution of degree change (see the rest of the MM and SI~\ref{SI:analytical}). The purely `empirical' approach consists of finding the value of $\eta$ that maximizes the mean edge activation probability, where this mean is computed empirically over a set of geometrically spaced values of $\eta$. For a rolling time window of width $\eta$ and overlap $\nu$, the activation probability of a single edge is computed as the count of changes from a window with zero events to a following window with more than one event, normalized by the number of rolling windows in the entire observation period of the dataset. Then, the mean edge activation probability is the average of activation probabilities over all edges in the empirical network.

}

{\small\subsection*{Time scale of global connectivity}

For a static configuration-model network with degree distribution $P(k)$ where each edge is independently and temporally active with probability $E_+(\eta) = 1 - E_0(\eta)$ within an aggregation window of length $\eta$, we expect an edge percolation phase transition at a critical window size $\eta_c$ characterized by the emergence of a largest connected component (LCC) of active links \cite{unicomb_dynamics_2021}. For $\eta < \eta_c$, active edges form dynamic but finite components with exponentially distributed sizes far smaller than network size as $N \to \infty$ (even when the static network itself has a giant component). For increasing $\eta > \eta_c$, we see instead a growing and dynamic active LCC that scales with network size, with a varying composition of nodes and edges percolating throughout the network. Following \cite{newman_spread_2002}, the average size $\langle s \rangle$ of an active connected component (excluding the active LCC) can be written as
\begin{equation}
\label{eq:av_size_mm}
\langle s \rangle = 1 + \frac{g_0'(1; \eta)}{1 - g_1'(1; \eta)} = 1 + \frac{ E_+(\eta) g_0'(1)}{1 - E_+(\eta) g_1'(1)},
\end{equation}
where $g_0(x) = \sum_{k = 0}^{\infty} P(k) x^{k}$ is the generating function of the static degree distribution, $g_0(x; \eta)$ is the generating function of the dynamic degree distribution; $g_1(x) = g_0'(x) / \langle k \rangle$ is the generating function of the static excess degree distribution, with $g_0'$ a derivative with respect to the dummy variable $x$, such that $g_0'(1) = \langle k \rangle$; and $g_1(x; \eta)$ is the generating function of the dynamic excess degree distribution.
Since Eq.~\ref{eq:av_size_mm} diverges for $\eta_c$ such that $E_+(\eta_c) g_1'(1) = 1$, the edge percolation critical point is given implicitly by
\begin{equation}
\label{eq:perc_scale1}
E_0(\eta_c) = 1 - \frac{\langle k \rangle}{\langle k^2 \rangle - \langle k \rangle},
\end{equation}
with $\langle k^n \rangle = \sum_{k = 0}^{\infty} k^n P(k)$ the n-th moments of the static degree distribution.
Eq.~\ref{eq:perc_scale1} relates properties of inter-event times on the left with properties of the aggregated network on the right. As static degree heterogeneity increases ($\langle k^2 \rangle \to \infty$), $\eta_c$ vanishes and we expect an active LCC for any window size.

We study archetypal examples of homogeneous and heterogeneous degree distributions.
For the homogeneous case, we consider Poisson distributed degrees. We show in SI~\ref{SI:analytical}, that the size of the LCC is governed by the following implicit equation, which we solve numerically:
\begin{equation}
    S_{\eta} = 1 - e^{-\langle k \rangle (1 - E_0(\eta)) S_{\eta}}.
\end{equation}
For the heterogeneous case, we consider a Zipfian power-law distribution of degrees. In this case, the size of the LCC solves the following:
\begin{equation}
    \begin{cases}
        u (u - E_0(\eta)) \zeta(\gamma - 1) - (1-E_0(\eta)) Li_{\gamma - 1}(u)& = 0\\
        S_{\eta} &= 1 - \frac{Li_{\gamma}(u)}{\zeta (\gamma)},
    \end{cases}
\end{equation}
where $\zeta(\gamma)$ is the Riemann Zeta function and $Li_\gamma(x)$ is the Polylogarithm of order $\gamma$.

}

{\small \subsubsection*{Data and code availability}
Code is publicly available at \url{https://github.com/gdm2019/temporal-dynamical-networks}. For data availability see \srefsi{SI:datasets}.

\subsubsection*{Acknowledgments}
G.d.M. and G.I. thank Javier Ureña-Carrión for valuable suggestions. G.d.M. acknowledges funding from a Magnus Ehrnroothin s\"{a}\"{a}ti\"{o} postdoctoral fellowship. M.K. acknowledges support from the MOMA WWTF project; the BEQUAL NKFI-ADVANCED grant (No. 153172); and the COLINE DUT European Partnership project (F-DUT-2023-0037).

\subsubsection*{Author contributions}
G.d.M., M.K., and G.I. conceived, designed, and developed the study.  G.d.M. analyzed empirical data.  G.I. derived analytical results.  G.d.M. and G.I. performed numerical simulations and model fitting.  G.d.M., M.K., and G.I. wrote the paper.

\subsubsection*{Competing interest statement}
All authors declare no competing interest.

\printbibliography

\newpage

\input{supplementary}

\end{document}

%% file: supplementary.tex
\begin{center}
{\LARGE Supplementary Information for}\\[0.7cm]
{\Large \textbf{Universal time scales linking structure and dynamics in temporal networks}}\\[0.5cm]

{\large G. de Meijere$^*$, M. Karsai, G. Iñiguez$^*$}\\[0.7cm]
{\small $^*$Corresponding author email: giulia.demeijere@tuni.fi, gerardo.iniguez@tuni.fi}\\[1cm]
\end{center}

\addtocontents{toc}{\protect\setstretch{0.1}}
\tableofcontents

\newpage

\FloatBarrier
\section{Temporal network data}
\label{SI:datasets}

In this section, we describe the 39 social and informational temporal network datasets that we gathered to form the corpus for our analysis.
We considered all the datasets present in the Netzschleuder open repository \cite{netzschleuder} that have a `time' edge attribute.
For datasets where multiple language versions exist (\texttt{edit\textunderscore wiktionary}, \texttt{edit\textunderscore wikinews}, \texttt{edit\textunderscore wikibooks}), we focused on the three language datasets with the largest number of edges.

\subsection{Data pre-processing}
\label{sec:preproc}

For some datasets, we manually pre-processed the data by selecting only a portion of the available observation period.
The motivation for limiting ourselves to part of the observation period is in some cases (several SocioPatterns datasets) grounded in the presence of distinct periods of data collection (relatively long periods of time where none of the edges is active). We split these datasets into distinct subdatasets for each of the distinct observation periods. We only show results for the subdataset that contains the largest number of edges (after pre-processing on edge weight, see below).
For other datasets, the reason to focus on a narrower observation period is due to significant heterogeneity in the levels of edge activity across the observation period. We thus remove the initial or final periods when they are characterized by significantly lower edge activity than in the rest of the observation period.

\begin{figure}
    \centering
    \includegraphics[width=15cm]{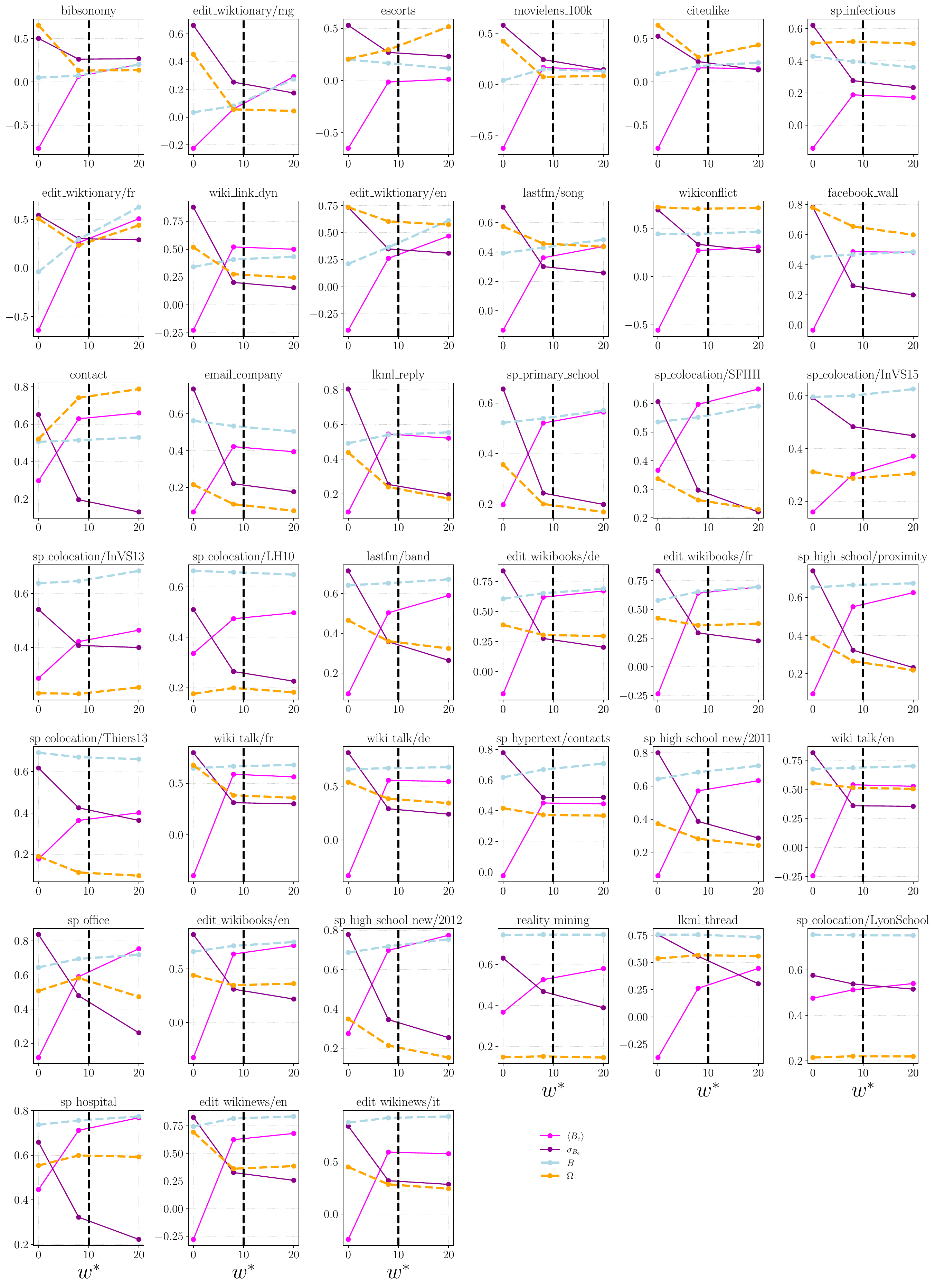} 
    \caption{\textbf{Effect of the weight cutoff $w^*$ on the burstiness, and openness of the empirical networks.} Mean and standard deviation of the edge burstiness, system burstiness, and openness for increasing cutoff $w^*$ on the edge weight, for the 39 datasets. The datasets are ordered by increasing system (edge) burstiness. While edge properties (continuous lines) stabilize for most datasets around the cut-off value of $w^* = 10$, system properties (dashed lines) remain roughly constant.}
    \label{fig:w_star_effect}
\end{figure}

Moreover, to reduce noise in the edge properties, the edges associated with a weight strictly lower than $w^*$ are removed from the empirical networks. 
We choose $w^* = 10$ as the cut-off point for all the datasets because the mean edge burstiness (and its standard deviation, see \sref{SI:basic_stats} for the definitions of burstiness) tends to stabilize around $w^*=10$, as demonstrated in \frefsi{fig:w_star_effect}, for all datasets.
At the same time, the system burstiness remains fairly stable.
Additionally, \frefsi{fig:w_star_effect} shows that although the agreement between the complementary cumulative distribution of the IET and the distribution of residual waiting times, measured through the openness $\Omega$ (see \sref{SI:basic_stats} for its definition) remains generally unchanged across values of $w^*$, it sometimes improves (yielding a lower $\Omega$). 
After the pre-processing based on edge weight, we filtered out all datasets that were left with a number of edges $E$ lower than $E^* = 100$.

Overall, the three steps of pre-processing (observation period, edge weight, number of edges) rule out 17 datasets, including all the technological and economic datasets.
Additionally, we do not show results for the \texttt{dblp\textunderscore coauthor} dataset which has a temporal granularity of 1 year and for the \texttt{edit\textunderscore wikinews/sr} dataset where edges die out very soon after being formed.

\subsection{List of datasets}
\label{sec:list_data}

Here below are brief descriptions of the 39 datasets considered, 15 of which are bipartite networks.

\subsubsection{Unipartite temporal networks}

\paragraph{Physical proximity - \texttt{sp\textunderscore colocation}:}
Network of physical proximity (co-location) across various contexts from the SocioPatterns datasets
\cite{sociopatterns, sp_colocation}.
Physical proximity was measured through wearable sensors containing RFID devices that exchange low-power radio packets. 

\begin{itemize}
\item /LH10: The data was gathered over 3 days in a hospital in Lyon (France), in 2010. We split the dataset into three datasets, for each of the three days. We show results for the first day only.
    \item /InVS13: The data was gathered on a workplace over 2 weeks in 2013. We split the dataset into ten datasets, for each of the ten working days. We show results for the first day only.
    \item /InVS15: The data was gathered on a workplace over 2 weeks in 2015. We split the dataset into ten datasets, for each of the ten working days. We show results for the seventh day only.
    \item /LyonSchool: The data was gathered in a primary school in Lyon (France), over 2 days in 2009. We split the dataset into two datasets, for each of the two days. We show results for the first day only.
    \item /Thiers13: The data was gathered in a high school in Marseilles (France), over 1 week in 2013. We split the dataset into five datasets, for each of the 5 days of school. We show results for the third day only.
    \item /SFHH: The data was gathered during a scientific conference in Nice (France), over 2 days in 2009. We split the dataset into two datasets, for each of the two days. We show results for the first day only.
\end{itemize}

\paragraph{Physical proximity in a hospital - \texttt{sp\textunderscore hospital}:}
Dataset of physical proximity (face to face interactions) in the geriatric unit of a university hospital in Lyon (France). The data was gathered over 4 days in 2010 by the SocioPatterns collaboration \cite{sp_hospital}. 
Close-range interactions are recorded with a spatial resolution of about 1.5 meters and a temporal resolution of 20 seconds. 
We here-after refer to this way of measuring short-range proximity as the `SocioPatterns protocol'. The data recorded physical contacts among healthcare workers, among patients, and between patients and health-care workers. The study included 46 health-care workers and 29 patients. Most contacts were detected during daytime.

\paragraph{Physical proximity in an office - \texttt{sp\textunderscore office}:} Dataset of physical proximity (face to face interactions) in an office building in France measured over two weeks in 2013 \cite{sp_office}. Physical proximity is recorded with the Sociopatterns protocol.
We split the dataset into two datasets, for each of the two weeks. We show results for the first week only.

\paragraph{Physical proximity at a conference - \texttt{sp\textunderscore hypertext/contacts}:} 
Dataset of physical proximity (face to face interactions) among attendees of the ACM Hypertext 2009 conference, interacting over 2.5 days \cite{sp_hypertext_infectious}. Conference attendees volunteered to wear radio badges that monitored their face-to-face proximity. 
We split the dataset into three datasets, for each of the three days. We show results for the second day only.

\paragraph{Physical proximity in a high school - \texttt{sp\textunderscore high\textunderscore school\textunderscore new}:}
Dataset of physical proximity (face to face interactions) in a high school in Marseilles (France)
\cite{sp_high_school_new}.
Physical proximity is recorded with the Sociopatterns protocol.

\begin{itemize}
    \item /2011: Data recorded close-range interactions between the students of three classes over 4 days in 2011. We split the dataset into four datasets, for each of the four days. We show results for the first day only.
    \item /2012: Data recorded close-range interactions between the students of 5 classes over 7 days (from a Monday to the Tuesday of the following week) during 2012. We split the dataset into two datasets, for each of the two weeks. We show results for the first week only.
\end{itemize}

\paragraph{Physical proximity in a primary school - \texttt{sp\textunderscore primary\textunderscore school}:} 
Dataset of physical proximity (face to face interactions) between students and teachers at a primary school in Lyon (France) over two consecutive days in 2009 \cite{sp_primary_school}.
Physical proximity is recorded with the Sociopatterns protocol.
We split the dataset into two datasets, for each of the two days. We show results for the second day only.

\paragraph{Physical proximity in a high school - \texttt{sp\textunderscore high\textunderscore school/proximity}:} 
Network of physical proximity (face to face interactions) between the students of nine classes in a high school, over 5 days in 2013
\cite{sp_high_school_proximity}. Physical proximity is recorded with the Sociopatterns protocol.
We split the dataset into five datasets, for each of the five days. We show results for the second day only.

\paragraph{Physical proximity at an event - \texttt{sp\textunderscore infectious}:}
Network of physical proximity (face to face interactions) collected during an event at the artscience exhibition INFECTIOUS: STAY AWAY that took place at the Science Gallery, in Dublin (Ireland) in 2009 \cite{sp_hypertext_infectious}. Physical proximity is recorded with the Sociopatterns protocol.

\paragraph{Physical proximity in a university - \texttt{reality\textunderscore mining}:}
Dataset of physical proximity (Bluetooth proximity) between students at the Massachusetts Institute of Technology (MIT), as measured by personal mobile phones between September 2004 and May 2005 \cite{reality_mining}.
Physical proximity is recorded when two devices make a Bluetooth handshake with a temporal resolution of 5 minutes and a spatial resolution of 7-10 meters.

\paragraph{Physical proximity in a university - \texttt{contact}:}
Network of physical proximity (WiFi proximity) among freshmen students at U.C. San Diego in 2002. 
Physical proximity is inferred from wireless network availability traces (WiFi) collected from HP Jornada devices that the freshmen use as their personal digital assistants.
\cite{contact2, contact}.
We consider the first 3 days of data.

\paragraph{Email communication among Linux developers - \texttt{lkml\textunderscore reply}:}
Network of emails exchanged between individuals (identified by their email addresses) in the Linux kernel mailing list
\cite{lkml}.

\paragraph{Email communication in a company - \texttt{email\textunderscore company}:}
Network of emails sent among employees of a mid-sized manufacturing company \cite{email_company}.

\paragraph{Communication on Facebook - \texttt{facebook\textunderscore wall}:}
Network of friendships and interactions (wall posts) for a subset of the Facebook social network in 2009, recorded between October 2004 and January 2009
\cite{facebook_wall}.

\paragraph{Conflicts between Wikipedia editors - \texttt{wikiconflict}:}
Network of editorial conflicts among editors of the English Wikipedia
\cite{wikiconflict}.

\paragraph{Online communication between Wikipedia editors - \texttt{wiki\textunderscore talk}:} 
Network of communication between Wikipedia editors using the editors' talk pages \cite{lkml}.
\begin{itemize}
    \item /de: The data was extracted from the German Wikipedia. We consider the last 10 years of data. 
    \item /fr: The data was extracted from the French Wikipedia. We consider the last 10 years of data. 
    \item /en: The data was extracted from the English Wikipedia.
\end{itemize}

\paragraph{Hyperlinks between Wikipedia pages - \texttt{wiki\textunderscore link\textunderscore dyn}:} 
Network of hyperlinks between Wikipedia articles starting from 2011
\cite{wikilinkdyn}.
We consider the last 6 years of data.

\subsubsection{Bipartite temporal networks}

\paragraph{Online purchase of sexual intercourse - \texttt{escorts}:}
Bipartite buyer-seller network of purchase of sexual intercourse with a female escort on an online Brazilian forum
\cite{escorts}. The data is recorded with a daily temporal resolution and over a period of six years. Edges indicate the rating of sexual intercourse and the two types of nodes are sex buyers and sex sellers. The dataset covers the beginning of the online forum, spanning the period September 2002 through October 2008.

\paragraph{Online rating of movies - \texttt{movielens\textunderscore 100k}:}
Bipartite user-movie network of movie ratings on the Movielens online movie recommendation service \cite{movielens_100k}. Edges indicate movie ratings and the two types of nodes are the service users and the movies.

\paragraph{Online edits on Wikipedia pages - \texttt{edit\textunderscore wikinews}:}
Bipartite user-page network of edits of Wikipedia pages by editors \cite{wiki}.
\begin{itemize}
    \item /it: The data was extracted from the Italian Wikipedia.
    \item /en: The data was extracted from the English Wikipedia.
\end{itemize}

\paragraph{Online edits on Wikipedia pages - \texttt{edit\textunderscore wikibooks}:} 
Bipartite user-page network of edits of Wikipedia pages about books
\cite{wiki}.
\begin{itemize}
    \item /fr: The data was extracted from the French Wikipedia. We consider the last 12 years of data. 
    \item /de: The data was extracted from the German Wikipedia. We consider the last 12 years of data. 
    \item /en: The data was extracted from the English Wikipedia. We consider the last 11 years of data.
\end{itemize}

\paragraph{Listenings of music on an online platform - \texttt{lastfm}:}
Bipartite network of listenings on the music website last.fm.
\cite{lastfm}

\begin{itemize}
    \item /band: The two types of nodes are users and music bands. We consider the last 3 years of data.
    \item /song: The two types of nodes are users and songs. We consider the last 5 years of data.
\end{itemize}

\paragraph{Contributions of Linux developers - \texttt{lkml\textunderscore thread}:} 
Bipartite user-thread network of contributions of individuals in the Linux kernel mailing list (identified by an email address) to threads
\cite{lkml}.
We consider the last 8 years of data.

\paragraph{Tag assignment to scientific publications on online platform - \texttt{citeulike}:}
Bipartite tag-publication network of tag assignments to publications by users of Citeulike, a defunct bookmarking platform for scientific papers
\cite{citeulike}.

\paragraph{Tag assignment to publications on online platform - \texttt{bibsonomy}:} 
Bipartite tag-publication network of tag assignments to publications by users of Bibsonomy, a collaborative social bookmarking and publication sharing system
\cite{bibsonomy}.
We consider the last  4 years of data.

\paragraph{Online edits on Wikipedia pages - \texttt{edit\textunderscore wiktionary}:} 
Bipartite user-page network of edits of Wikipedia pages by editors \cite{wiki}.
\begin{itemize}
    \item /fr: The data was extracted from the French Wikipedia. We consider the last 11 years of data. 
    \item /mg: The data was extracted from the Malagasy Wikipedia. We consider the last 6 years of data. 
    \item /en: The data was extracted from the English Wikipedia.
\end{itemize}

\begin{table}
    \centering
    \small
    \begin{tabular}{lcccccccc
}
         \textbf{Name}&  \textbf{Type}&  \textbf{Period $T$}&\textbf{Nodes $N$}&  \textbf{Events $V$}& \textbf{Edges $E$}\\
          \texttt{bibsonomy}& I. & 4.0 y & 124 (67, 57) & 1491 & 111 \\
          \texttt{edit\textunderscore wiktionary/mg}& I. & 6.0 y & 67228 (66963, 265) & 1035664 & 69212 \\
         \texttt{escorts} & S., offline & 5.4 y & 219 (116, 103) & 2307 & 151  \\
         \texttt{movielens\textunderscore 100k} & I. & 3.0 y & 324 (162, 162) & 3898 & 288 \\
          \texttt{citeulike}& I. & 3.3 y & 231 (102, 129) & 3221 & 208 \\
         \texttt{sp\textunderscore infectious}& S., offline & 2.6 m & 9074 & 338995 & 9723 \\
          \texttt{edit\textunderscore wiktionary/fr}& I.& 11.0 y & 29144 (28193, 951) & 723221 & 39477 \\
          \texttt{wiki\textunderscore link\textunderscore dyn}& I. &6.0 y & 8587 & 309520 & 14653 \\
          \texttt{edit\textunderscore wiktionary/en}& I. & 16.3 y & 82706 (2864, 79842) & 2767701 & 134583 \\
          \texttt{lastfm/song}& S., online & 4.3 y & 179235 (885, 178350) & 9911013 & 430004 \\
         \texttt{wikiconflict}& S., conflict & 7.2 y & 7834 & 255635 & 13798 \\
         \texttt{facebook\textunderscore wall}& S., online & 4.3 y & 12884 & 450921 & 17636 \\
         \texttt{contact} & S., offline & 3.0 d & 67 & 24676 & 833  \\
          \texttt{email\textunderscore company}& S., comm. & 8.9 m & 141 & 76476 & 1138 \\
         \texttt{lkml\textunderscore reply}& S., comm. & 8.0 y &  3526 & 742752 & 19951 \\
         \texttt{sp\textunderscore primary\textunderscore school}& S., offline & 8.5 h & 237 & 53225 & 1541 \\
         \texttt{sp\textunderscore colocation/SFHH}& S., offline & 11.2 h & 396 & 794208 & 29099 \\         
         \texttt{sp\textunderscore colocation/InVS15}&S., offline  & 12.0 h & 218 & 352254 & 5764 \\
         \texttt{sp\textunderscore colocation/InVS13}&S., offline  & 11.4 h & 94 & 93507 & 1080 \\
         \texttt{sp\textunderscore colocation/LH10}& S., offline & 22.0 h & 66 & 51012 & 507 \\
         \texttt{lastfm/band}& S., online & 3.0 y & 49163 (943, 48220) & 14804440 & 207728 \\
         \texttt{edit\textunderscore wikibooks/de} & I. & 12.0 y & 8072 (6206, 1866) & 220997 & 8343 \\  
         \texttt{edit\textunderscore wikibooks/fr}& I. & 12.0 y & 5130 (1081, 4049) & 171454 & 5188 \\   
         \texttt{sp\textunderscore high\textunderscore school/proximity} & S., offline & 9.0 h & 300 & 42215 &  619 \\
         \texttt{sp\textunderscore colocation/Thiers13} & S., offline & 9.7 h & 327 & 3688551 & 11403 \\
          \texttt{wiki\textunderscore talk/fr}& S. comm. & 10.0 y & 10636 & 1469645 & 35345 \\
          \texttt{wiki\textunderscore talk/de}& S. comm. & 10.0 y & 17868 & 3724010 & 62368 \\
         \texttt{sp\textunderscore hypertext/contacts}& S., offline & 11.6 h & 91 & 5083 & 171 \\
         \texttt{sp\textunderscore high\textunderscore school\textunderscore new/2011}& S., offline & 10.6 h &  103 & 8556 & 189 \\
          \texttt{wiki\textunderscore talk/en}& S., comm. & 14.3 y & 81758 & 11493842 & 258393 \\
         \texttt{sp\textunderscore office}& S., offline & 4.4 d & 70 & 3599 & 114 \\
         \texttt{edit\textunderscore wikibooks/en}& I. & 11.0 y &  27798 (11243, 16555) & 862021 & 26839 \\
         \texttt{sp\textunderscore high\textunderscore school\textunderscore new/2012} & S., offline & 4.4 d &  169 & 28852 & 483 \\
          \texttt{reality\textunderscore mining} & S., offline  & 7.6 m & 94 & 1083906 & 1664 \\
          \texttt{lkml\textunderscore thread}& S., comm. & 8.0 y & 14336 (2846, 11490) & 329328 & 19338 \\
          \texttt{sp\textunderscore colocation/LyonSchool} & S., offline & 8.8 h & 237 & 3503893 & 21624 \\
          \texttt{sp\textunderscore hospital}& S., offline  & 4.0 d & 72 & 30065 & 505 \\
          \texttt{edit\textunderscore wikinews/en}& I. & 12.7 y & 11520 (9964, 1556) & 721747 & 15085 \\
         \texttt{edit\textunderscore wikinews/it}& I. & 12.3 y &  1554 (1317, 237) & 773193 & 1756 \\    
    \end{tabular}
    \caption{Basic information on the 39 datasets: type of network, total observation period, number of nodes (total and for each of the two node groups, in case of bipartite networks), number of events and number of distinct undirected edges. \textit{S.} stands for Social, and \textit{I.} for informational. \textit{comm.} is short for communication. The dataset corpus spans a wide range of dataset types (social online/offline/communication or informational, and uni- or bi-partite), observation periods, as well as network size and volume.
    }
    \label{tab:net_info}
\end{table}

In order to analyze all datasets in the same way, we consider all networks to be undirected. \trefsi{tab:net_info} summarizes a few basic properties of the 39 pre-processed networks, including the entire observation period (ranging from 8.5 hours to 16.3 years), the number of nodes (from 66 to 179235), the number of events (from 1491 to 14804440) and the number of distinct undirected edges present in the dataset (from 111 to 430004).

\subsection{Basic statistics}
\label{SI:basic_stats}

\begin{table}
    \centering
    \scriptsize
    \begin{tabular}{lccccccccc}
         \textbf{Name}& \textbf{$\langle k \rangle$} & \textbf{$\langle d \rangle$} & \textbf{$\langle w \rangle$} & \textbf{$\langle \tau \rangle$} & \textbf{$B$} & \textbf{$\langle B_e \rangle$} & $\beta$ & $\rho$ & $\Omega$\\
          \texttt{bibsonomy}& 1.79 & 24.05 & 13.43 & 2.7 m & 0.08 & 0.07 & 1.06 & -1.18 & 0.07\\
          \texttt{edit\textunderscore wiktionary/mg}& 2.06 & 30.81 & 14.96 & 3.6 m & 0.13 & 0.10 & 1.00 & -1.08 & 0.03\\
          \texttt{escorts}& 1.38 & 21.07 & 15.28 & 1.4 m & 0.16 & 0.01 & 1.08 & -1.33 & 0.54\\
          \texttt{movielens\textunderscore 100k}& 1.78 & 24.06 & 13.53 & 2.3 m & 0.17 & 0.20 & 1.06 & -1.02 & 0.04\\
          \texttt{citeulike}& 1.80 & 27.89 & 15.49 & 1.7 m & 0.20 & 0.17 & 1.08& -1.13 & 0.27\\
          \texttt{sp\textunderscore infectious}& 2.14 & 74.72 & 34.87 & 1.2 min & 0.39 & 0.19 & 0.82 & -2.59 & 0.52\\
          \texttt{edit\textunderscore wiktionary/fr}& 2.71 & 49.63 & 18.32 & 3.0 m & 0.40 & 0.32 & 1.07 & -1.35 & 0.30\\
          \texttt{wiki\textunderscore link\textunderscore dyn}& 3.41 & 72.09 & 21.12 & 1.9 m & 0.41 & 0.51 & 1.01 & -1.07 & 0.24\\
          \texttt{edit\textunderscore wiktionary/en}& 3.25 & 66.93 & 20.57 & 2.1 m & 0.42 & 0.28 &1.07 & -1.20 & 0.59\\
          \texttt{lastfm/song}& 4.80 & 110.59 & 23.05 & 3.6 w & 0.44 & 0.38 & 0.99& -1.16 & 0.44\\
         \texttt{wikiconflict}& 3.52 & 65.26 & 18.53 & 2.9 w & 0.45 & 0.28 & 1.03 & -1.26 & 0.70\\
         \texttt{facebook\textunderscore wall}& 2.74 & 70.00 & 25.57 & 2.4 w & 0.47 & 0.49 & 1.06 & -1.18 & 0.64\\
         \texttt{contact}& 24.87 & 736.60 & 29.62 & 1.8 h & 0.52 & 0.64 & 0.96 & -1.05 & 0.08\\
         \texttt{email\textunderscore company}& 16.14 & 1084.77 & 67.20 & 3.3 d & 0.52 & 0.41 & 0.98 & -1.01 & 0.09\\
        \texttt{lkml\textunderscore reply}& 11.32 & 421.30 & 37.23 & 1.6 m & 0.55 & 0.54 & 1.02 & -1.02 & 0.22\\
        \texttt{sp\textunderscore primary\textunderscore school}& 13.00 & 449.16 & 34.54 & 9.3 min & 0.55 & 0.53 & 1.00 & -1.08 & 0.08\\
        \texttt{sp\textunderscore colocation/SFHH}& 146.96 & 4011.15 & 27.29 & 10.5 min & 0.56 & 0.61 &0.98 & -1.09 & 0.26\\
         \texttt{sp\textunderscore colocation/InVS15}& 52.88 & 3231.69 & 61.11 & 5.4 min & 0.60 & 0.32 & 0.98 & -1.06 & 0.25\\
         \texttt{sp\textunderscore colocation/InVS13}& 22.98 & 1989.51 & 86.58 & 7.0 min & 0.65 & 0.44 & 0.96 & -1.01 & 0.06\\
         \texttt{sp\textunderscore colocation/LH10}& 15.36 & 1545.82 & 100.62 & 3.6 min & 0.65 & 0.48 & 0.96 & -1.21 & 0.46\\          
         \texttt{lastfm/band}& 8.45 & 602.26 & 71.27 & 6.4 d & 0.66 & 0.53 & 1.00 & -1.06 & 0.34\\
          \texttt{edit\textunderscore wikibooks/de}& 2.07 & 54.76 & 26.49 & 4.2 w & 0.66 & 0.64 & 1.01 & -1.29 & 0.30\\ 
          \texttt{edit\textunderscore wikibooks/fr}& 2.02 & 66.84 & 33.05 & 3.8 w & 0.66 & 0.66 & 0.97& -1.23 & 0.36\\
         \texttt{sp\textunderscore high\textunderscore school/proximity}& 4.13 & 281.43 & 68.20 & 4.5 min & 0.67 & 0.57 & 0.96 & -1.06 & 0.18\\ 
         \texttt{sp\textunderscore colocation/Thiers13}& 69.74 & 22559.94 &  323.47 & 1.3 min & 0.67 & 0.38 & 0.99 & -1.00 & 0.08\\
         \texttt{wiki\textunderscore talk/fr}& 6.65 & 276.35 & 41.58 & 2.5 w & 0.67 & 0.59 & 1.02& -1.11 & 0.37\\
         \texttt{wiki\textunderscore talk/de}& 6.98 & 416.84 & 59.71 & 2.3 w & 0.67 & 0.54 & 1.04 & -1.07 & 0.37\\
         \texttt{sp\textunderscore hypertext/contacts}& 3.76 & 111.71 & 29.73 & 8.3 min & 0.67 & 0.48 & 0.92 & -1.32 & 0.31\\
         \texttt{sp\textunderscore high\textunderscore school\textunderscore new/2011}& 3.67 & 166.14 & 45.27 & 5.4 min & 0.69 & 0.55 & 0.85 & -1.21 & 0.21\\
         \texttt{wiki\textunderscore talk/en}& 6.32 & 281.17 & 44.48 & 2.1 w & 0.69 & 0.54 &1.03 & -1.13 & 0.51\\
         \texttt{sp\textunderscore office}& 3.26 & 102.83 & 31.57 & 1.4 h & 0.70 & 0.64 & 0.99 & -1.29 & 0.27\\
         \texttt{edit\textunderscore wikibooks/en}& 1.93 & 62.02 & 32.12 & 2.1 w & 0.72 & 0.66 & 1.01 & -1.28 & 0.35\\
         \texttt{sp\textunderscore high\textunderscore school\textunderscore new/2012}& 5.72 & 341.44 & 59.73 & 58.1 min & 0.73 & 0.71 & 1.00 & -1.10 & 0.15\\
         \texttt{reality\textunderscore mining}& 35.40 & 23061.83 & 651.39 & 16.0 h & 0.74 & 0.54 & 0.94 & - 0.90 & 0.12\\
        \texttt{lkml\textunderscore thread}& 2.70 & 45.94 & 17.03 & 4.7 d & 0.75 & 0.33 & 1.01 & -1.53 & 0.56\\   
        \texttt{sp\textunderscore colocation/LyonSchool}& 182.48 & 29568.72 & 162.04 & 2.0 min & 0.75 & 0.51 & 0.96 &-1.02 & 0.12\\
        \texttt{sp\textunderscore hospital}& 14.03 & 835.14 & 59.53 & 32.9 min & 0.76 & 0.73 & 0.96 & -1.23 & 0.35\\
        \texttt{edit\textunderscore wikinews/en}& 2.62 & 125.30 &  47.85 & 5.6 d & 0.82 & 0.65 & 0.89 & -1.31 & 0.37\\
        \texttt{edit\textunderscore wikinews/it}& 2.26 & 995.10 & 440.31 & 16.0 h & 0.93 & 0.59 & 0.27& -1.44 & 0.27\\
\end{tabular}
    \caption{Basic statistics of 39 datasets from the Netzschleuder repository: mean (static) degree $k$, mean (static) weight $w$, mean inter-event time, system burstiness, mean edge burstiness, the fitted exponent $\beta$ in the relationship $\langle d \rangle = \langle w \rangle \cdot k^\beta$ between node strength $d$ and node degree $k$, the fitted exponent $\rho$ in the relationship $\langle \tau \rangle = T \cdot \langle w \rangle^\rho$ between average edge inter-event time $\langle \tau \rangle$ and average edge weight $\langle w \rangle$, and the openness $\Omega$. The datasets are ordered by increasing system (edge) burstiness.}
    \label{tab:net_statistics}
\end{table}

In this section, we characterize the 39 datasets of our corpus through a few standard statistics summarized
in \trefsi{tab:net_statistics} for each dataset.

\begin{figure}
    \centering
    \includegraphics[width=15cm]{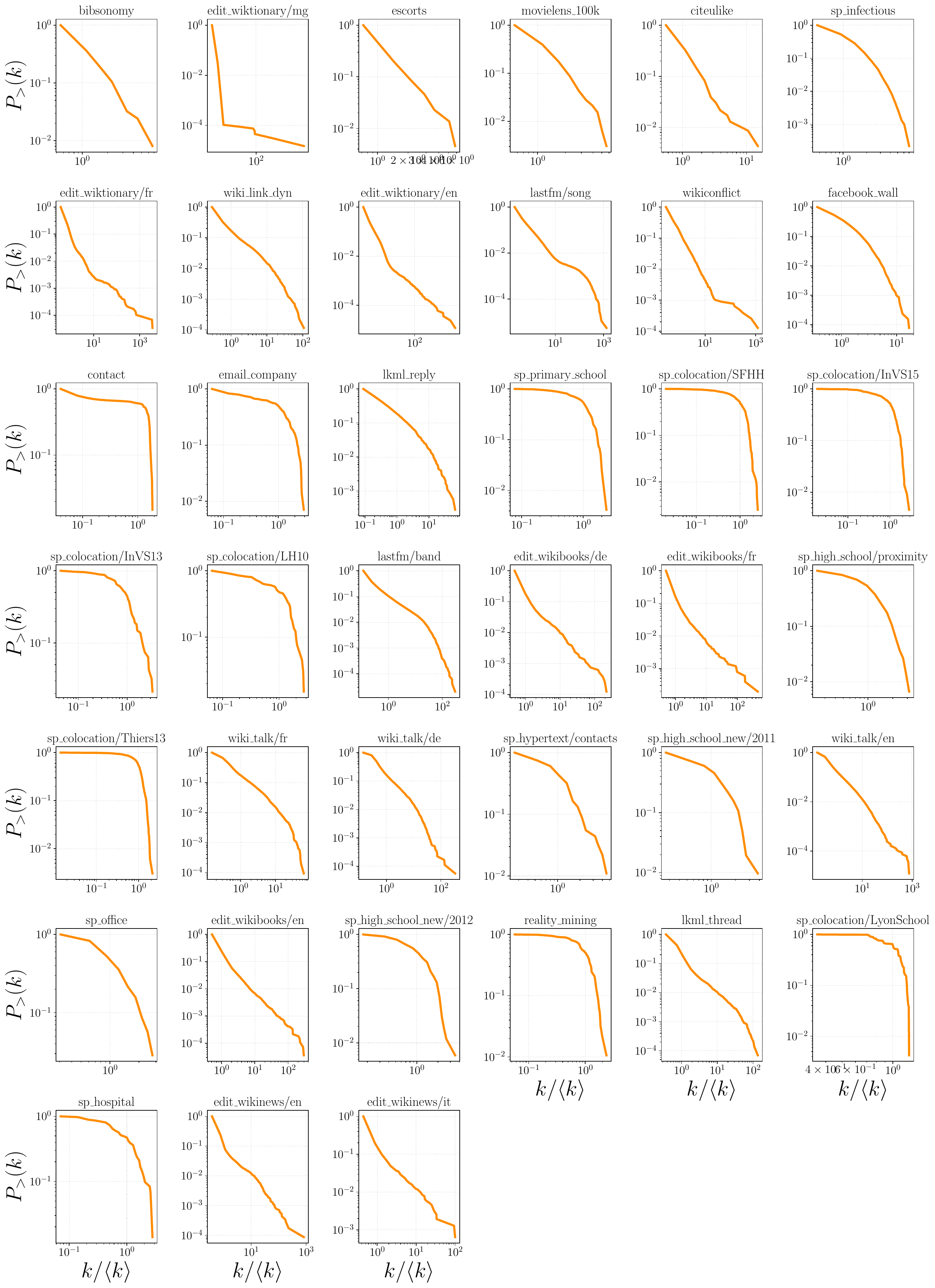}     \caption{Complementary cumulative distribution function of the fully-aggregated node degree, for the 39 datasets. The datasets are ordered by increasing system (edge) burstiness. We observe some level of heterogeneity in node degree within and across systems.}
    \label{fig:node_degrees}
\end{figure}

\begin{figure}
    \centering
    \includegraphics[width=15cm]{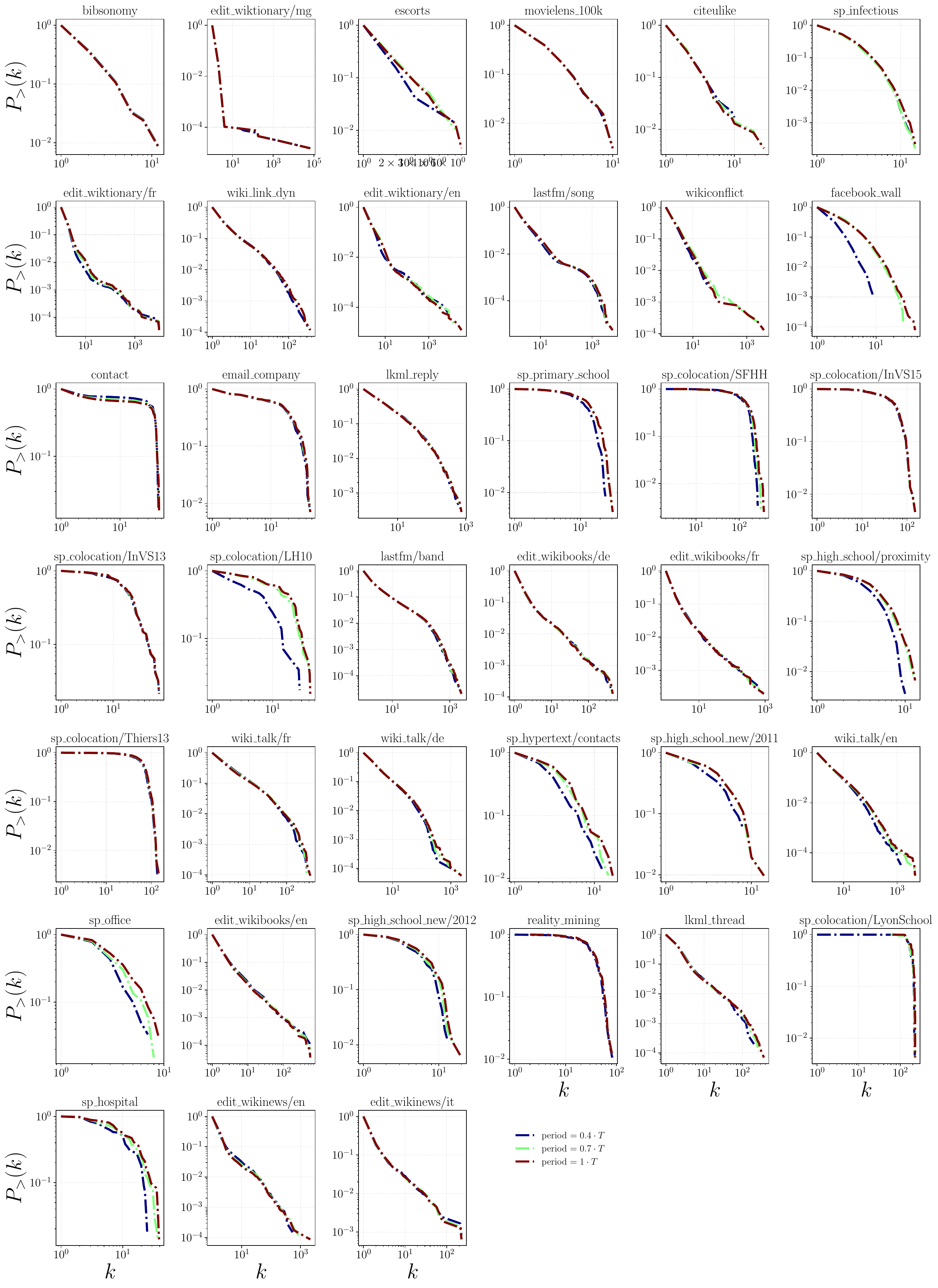} 
    \caption{Complementary cumulative distribution function of the fully-aggregated node degree, for the 39 datasets and for three different choices of the observation period. The datasets are ordered by increasing system (edge) burstiness. The degree distribution appears to be relatively stable.}
    \label{fig:node_degrees_windows}
\end{figure}

\paragraph{\textbf{Node degree:}} 
We define the static (fully aggregated) degree $k$ of a node as the number of undirected connections that exhibit at least one event during the entire observation window $T$ of the temporal network dataset.
In \frefsi{fig:node_degrees}, we show the complementary cumulative distribution of the static degree of nodes. The average static degree ranges from $1.4$ for the \texttt{escorts} dataset to $182.5$ for the \texttt{sp\textunderscore colocation/LyonSchool} dataset. Distributions of static degree at most cover three orders of magnitude, which mainly happens for the three studied language versions of the \texttt{edit\textunderscore wikibooks} datasets.
In \frefsi{fig:node_degrees_windows} we also show that the shape of the distribution remains remarkably stable across observation periods considered.

\begin{figure}
    \centering
    \includegraphics[width=15cm]{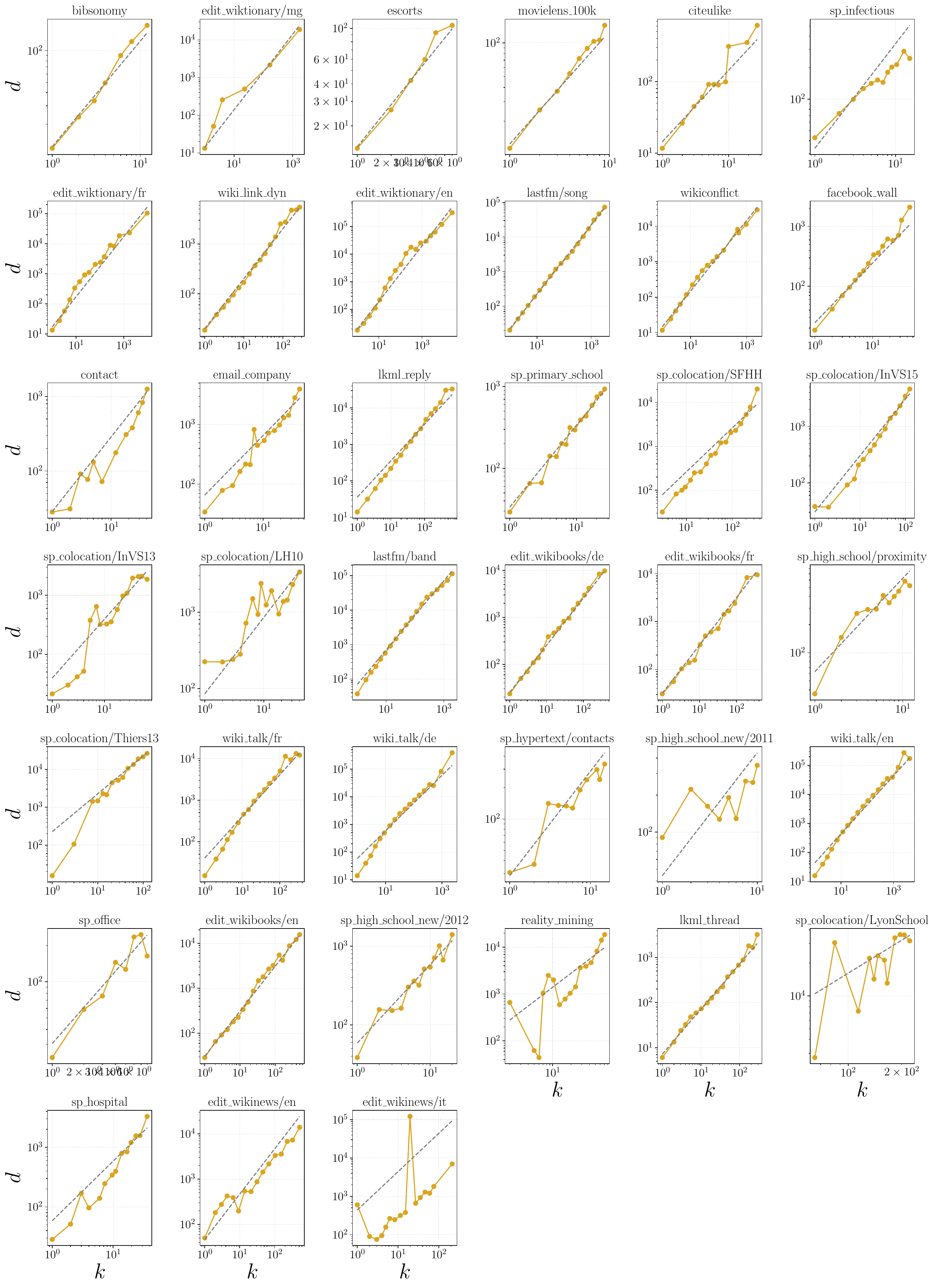} 
    \caption{Correlations between fully-aggregated node strength $d$ and fully-aggregated node degree $k$, in the 39 datasets. The dashed line represents the linear increase with slope $\langle w \rangle$. Most datasets exhibit a linear increase strikingly similar to the uncorrelated expectation. The datasets are ordered by increasing system (edge) burstiness.}
    \label{fig:strength_degree}
\end{figure}

\paragraph{\textbf{Node strength-degree correlations:}}
We define the static (fully aggregated) strength $d$ of a node as the number of events occurring along the edges incident to the node over the entire observation window $T$ of the temporal network dataset. 
In \frefsi{fig:strength_degree}, we plot the behavior of node strength as a function of node degree. The empirical behavior is compared to the linear increase of strength with degree which is expected in the absence of correlations. The data behaves surprisingly close to linearly, suggesting that the correlations between strength and degree are in fact weak.
Assuming a dependence $\langle d \rangle = \langle w \rangle \cdot k^\beta$ of the strength on the degree, with $\langle w \rangle$ the average edge weight, we fit the exponent $\beta$. The average fitted exponent $\beta$ across the 39 datasets of the corpus is very close to the uncorrelated value $1$ ($0.97$), with a relatively small standard deviation ($0.13$).
 
\begin{figure}
    \centering
    \includegraphics[width=15cm]{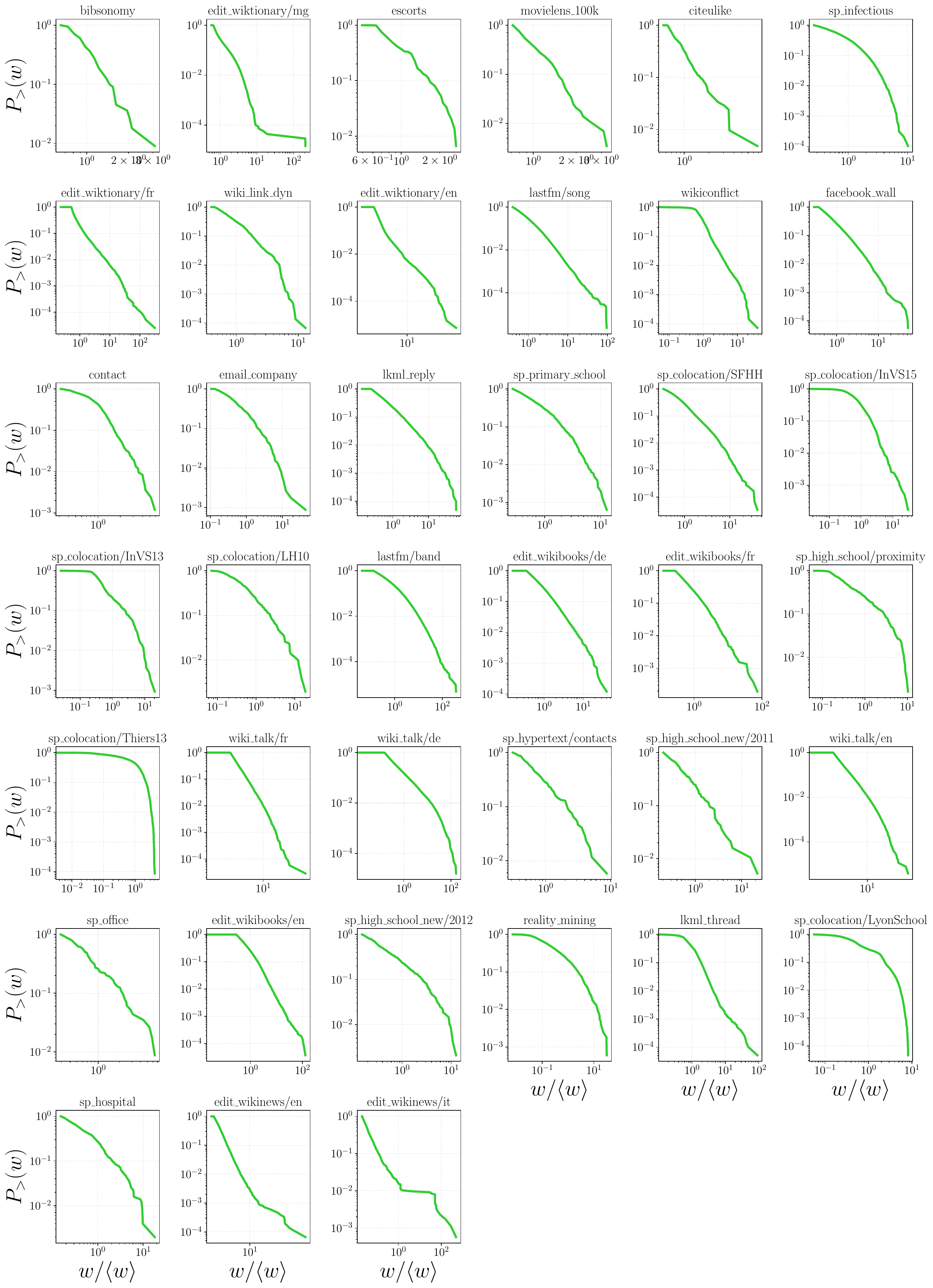} 
    \caption{Complementary cumulative distribution function of the (fully-aggregated) edge weights $w$, for the 39 datasets. The datasets are ordered by increasing system (edge) burstiness.}
    \label{fig:edge_weights}
\end{figure}

\paragraph{\textbf{Edge weight:}} 
We define the static (fully aggregated) weight $w$ of an undirected edge as the number of events occurring along the edge over the entire observation window $T$ of the temporal network dataset. 
In \frefsi{fig:edge_weights}, we plot the complementary cumulative distribution of weights.
The average weight ranges from $13.43$ for the \texttt{bibsonomy} dataset to $651.39$ for the \texttt{reality\textunderscore mining} dataset.

\begin{figure}
    \centering
    \includegraphics[width=15cm]{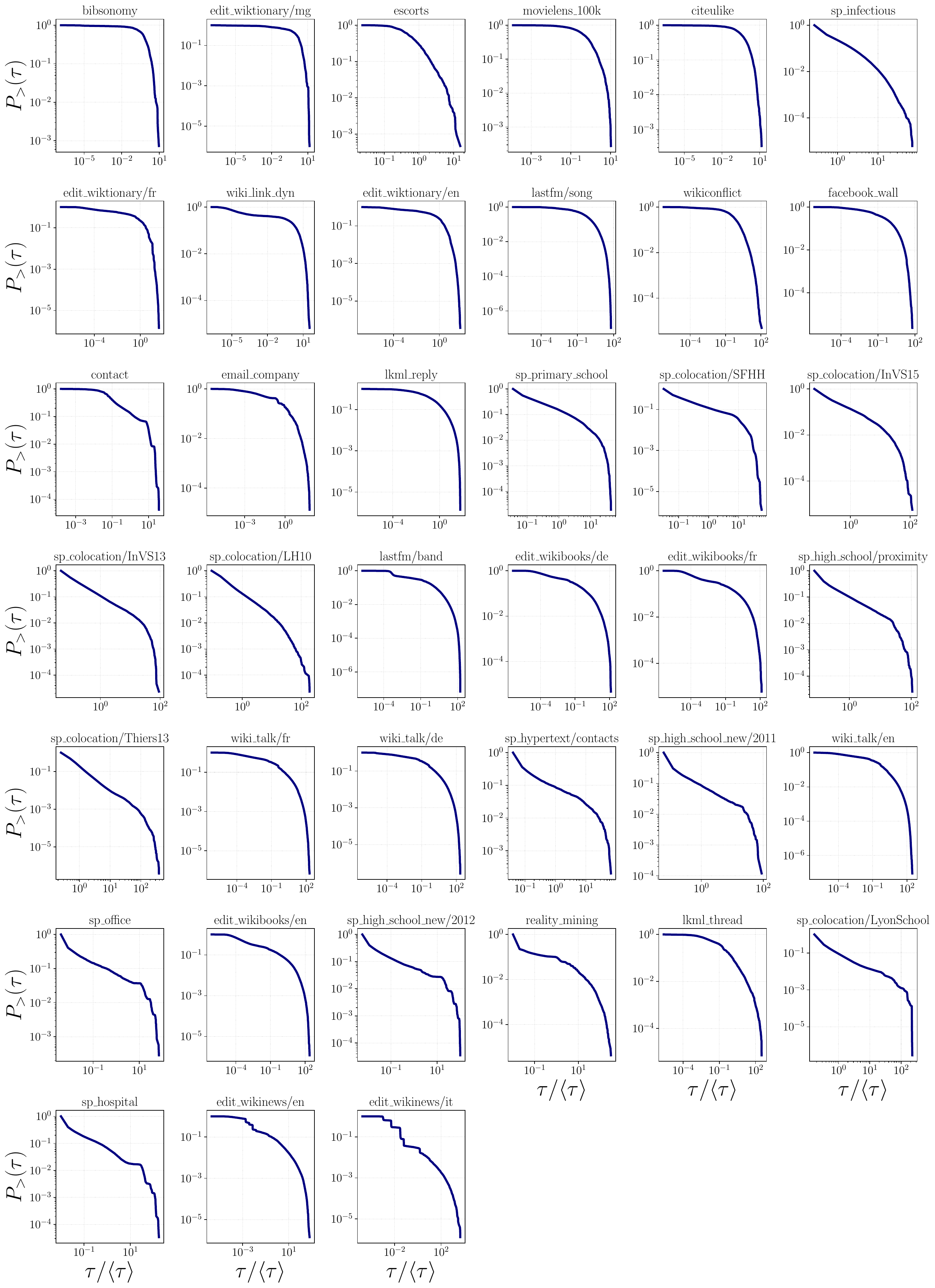} 
    \caption{Complementary cumulative distribution function of the edge inter-event times (concatenated interevent times computed along undirected edges), for the 39 datasets. While most of the offline social interaction datasets exhibit a power-law regime and an exponential tail, the remaining datasets hardly exhibit any power-law regime. The datasets are ordered by increasing system (edge) burstiness.}
    \label{fig:IETs}
\end{figure}

\begin{figure}
    \centering
    \includegraphics[width=15cm]{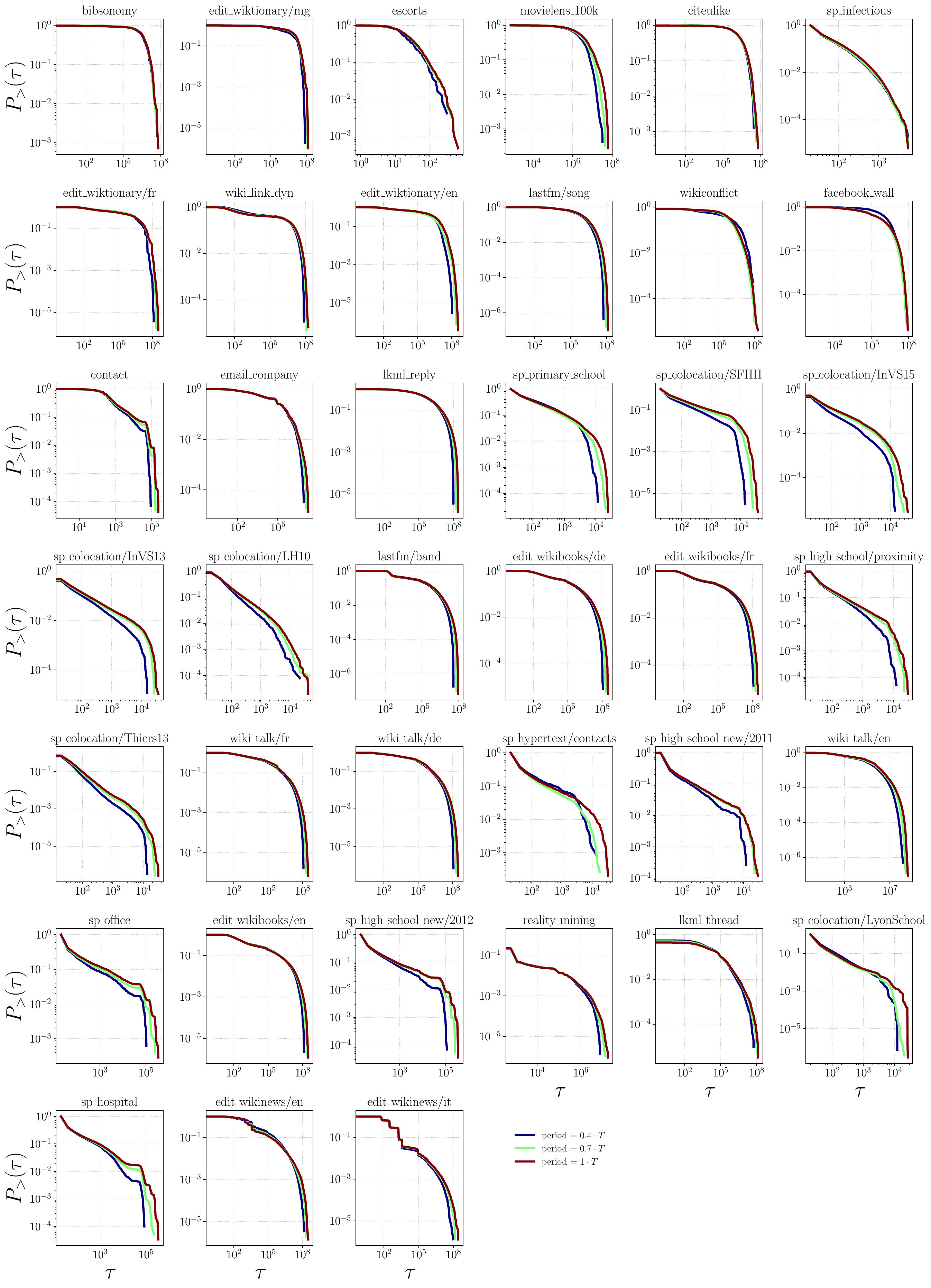} 
    \caption{Complementary cumulative distribution function of the edge inter-event times (concatenated interevent times computed along undirected edges), for the 39 datasets and for three different choices of the observation period. The shape of the IET distribution remains fairly robust to the choice of the observation period. The datasets are ordered by increasing system (edge) burstiness.}
    \label{fig:IETs_period}
\end{figure}

\paragraph{\textbf{Inter-event time:}} 
At the edge level, we define an inter-event time (IET) as the time between successive events along an undirected edge of the temporal network dataset. 
The average edge IET ranges from $1.2$ minutes for the \texttt{sp\textunderscore infectious} dataset to $3.6$ months for the \texttt{edit\textunderscore wiktionary/mg} dataset.
In \frefsi{fig:IETs}, we plot the complementary cumulative distribution of edge inter-event times $\tau$.
While most of the offline social interaction datasets exhibit a power-law regime and an exponential tail, the remaining datasets hardly exhibit any power-law regime. 
In \frefsi{fig:IETs_period}, we also show that the shape of the edge IET distribution remains fairly robust to the choice of the observation period.

\begin{figure}
    \centering
    \includegraphics[width=15cm]{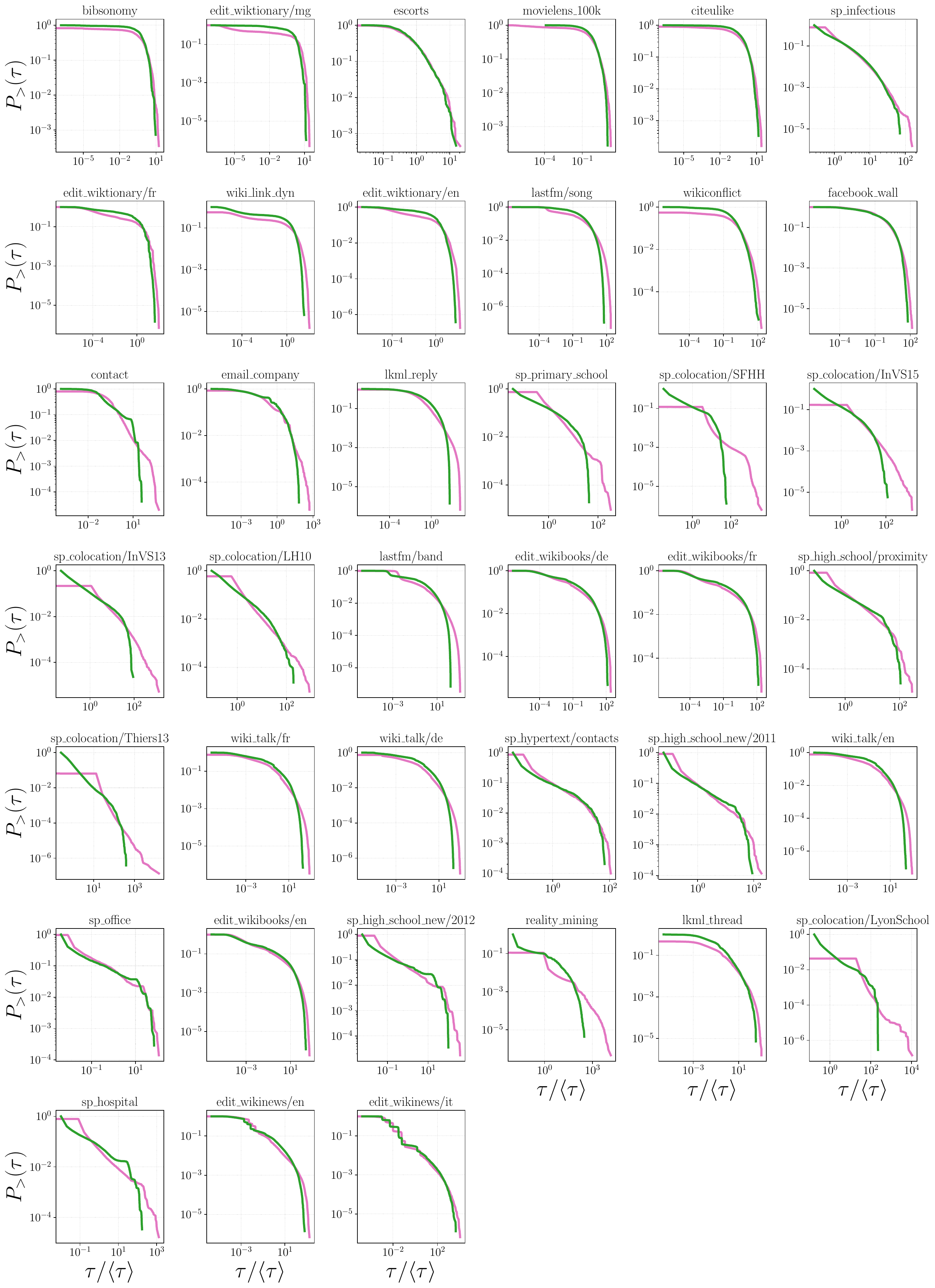} 
    \caption{Comparison between the complementary cumulative distribution of edge-level and node-level IETs (the node IET is renormalised by the average IET at the node level). The distributions of IETs vary with the topological coarse-graining from edge to node level. However, the shapes of the distributions are roughly maintained. pink: node-level, green: edge-level. The datasets are ordered by increasing system (edge) burstiness.}
    \label{fig:edge_vs_node_iet}
\end{figure}

Coarse-graining from the edge level to the node level, we can define the node inter-event time as the time between two successive ordered events along any of the edges of a node.
This coarse-graining affects the IET distribution, as illustrated in \frefsi{fig:edge_vs_node_iet}.

\begin{figure}
    \centering
    \includegraphics[width=15cm]{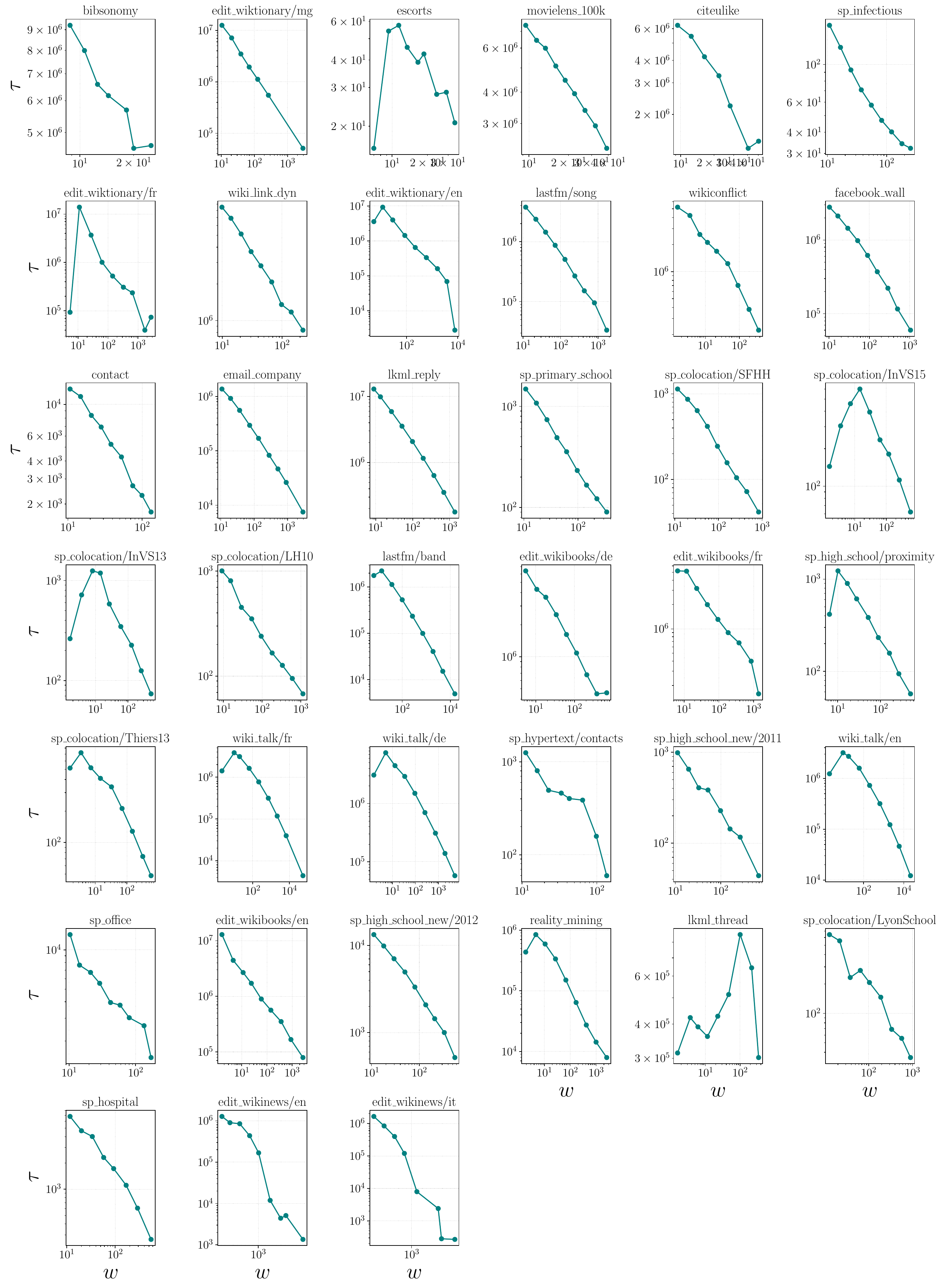} 
    \caption{Correlations between the average IET and the weight, for the 39 datasets. Across datasets in the corpus, the dependence is not too far from the uncorrelated expectation of $\langle \tau \rangle = T/\langle w \rangle$. The datasets are ordered by increasing system (edge) burstiness.}
    \label{fig:IET_weight}
\end{figure}

\paragraph{\textbf{Edge inter-event time-weight correlations:}} In \frefsi{fig:IET_weight}, we plot the decreasing mean edge inter-event time with edge weight. 
Assuming the relationship $\langle \tau \rangle = T \cdot (\langle w \rangle)^\rho$ between mean edge IET and weight, we can fit the exponent $\rho$. Across datasets in the corpus, the fitted $\rho$ is not too far from the uncorrelated expectation of $-1$. On average we indeed find a $-1.20$ fitted exponent, with a standard deviation of $0.26$. 

\paragraph{\textbf{Burstiness:}} 
\label{parSI:burstiness}

To compute the heterogeneity of the inter-event time distribution, we use the standard burstiness index~\cite{goh2008burstiness}:
\begin{equation}
    B = \frac{r - 1}{r + 1},
\end{equation}
where $r$ is the coefficient of variation $r:=\sigma_{\tau}/\langle \tau \rangle$, i.e. the ratio between the standard deviation of the inter-event time distribution and its mean.

\begin{figure}
    \centering
    \includegraphics[width=15cm]{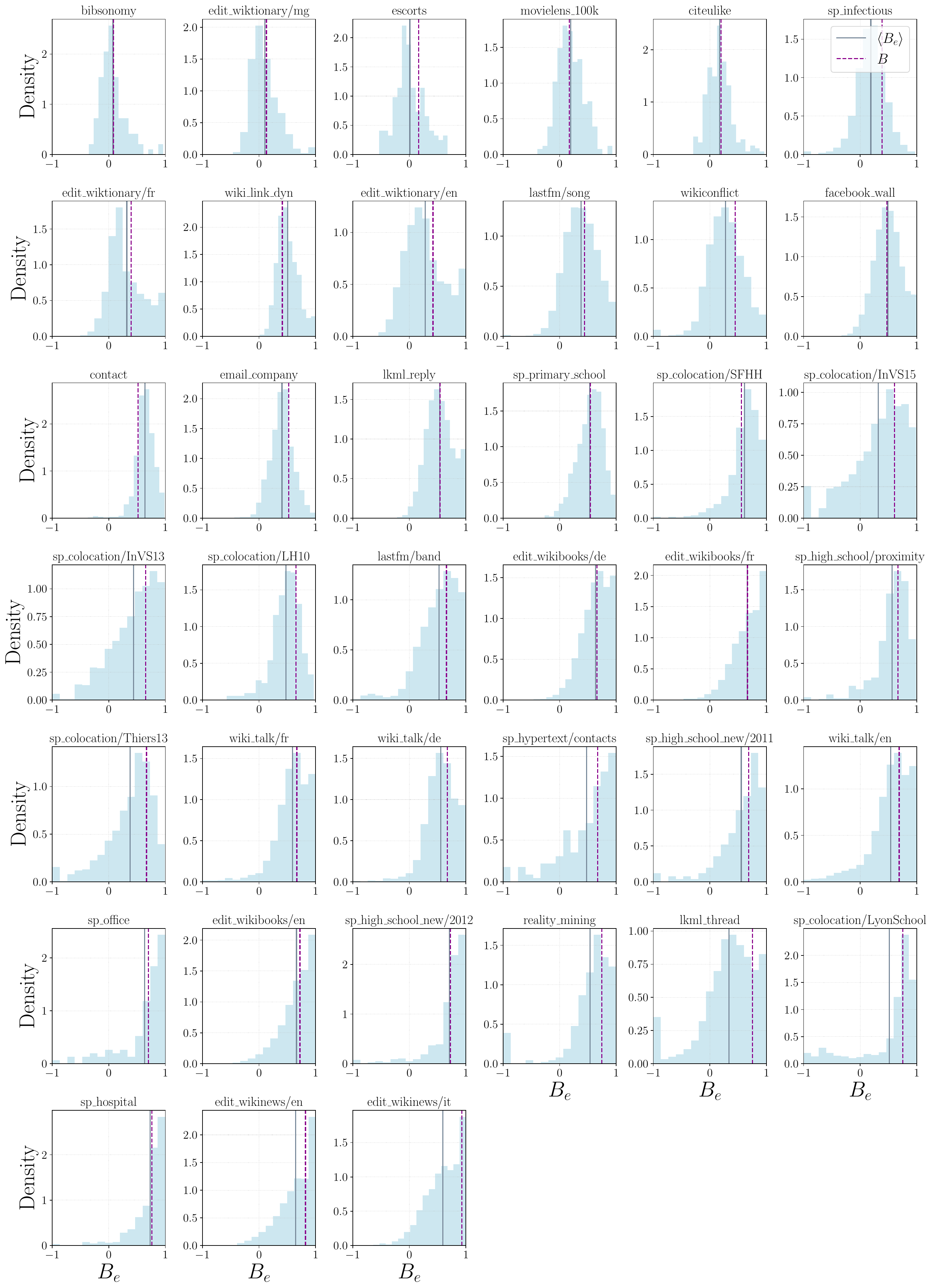} 
    \caption{Distribution of edge burstiness for the 39 datasets. Dashed purple line: system burstiness, grey solid line: average edge burstiness. The datasets are ordered by increasing system (edge) burstiness.}
    \label{fig:edge_burstiness}
\end{figure}

The same functional form can be applied to different topological scales, depending on what data is used to measure the coefficient of variation. Focusing on a single edge and its IET distribution, one may compute its burstiness index $B_e$. The same could be done at a single node level.
One can however also consider the system-wide burstiness $B$ by taking into account all the edge (or node) IETs in the network.

Although at the system level the volume of events is suffiently large to justify using the standard definition of burstiness, when we compute the burstiness at the single edge level, we must correct for finite-size effects with the following adjusted index~\cite{Kim2016}:
\begin{equation}
    B_e = B_w(r) = \frac{\sqrt{w+1} r - \sqrt{w-1}}{(\sqrt{w+1} - 2) r - \sqrt{w-1}},
\end{equation}
where $w$ is the number of inter-event times considered (or the edge weight).
In general, the mean edge burstiness $\langle B_e \rangle$ does not coincide with the system (edge) burstiness $B$:
\begin{equation}
    \left(\langle B_e \rangle := \frac{1}{E} \sum_{e = 1}^E B_w\left(\frac{\sigma_e}{\mu_e}\right)\right) \neq \left(B := B_w\left(\frac{\sigma_s}{\mu_s}\right) \right),
\end{equation}
where 
\begin{align}
    \mu_s = \frac{\sum_{e=1}^E \sum_{i=1}^{w_e-1} \tau_i}{\sum_{e=1}^E (w_e-1)}&, \quad \sigma_s^2 = \frac{\sum_{e=1}^E \sum_{i=1}^{w_e-1} (\tau_i - \mu_s)^2}{\sum_{e=1}^E (w_e-1)}\\
    \mu_e = \frac{\sum_{i=1}^{w_e-1} \tau_i}{w_e}&, \quad \sigma_e^2 = \frac{\sum_{i=1}^{w_e-1} (\tau_i - \mu_e)^2}{w_e}.
\end{align}

In \frefsi{fig:edge_burstiness}, we plot the binned distribution of the edge burstiness $B_e$ and the system edge burstiness $B$. In general, the distribution of edge burstiness is centered around its mean $\langle B_e \rangle$ which rarely coincides with the system burstiness. 
Coarse-graining from edge to node level would likely exhibit a similar distribution of single node burstiness compared to the system-wide node burstiness, although with a narrower variance due to larger node weights.
At the edge level, the average distance between measures of the system (edge) burstiness and of the mean edge burstiness across datasets is $0.11$.
Across the corpus, the system (edge) burstiness ranges from $0.08$ for the \texttt{bibsonomy} datasets to $0.93$ for the \texttt{edit\textunderscore wikinews/it} dataset.

\paragraph{Openness:}
\label{parSI:openness}
We define the openness of a dataset as the Cramér–von Mises $L_1$ distance~\cite{anderson1962distribution} between the distribution of residual times or time of first event $P(\tau_R)$ and the complementary cumulative distribution of IETs $P_{>}(\tau)$.
The openness can be computed at the edge-level (default) as well as at the node-level.

\begin{figure}
    \centering
    \includegraphics[width=15cm]{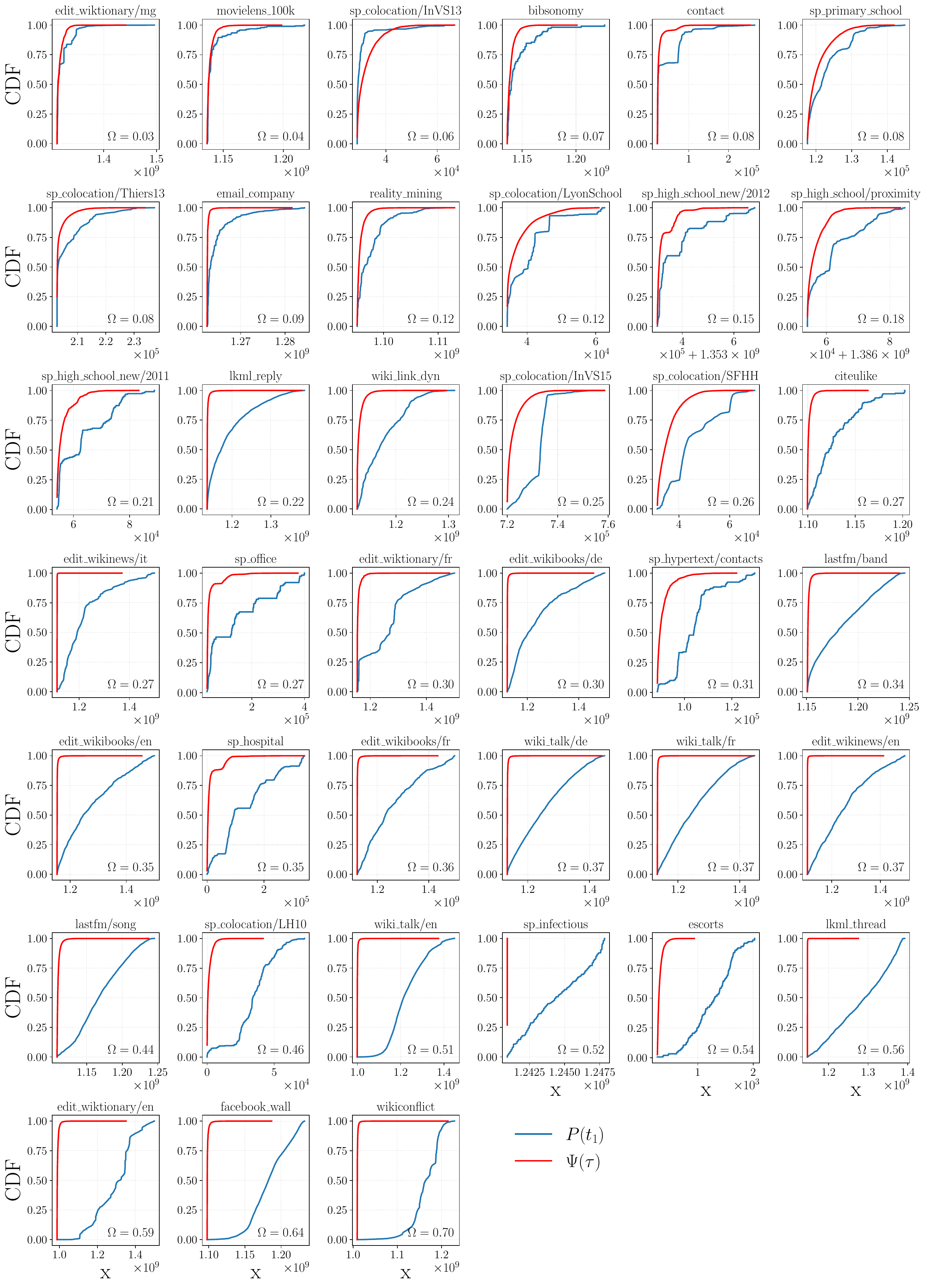} 
    \caption{Comparison between the Cumulative distributions of $P_{>}(\tau)$ (red) and $P(t_1)$ (blue) for the 39 datasets, with their measured openness $\Omega$. The datasets are ordered by increasing openness.}
    \label{fig:residual_ccdf_iet}
\end{figure}

In \frefsi{fig:residual_ccdf_iet}, we show the comparison between the distribution of residual times and the complementary cumulative distribution $P_{>}(\tau)$ computed at the edge-level, with the resulting measured openness $\Omega$. Across the corpus, the measure of openness ranges from $0.03$ for the \texttt{edit\textunderscore wiktionary/mg} dataset to $0.70$ for the \texttt{wikiconflict} dataset.
Since a node's residual time is the minimum of its edges' residual times, we expect the openness measure potentially to decay with topological coarse-graining from edge to node-level.

\begin{figure}
    \centering
    \includegraphics[width=15cm]{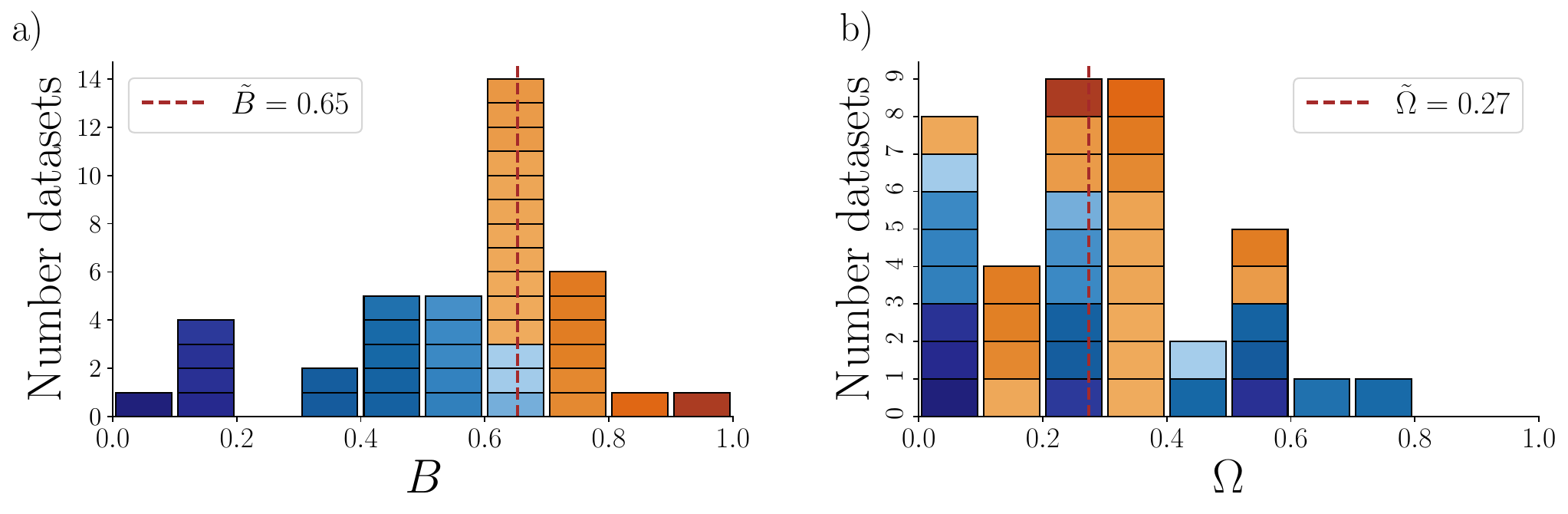} 
    \caption{Distribution of datasets along each of the two dimensions: burstiness and openness (edge-level). The median values separating the groups of high/low burstiness/openness are shown as dashed vertical lines. The colormap is the same as in Fig. 1 (main): each color maps to a single dataset, through its burstiness index. }
    \label{fig:two_dimensions}
\end{figure}

\paragraph{Groups of datasets.}

The corpus of datasets spans most of the positive (system-wide, edge-level) burstiness range $[0, 1]$, with a median value of $\tilde B = 0.65$. The datasets also cover a large portion of the $[0,1]$ interval of openness, with a median value of $\tilde \Omega = 0.27$. The binned distributions of datasets along these two measures is provided in \frefsi{fig:two_dimensions}. Each block describes a dataset colored by its burstiness index, with orange (blue) datasets corresponding to more (less) bursty ones. The datasets are stacked on top of each other within a bin by increasing burstiness index. We do not observe any particular correlation betweeen openness and burstiness, with very poissonian or bursty datasets appearing both at low and high openness values.

We divide the corpus of 39 datasets into 4 groups depending on their level of burstiness (system-wide, edge-level) and of openness (edge-level) relative to the median of the corpus. 
The `High-burstiness, High-Openness' group contains
\begin{itemize}[noitemsep, topsep=0pt]
    \item \texttt{lastfm/band},
    \item \texttt{edit\textunderscore wikibooks/de}, 
    \item \texttt{edit\textunderscore wikibooks/fr}, 
    \item \texttt{wiki\textunderscore talk/fr}, 
    \item \texttt{wiki\textunderscore talk/de}, 
    \item \texttt{sp\textunderscore hypertext/contacts}, 
    \item \texttt{wiki\textunderscore talk/en}, 
    \item \texttt{sp\textunderscore office}, 
    \item \texttt{edit\textunderscore wikibooks/en}, 
    \item \texttt{lkml\textunderscore thread}, 
    \item \texttt{sp\textunderscore hospital}, 
    \item \texttt{edit\textunderscore wikinews/en}.
\end{itemize}
Their burstiness index is greater than the median of the corpus $B > 0.65$ and their openness is greater than the median $\Omega > 0.27$.

The `Low-Burstiness, High-Openness' group of datasets contains
\begin{itemize}[noitemsep, topsep=0pt]
    \item \texttt{escorts}, 
    \item \texttt{sp\textunderscore infectious}, 
    \item \texttt{edit\textunderscore wiktionary/fr},
    \item \texttt{edit\textunderscore wiktionary/en},
    \item \texttt{lastfm/song},
    \item \texttt{wikiconflict},
    \item \texttt{facebook\textunderscore wall},
    \item \texttt{sp\textunderscore colocation/LH10}.
\end{itemize}
Their burstiness index is smaller than the median of the corpus $B \leq 0.65$ and their openness is larger than the median $\Omega > 0.27$.

The `High-Burstiness, Low-Openness' group of datasets contains
\begin{itemize}[noitemsep, topsep=0pt]
    \item \texttt{sp\textunderscore high\textunderscore school/proximity},
    \item \texttt{sp\textunderscore colocation/Thiers13},
    \item \texttt{sp\textunderscore high\textunderscore school\textunderscore new/2011},
    \item \texttt{sp\textunderscore high\textunderscore school\textunderscore new/2012},
    \item \texttt{reality\textunderscore mining},
    \item \texttt{sp\textunderscore colocation/LyonSchool},
    \item \texttt{edit\textunderscore wikinews/it}.
\end{itemize}
Their burstiness index is larger than the median of the corpus $B > 0.65$ and their openness is smaller than the median $\Omega \leq 0.27$.

The `Low-Burstiness, Low-Openness group datasets contains
\begin{itemize}[noitemsep, topsep=0pt]
    \item \texttt{bibsonomy},
    \item \texttt{edit\textunderscore wiktionary/mg},
    \item \texttt{movielens\textunderscore 100k},
    \item \texttt{citeulike},
    \item \texttt{wiki\textunderscore link\textunderscore dyn},
    \item \texttt{contact},
    \item \texttt{email\textunderscore company},
    \item \texttt{lkml\textunderscore reply},
    \item \texttt{sp\textunderscore primary\textunderscore school},
    \item \texttt{sp\textunderscore colocation/SFHH},
    \item \texttt{sp\textunderscore colocation/InVS15},
    \item \texttt{sp\textunderscore colocation/InVS13}.
\end{itemize}
Their burstiness index is smaller than the median of the corpus $B \leq 0.65$ and their openness is smaller than the median $\Omega \leq 0.27$.

The low/high burstines $B$ groups appear in the colormap for the datasets in Fig. 1, 2, and 3 of the main: low burstiness datasets are colored in shades of blue while high burstiness datasets are colored in shades of orange. The low/high openness $\Omega$ groups appear in Fig. 2e of the main.

\FloatBarrier
\section{Event aggregation in temporal networks}

In this chapter, we review the effects of aggregation on all the datasets of our corpus. 
Most results presented in this section are the same as those shown in the main text, but here they are reported for all datasets rather than for only selected ones.

\subsection{Time scale of maximal dynamicity}
\label{SI:Pplus}

\paragraph{Edge level.}

\begin{figure}
    \centering
    \includegraphics[width=15cm]{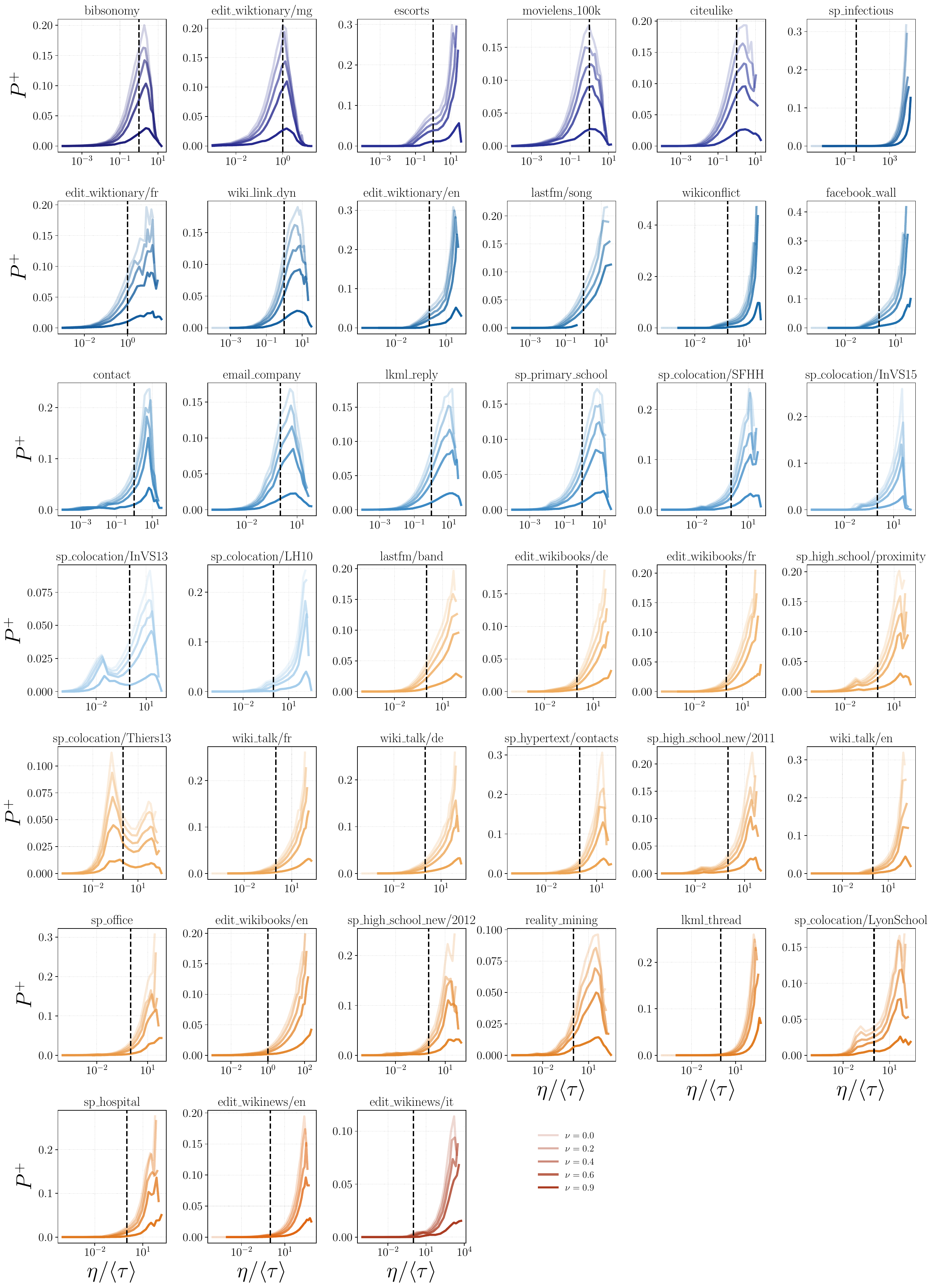} 
    \caption{Mean edge activation probability as a function of the rescaled width of aggregation window $\eta$, for various choices of the fractional overlap $\nu$, for all datasets of the corpus. The datasets are ordered by increasing system (edge) burstiness.}
    \label{fig:Pplus_eta}
\end{figure}

\begin{figure}
    \centering
    \includegraphics[width=15cm]{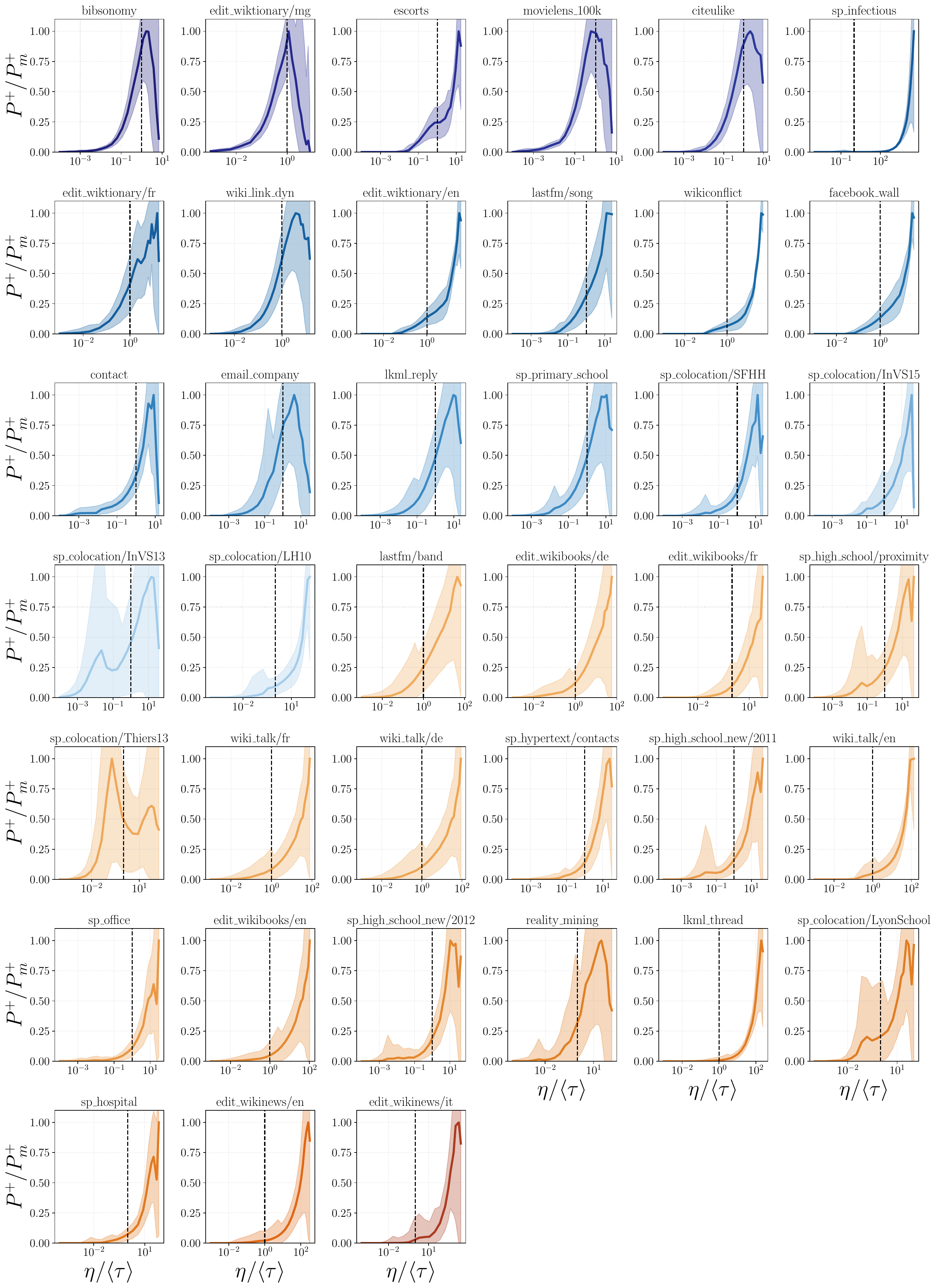} 
    \caption{Mean edge activation probability (rescaled by its maximum across values of $\eta$) as a function of the rescaled width of aggregation window $\eta$, for all datasets of the corpus. Shaded areas: standard deviation of the activation probability. The datasets are ordered by increasing system (edge) burstiness. We consider $\nu=0.2$.}
    \label{fig:Pplus_eta_std}
\end{figure}

In this paragraph, we show empirical measures of the edge activation probability for different choices of aggregation. When computing the activation probability of a dataset we systematically use an upper bound in the number of windows, to keep computational times under control.
Indeed, for small widths $\eta$ of the aggregation window, the number of windows to be considered can get very large although the activation probability is usually simply very close to zero (with a negligible standard deviation). 
In these cases, we find that the number of windows required to get a good empirical estimation of the activation probability can be bounded from above, without affecting the results. We take the maximum number of windows to be $1000$.

\begin{figure}
    \centering
    \includegraphics[width=9cm]{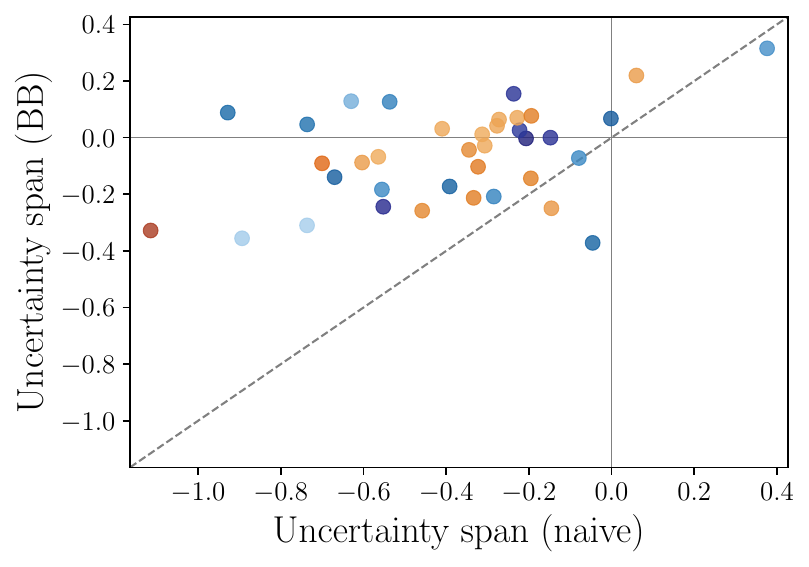} 
    \caption{Across datasets, difference between the relative uncertainty at high overlap $\nu = 0.9$ and at low overlap $\nu=0$ (uncertainty span), for a naive approach and a block bootstrap (BB) approach. While a majority of datasets exhibits negative uncertainty spans in the naive approach, the BB cancels most of this trend and significantly reduces the uncertainty spans. Color follows burstiness.}
    \label{fig:CI_vs_overlap}
\end{figure}

In \frefsi{fig:Pplus_eta}, we plot the empirical mean edge activation probability $P^+$ as a function of the rescaled window size $\eta$ (by the mean IET) for all datasets, and for various choices of the fractional overlap $\nu$. This plot is the analogue of Fig. 1c of the main.
The overlap increases the number of windows that can be considered in the fixed observation period. This increase in the number of windows decreases the amplitude of the activation probability for every value of $\eta$, but this decay is independent of $\eta$. As a consequence, the optimally dynamic aggregation scale remains unchanged under variations of $\nu$. While the maximum relative confidence interval with respect to the amplitude of the activation probability (`relative uncertainty') tends to decrease with increasing overlap when considering a standard bootstrap (naive) approach, accounting for correlated windows through a block bootstrap confidence interval cancels most of this overlap-dependence. To support this claim, \frefsi{fig:CI_vs_overlap} shows that the naive approach gives a vast majority of decreasing relative uncertainty from $\nu = 0$ to $\nu = 0.9$ (negative uncertainty span), while the block bootstrap (BB) approach does not. Overall, the overlap thus mainly affects the amplitude of the activation probability.

For the sake of the readability of Fig. 1c of the main (and \frefsi{fig:Pplus_eta}), we have decided to only display the mean edge activation probability. To get a sense of the variability of activation probability across edges (and how it varies with the choice of $\eta$), we show in \frefsi{fig:Pplus_eta_std} the mean edge $P^+$ together with its standard deviation, represented as a shaded area, for $\nu = 0.2$. 

\paragraph{Effect of openness.}

\begin{figure}
    \centering
    \includegraphics[width=11cm]{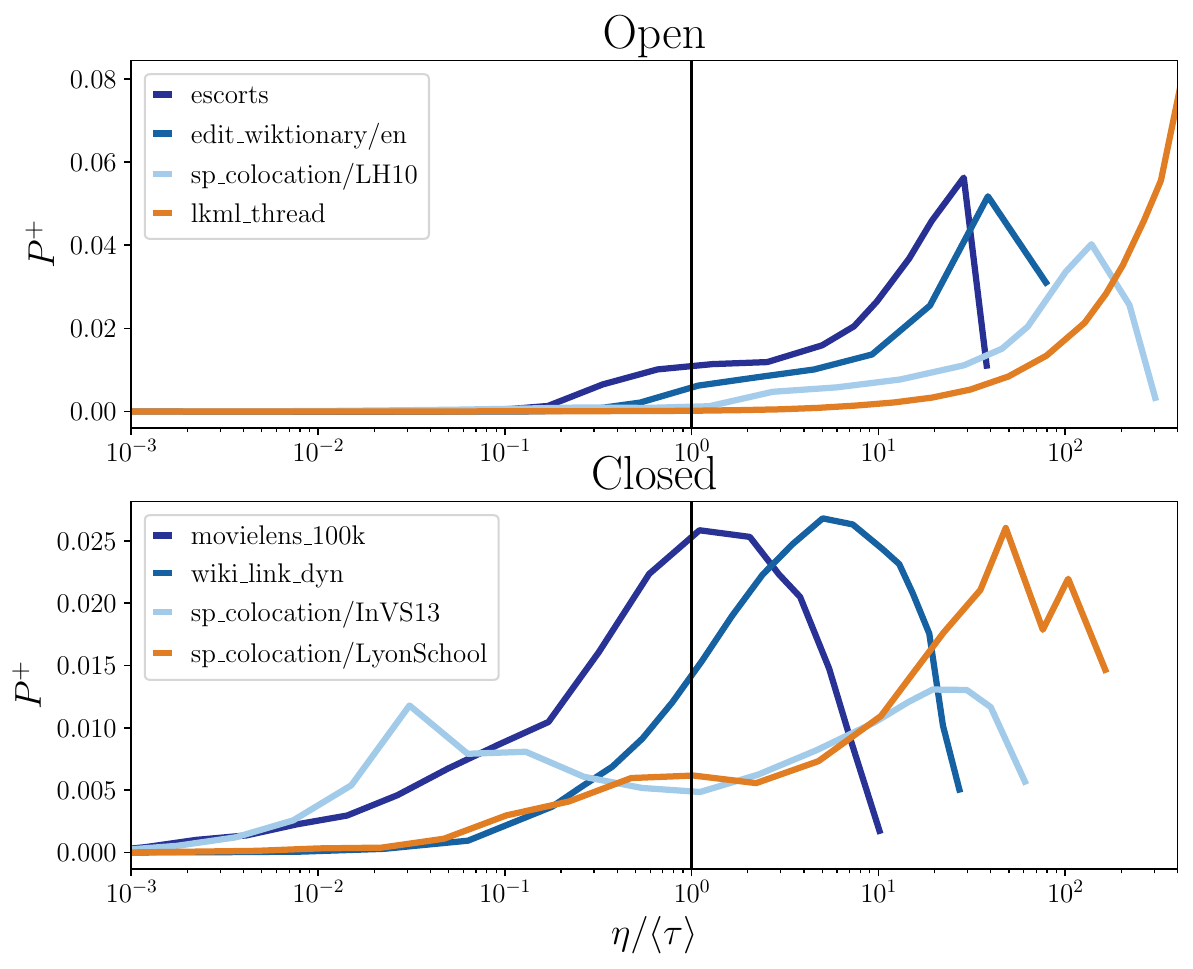}    
     \caption{\textbf{Effect of the openness on the dependence of activation probability on the aggregation scale $\eta$, for a fixed burstiness index}. Activation probability as a function of the rescaled $\eta$ for a selection of 8 datasets forming 4 pairs. Each pair is characterized by comparable burstiness indices but very distinct measures of openness: one of the pair's datasets is in the `High-Openness' group and the other is in `Low-Openness'. Fractional overlap considered: $\nu = 0.6$. The colormap is identical to the one used in Fig. 1 (main): each line color maps to a single dataset, through its burstiness index.}
    \label{fig:Pplus_eta_open_closed}
\end{figure}

In \frefsi{fig:Pplus_eta_open_closed}, we show how the activation probability differs between four pairs of an open and a closed system that have a similar burstiness index ($B =0.16-0.17$, $0.41-0.42$, $0.65$ or $0.75$, respectively).
For a given value of burstiness, the activation probability of an open dataset is systematically shifted towards wider aggregation windows compared to the corresponding `Low-Openness' dataset. As a consequence, for `Open' datasets, it is also more likely that a decay of activation probability for $\eta > \eta^*$ is not observed before the upper bound of $\eta$ (due to the finite observation window $T$) is reached.

\paragraph{Projected bipartite temporal networks.}

So far, we have treated bipartite temporal networks as such. However, one may wonder how a projection of the bipartite networks on either node layers behaves in terms of activation probability.
Here, we propose a simple way of projecting bipartite temporal networks on either layer to explore the robustness of our results.

Let's assume we want to project the bipartite network $G$ on the node layer $\mathcal{L}_1$. We consider that two nodes $n_1$ and $n_2$ on $\mathcal{L}_1$ are connected if they share a neighbor $n_3$ on the other layer $\mathcal{L}_2$. 
Although projections would generally involve aggregating the data over time windows \cite{wu2014temporal}, we here want to deal with data as raw as possible, before performing any type of aggregation. 
We consider events on the new edge $n_1-n_2$ to take place every time the original edges $n_1-n_3$ and $n_2-n_3$ exhibit an event. 
This projection not only affects the degree distribution of the projected networks but also the IET distribution.
Although a directed network representation would be more accurate, we assume that an event on $n_1-n_3$ contributes to the same edge $n_1-n_2$ as an event on $n_2-n_3$.

For the \texttt{bibsonomy} and \texttt{escorts} networks, the Netzschleuder repository contains node attributes which enable to distinguish and identify the two distinct node layers (tags/publications and male/female, respectively). However, the remaining bipartite datasets in our corpus do not have such node attributes. Despite being unable to identify the nature of each node layers, we are able to reconstruct the two distinct sets of nodes from the connected network structure.

\begin{figure}
    \centering
    \includegraphics[width=15cm]{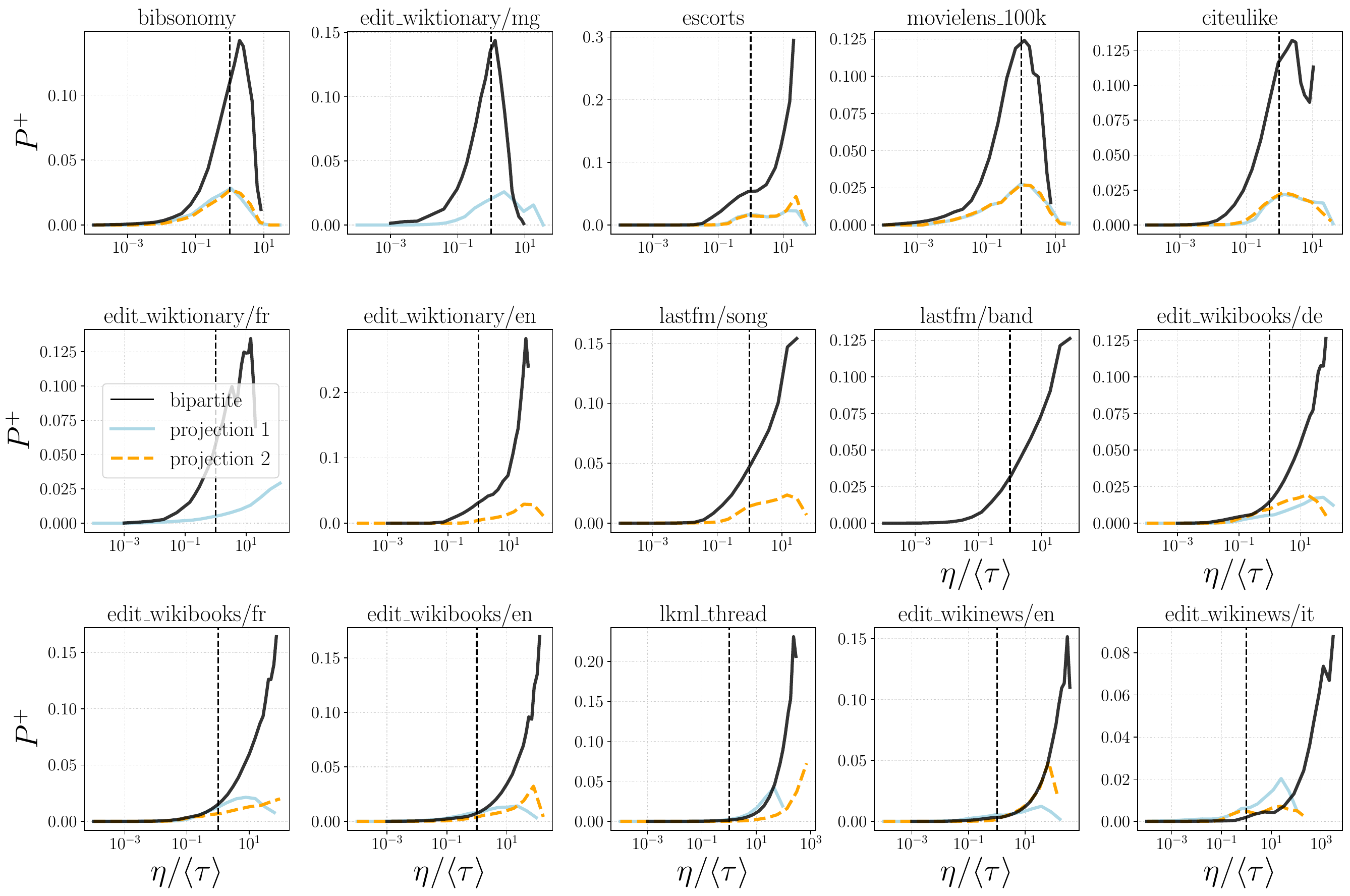} 
    \caption{Activation probability at the edge level, as a function of the rescaled width of aggregation window $\eta$, for the unipartite treatment (black line) and the projections on node-layers (solid light blue line and dashed orange line) that contain fewer than 20k nodes, for all datasets of the corpus. The datasets are ordered by increasing system (edge) burstiness.}
    \label{fig:proj}
\end{figure}

In \frefsi{fig:proj}, we show the mean edge activation probability as a function of $\eta$ for all bipartite networks in the corpus, comparing the unprojected network (black solid line) to the projections on either layers (light blue or orange lines). Although the behavior can vary quantitatively, the qualitative behavior appears to be robust.

\paragraph{Node level.}

\begin{figure}
    \centering
    \includegraphics[width=15cm]{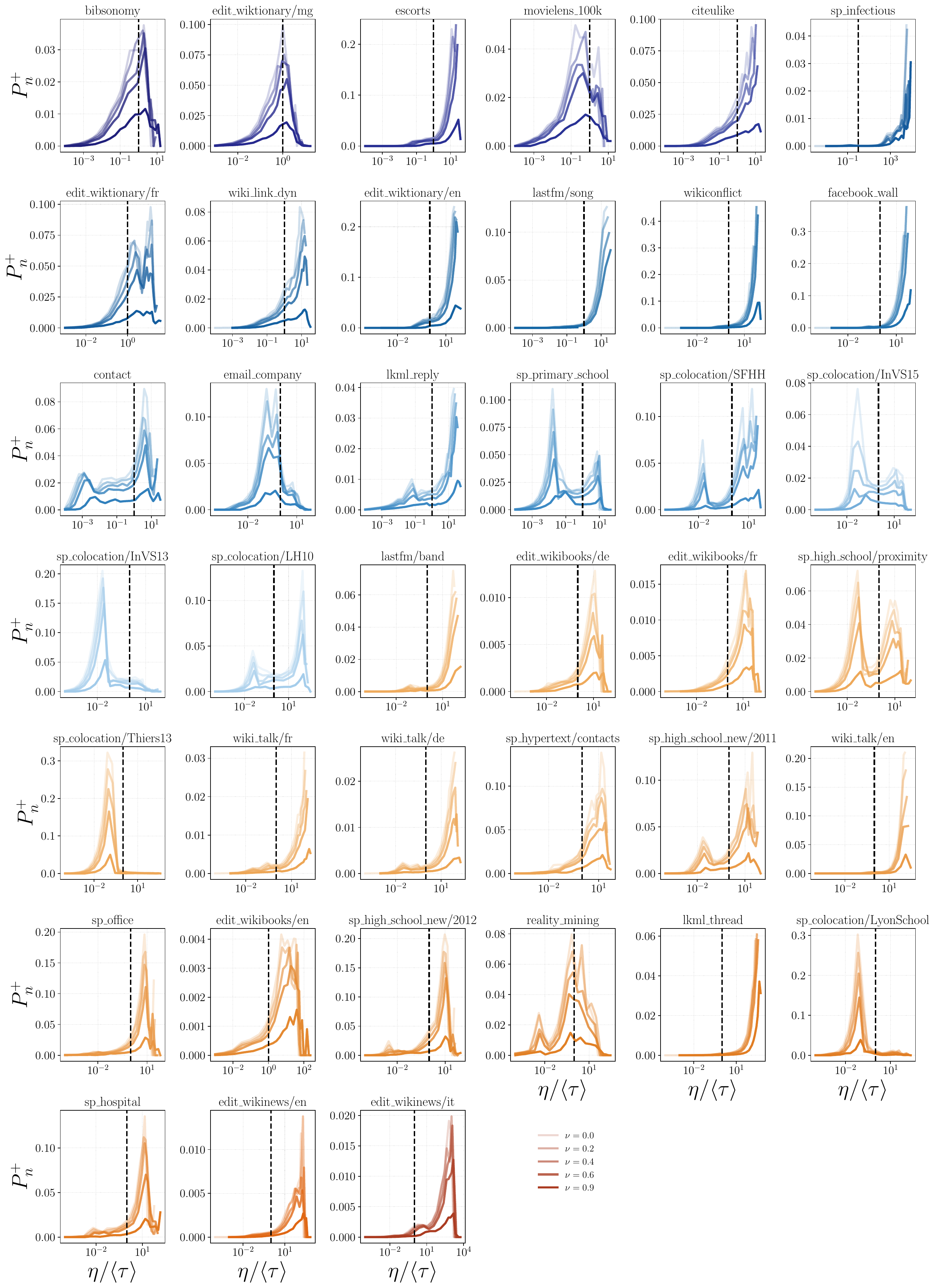} 
    \caption{Activation probability at the node level, as a function of the rescaled width of aggregation window $\eta$, for various choices of the fractional overlap $\nu$, for all datasets of the corpus. The qualitative behavior of the activation probability is identical after topological coarse-graining from edge- to node-level. The datasets are ordered by increasing system (edge) burstiness.}
    \label{fig:Pplus_node_eta}
\end{figure}

In Fig. 3a of the main, we move from the edge-level to the node-level activation probability. 
For a given node, we consider all the events that it is involved in across all the undirected edges incident to it. 
Moving from the edge-level to the node-level series of event times affects the IET distribution, as demonstrated in \frefsi{fig:edge_vs_node_iet}.
The dependence of the node-level activation probability $P^+_n$ on the rescaled window size $\eta$ (by the mean IET), for all datasets in the corpus is demonstrated in \frefsi{fig:Pplus_node_eta}. The impact of the overlap on the activation probability is identical to the one observed at the edge-level, i.e. demonstrating an increasing activation probability for decreasing overlap and no impact of the overlap on the maximally dynamic scale.

\paragraph{Deactivation probability.}

\begin{figure}
    \centering
    \includegraphics[width=15cm]{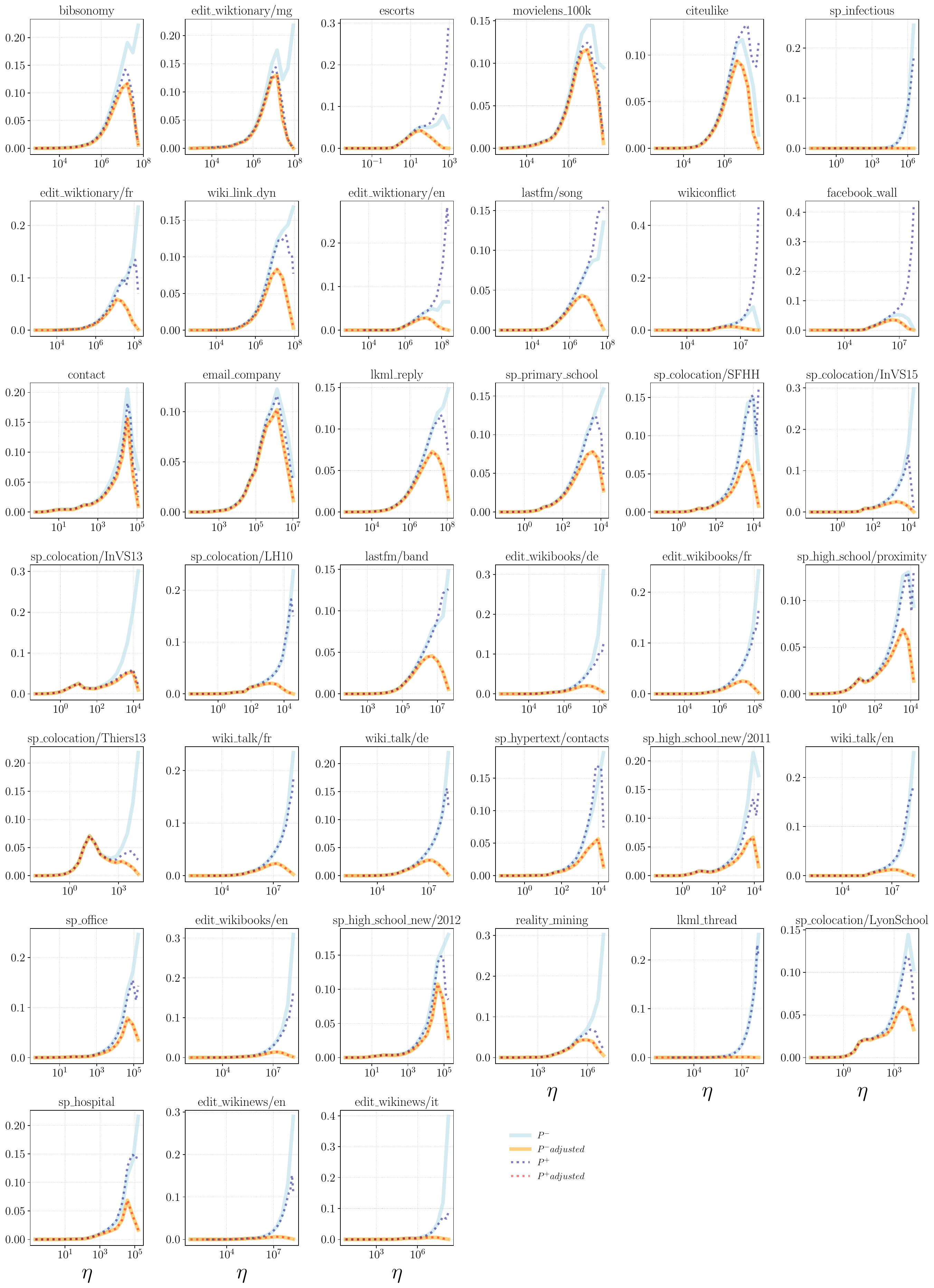} 
    \caption{\textbf{Activation–Deactivation Equivalence Under Openness Adjustment.} Activation and de-activation probabilities as functions of $\eta$. We also adjust the activation (deactivation) probability by removing one activation (deactivation) count for each edge which has a first (last) inactive window. After adjusting for openness, we observe a perfect match between the activation and deactivation probabilities. Fractional overlap $\nu = 0.4$.}
    \label{fig:deactivation}
\end{figure}

To address the expected analytical time-reversal symmetry between activation and de-activation probabilities we plot them against each other in \frefsi{fig:deactivation}, for $\nu = 0.4$. Although the two do not systematically match, especially for large values of $\eta$, we also show that adjusting each for openness, we obtain a perfect match. The adjustment involves removal of one activation (de-activation) count, for each edge that has a first (last) inactive window. This adjustment makes sure that the activation (de-activation) count due to exceedingly large (small) times of first (last) events, i.e. due to the openness of the system, are left out.

\paragraph{Summary table of relevant scales.}

The $\eta$ that maximizes the activation probability at the edge-level and at the node-level are reported in \trefsi{tab:scale}, as $\eta^* (data)$ and $\eta^*_n (data)$, respectively. 
When the found $\eta^*$ corresponds to the highest $\eta$ considered, we write it in squared brackets to indicate that we only find a lower bound to the maximally dynamic scale (a boundary-maximum) as a consequence of finite observation periods.

In the same table, we report the scales of maximal dynamicity $\eta^* (analytical)$ and of emergence of the LCC $\eta_c(analytical)$ obtained by plugging the empirical distributions of IET and static degree in the respective analytical equations (\eref{eq:opt_scale} and \eref{eq:perc_scale}).

For comparison, we also provide some scales of aggregation that have been used in the literature when studying the same networks (although not pre-processed in the same way). We see that most chosen scales of aggregation correspond to meaningful scales in terms of circadian or social rhythms. However, these do not always match the optimal scales of aggregation that we have identified.

\begin{sidewaystable}
    \centering
    \scriptsize 
    \begin{tabular}{lccccc}
         \textbf{Name}& \textbf{literature $\eta$}& \textbf{$\eta^*$ (data)} & \textbf{$\eta^*_n$ (data)} & \textbf{$\eta^*$ (analytical)}& \textbf{$\eta_c$ (analytical)}\\
          \texttt{bibsonomy} & 1 m \cite{Cattuto2007} & 4.4 m (3.2 m, 5.7 m)&  4.4 m (3.2 m, 5.7 m)& 3.0 m & 1.8 m\\
          \texttt{edit\textunderscore wiktionary/mg}&  -& 4.1 m (2.9 m, 5.6 m)&  4.1 m (2.9 m, 5.6 m)& 4.0 m & 1.0 s\\
         \texttt{escorts}& 800 d \cite{Rocha11}, 182 d \cite{RytherPhDthesis}& 1.6 y  (1.1 y, 2.1 y)&  1.6 y (1.1 y, 2.1 y)& 2.6 m & 5.4 m\\
         \texttt{movielens\textunderscore 100k} &  1 m \cite{Harper15}, 3 m \cite{zeng13}& 1.4 m (3.3 w, 2.4 m)& 1.4 m (3.3 w, 2.4 m)& 3.3 m & 2.3 m\\
          \texttt{citeulike} & ~1 w \cite{Kashoob12}& 3.5 m (2.5 m, 4.6 m)&  11.8 m (8.3 m, 1.3 y)& 2.3 m & 2.6 w\\
          \texttt{sp\textunderscore infectious} & 1.7 min, 3.3 min \cite{Malizia25} & [1.0 m]&  [1.0 m]& 6.4 min & 1.5 min\\
          \texttt{edit\textunderscore wiktionary/fr} & -& 3.6 y (3.0 y, 4.2 y)& 3.6 y (3.0 y, 4.2 y)& 6.3 m & 0.9 s\\
         \texttt{wiki\textunderscore link\textunderscore dyn} &  1s, 1 min, 1 h \cite{Almeida2007}, 1d, 1 m \cite{Prangnawarat17}& 7.0 m (5.0 m, 9.9 m)&  2.0 y (1.6 y, 2.3 y)& 4.5 m & 1.1 s\\
          \texttt{edit\textunderscore wiktionary/en}&  -& 5.3 y (4.3 y, 6.3 y)&  [6.3 y]& 4.2 m & 1.2 s\\
          \texttt{lastfm/song}& 1 w \cite{Reiter-Haas21} & 10.4 m (5.4 m, 1.7 y)& [1.7 y]& 2.0 m & 1.0 s\\
         \texttt{wikiconflict} & 1 w \cite{Miller19}, 91 d \cite{RytherPhDthesis}&  2.3 y (1.9 y, 2.8 y)& 2.3 y (1.9 y, 2.8 y)& 3.4 w & 0.8 s\\
         \texttt{facebook\textunderscore wall} & 1 w \cite{Miller19} & 1.4 y (1.1 y, 1.6 y)&  1.4 y (1.1 y, 1.6 y)& 1.4 m & 1.2 w\\
         \texttt{contact} & -& 14.5 h (11.1 h, 20.9 h)& 5.9 h (4.1 h, 7.8 h)& 6.2 h & 1.5 s\\
         \texttt{email\textunderscore company} & 1 w \cite{Miller19} & 2.0 w (1.0 w, 3.0 w)& 12.1 h (6.2 h, 23.6 h)& 1.4 w & 1.0 s\\
         \texttt{lkml\textunderscore reply} & 3 m \cite{Chen23} &   1.2 y (10.0 m, 1.6 y)& [3.1 y]& 6.0 m & 0.9 s\\
          \texttt{sp\textunderscore primary\textunderscore school} & 1.7 min, 3.3 min \cite{Malizia25}, 1 d \cite{Stehle11, Barrat13}, 2.5 min, 1 d \cite{Arregui24}& 1.7 h (1.3 h, 2.5 h)& 9.7 s (5.1 s, 18.5 s)& 41.7 min & 44.5 s\\
         \texttt{sp\textunderscore colocation/SFHH} & 15 min, 1 d \cite{Plaszczynski24}, 15 min \cite{Mancastroppa23}& 2.3 h (1.7 h, 3.3 h)& 2.3 h (1.7 h, 3.3 h)& 42.7 min & 19.7 s \\
         \texttt{sp\textunderscore colocation/InVS15} & 15 min \cite{Mancastroppa23}, 2.5 min, 1 d \cite{Arregui24}& 3.5 h (2.3 h, 4.6 h)& 16.4 s (8.2 s, 32.7 s)& 22.6 min & 20.1 s\\
         \texttt{sp\textunderscore colocation/InVS13} & 2.5 min, 1 d \cite{Arregui24} & 1.7 h (1.1 h, 2.2 h)& 9.3 s (4.7 s, 18.3 s)& 44.4 min & 20.1 s\\
        \texttt{sp\textunderscore colocation/LH10} & 15 min \cite{Mancastroppa23}& [8.5 h]& 6.2 h (4.0 h, 8.5 h)& 17.2 min & 19.9 s\\          
         \texttt{lastfm/band}& 1 w \cite{Reiter-Haas21} &   9.3 m (6.3 m, 1.2 y)& [1.2 y]& 1.5 w & 1.0 s\\
         \texttt{edit\textunderscore wikibooks/de} &  -& [4.6 y]&  11.6 m (7.8 m, 1.4 y)& 3.2 m & 1.1 s\\ 
         \texttt{edit\textunderscore wikibooks/fr} &   -& [4.6 y]& 1.1 y (10.1 m, 1.7 y)& 2.4 m & 1.0 s\\
         \texttt{sp\textunderscore high\textunderscore school/proximity} &   -& [3.5 h]& 13.0 s (6.5 s, 25.8 s)& 30.2 min & 2.8 min\\
         \texttt{sp\textunderscore colocation/Thiers13} & 1 d \cite{Mastrandrea15}, 15 min \cite{Mancastroppa23}, 2.5 min, 1 d \cite{Arregui24}& 15.1 s (7.1 s, 32.1 s)& 7.1 s (3.3 s, 15.1 s)& 11.1 min & 19.9 s\\
          \texttt{wiki\textunderscore talk/fr} &  -&   [3.8 y]& [3.8 y]& 2.6 w & 1.4 s\\
          \texttt{wiki\textunderscore talk/de} & -&   [3.8 y]& [3.8 y]& 2.5 w & 1.2 s\\
         \texttt{sp\textunderscore hypertext/contacts} &  -& 3.4 h (2.3 h, 4.5 h)& 2.3 h (1.7 h, 3.4 h)& 47.9 min & 8.7 min\\
         \texttt{sp\textunderscore high\textunderscore school\textunderscore new/2011} & 1 d \cite{Mastrandrea15}, 2.5 min, 1 d \cite{Arregui24} & [4.1 h]& 1.0 h (46.7 min, 1.5 h)& 44.6 min & 6.6 min\\
          \texttt{wiki\textunderscore talk/en} &  -& [5.5 y]&  [5.5 y]& 3.1 w & 1.1 s\\
         \texttt{sp\textunderscore office} &   -& [1.7 d]& 10.8 h (8.2 h, 15.9 h)& 9.2 h & 5.1 h\\
         \texttt{edit\textunderscore wikibooks/en} &  -& [4.2 y]& 1.5 y (1.2 y, 1.9 y)& 2.5 w & 1.1 s\\
          \texttt{sp\textunderscore high\textunderscore school\textunderscore new/2012} & 1 d \cite{Mastrandrea15}, 2.5 min, 1 d \cite{Arregui24} & 10.5 h (7.9 h, 15.6 h)& 7.9 h (5.3 h, 10.5 h)& 6.9 h & 41.2 min\\
          \texttt{reality\textunderscore mining} & 1h \cite{Nunes16} & 2.9 w (2.1 w, 1.0 m)& 2.2 d (1.0 d, 4.6 d)& 1.1 d & 10.0 min\\
          \texttt{lkml\textunderscore thread} & 3 m \cite{Chen23} & 2.5 y (2.0 y, 3.1 y)& [3.1 y]& 1.1 w & 1.4 s\\
          \texttt{sp\textunderscore colocation/LyonSchool} &  15 min \cite{Mancastroppa23}& 1.2 h (47.1 min, 1.6 h)& 8.3 s (4.0 s, 17.2 s)& 18.9 min & 19.8 s\\
          \texttt{sp\textunderscore hospital} & 2.5 min, 1 d \cite{Arregui24}& [1.5 d]& 6.7 h (4.5 h, 9.0 h)& 5.3 h & 20.2 s\\
          \texttt{edit\textunderscore wikinews/en} &   -& 4.0 y (3.1 y, 4.9 y)&  3.1 y (2.5 y, 4.0 y)& 2.6 w & 0.9 s\\
         \texttt{edit\textunderscore wikinews/it} & -& 3.3 y (1.9 y, 4.7 y)&  1.9 y (1.4 y, 3.3 y)& 2.1 w & 0.8 s\\
    \end{tabular}
    \caption{For the datasets where relevant sources could be found in the literature, we report the aggregation scale used in other papers to analyze the same datasets (n.b., without the same pre-processing). The second column contains the optimally dynamic aggregation scale for each dataset computed numerically at the edge level with $\nu = 0.2$. The third column reports the optimally dynamic aggregation scale for each dataset computed numerically at the node level with $\nu = 0.2$. In brackets, we include an estimate of the confidence interval conferred by the granularity of $\eta$. The fourth column contains the optimally dynamic aggregation scale computed from \eref{eq:opt_scale} using the empirical $P_{>}(\tau)$ of the corresponding datasets. In the fifth column we report the scale $\eta_c$ of emergence of the LCC computed from \eref{eq:perc_scale_exp} using the empirical $P_{>}(\tau)$ and $P(k)$ of the corresponding datasets. When the scale is indicated in squared brackets it means that no lower value of the mean edge activation probability was found for $\eta > \eta^*$. The datasets are ordered by increasing system (edge) burstiness.}
    \label{tab:scale}
\end{sidewaystable}

\subsection{Microscopic self-regulation and macroscopic stability}
\label{sec:micro_regulation}

\paragraph{Degree change vs degree.}

\begin{figure}
    \centering
    \includegraphics[width=15cm]{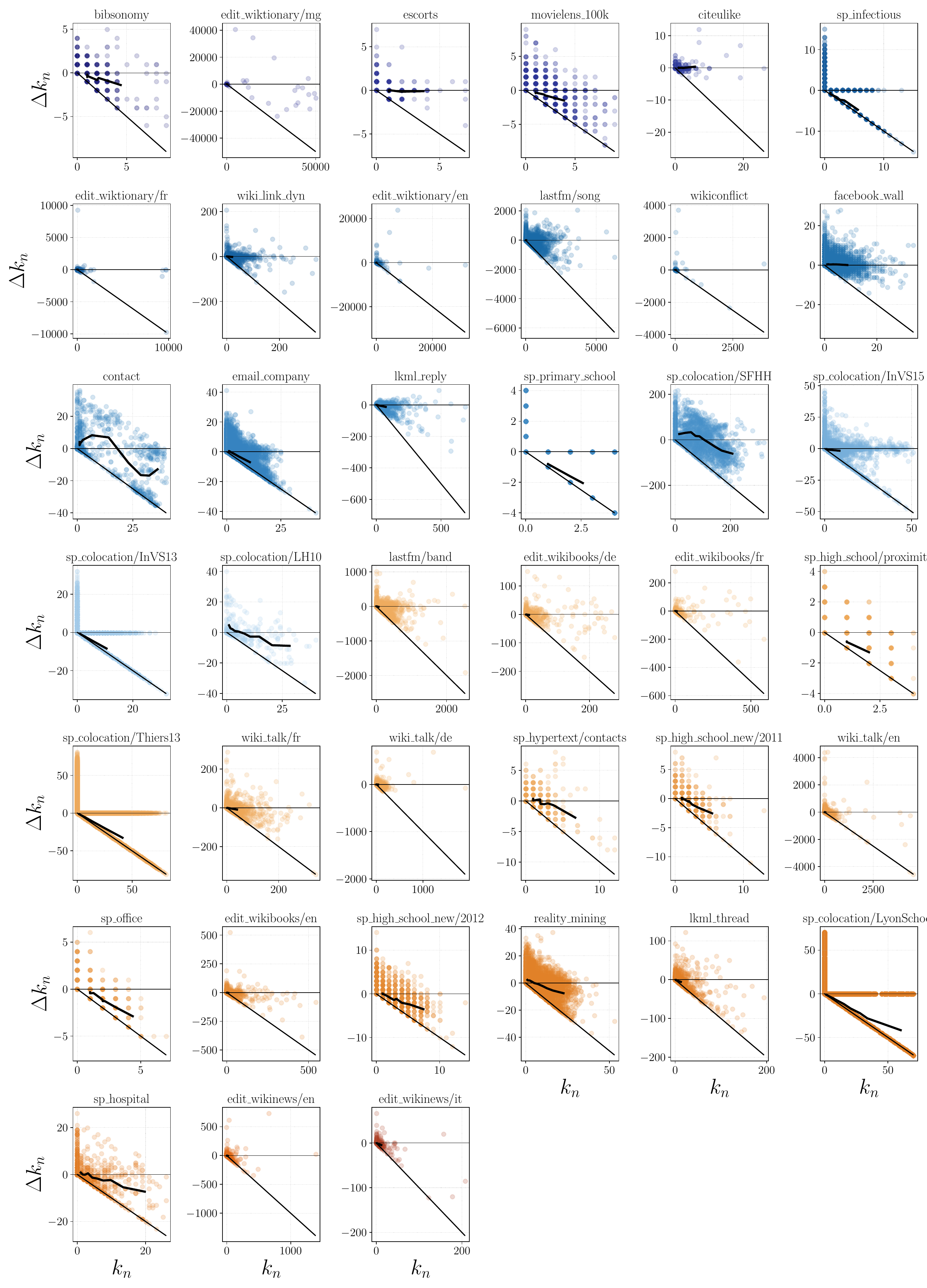} 
    \caption{Degree change vs degree for all the datasets in the corpus. When nodes have low degree they tend to gain degree in the next time step. When nodes have high degree they tend to lose degree in the next time step. Fractional overlap: $\nu = 0.2$. When the dataset exhibits an optimally dynamic aggregation scale at the node level, we consider $\eta = \eta^*_n$. The datasets are ordered by increasing system (edge) burstiness.}
    \label{fig:deltak_k}
\end{figure}

Figure 3b of the main displays the correlation between node degree across consecutive time windows, showing that nodes tend to gain (lose) degree in one time step when poorly (highly) connected.
All time-nodes in a dataset are associated with a dynamic degree $k_n$ and a degree change $\Delta k_n$. We show in \frefsi{fig:deltak_k} a tendency for decreasing degree change $\Delta k_n$ as the dynamic degree $k_n$ increases. This result is robust across all datasets with some variations in intensity and noise.
This plot is obtained for a given choice of the overlap $\nu$ and window size $\eta$. 

On the one hand, aggregating over wider time windows increases the probability that edges are active at a discrete time and thus increases the degree of nodes. Hence, higher values of $\eta$ extend the ranges of $\Delta k_n$ and $k_n$ in these figures. Additionally, when the windows are either too narrow or too wide, the similarity of consecutive time windows indirectly (through the activation probability) increases and the range of $\Delta k_n$ decreases, making the resulting slope in the ($\Delta k_n$,$k_n$) plane approach $0$.
On the other hand, the fractional overlap $\nu$ directly affects the similarity between consecutive time windows. As a result, for small fractional overlaps the slope approaches $-1$ (a node-time loses its entire degree in a single time step) and for large overlaps, the slope approaches $0$. In the next paragraph, we show that the slope undergoes different regimes for positive and negative overlaps.

\paragraph{Fitted slope of degree change vs degree, against the fractional overlap.}

\begin{figure}
    \centering
    \includegraphics[width=15cm]{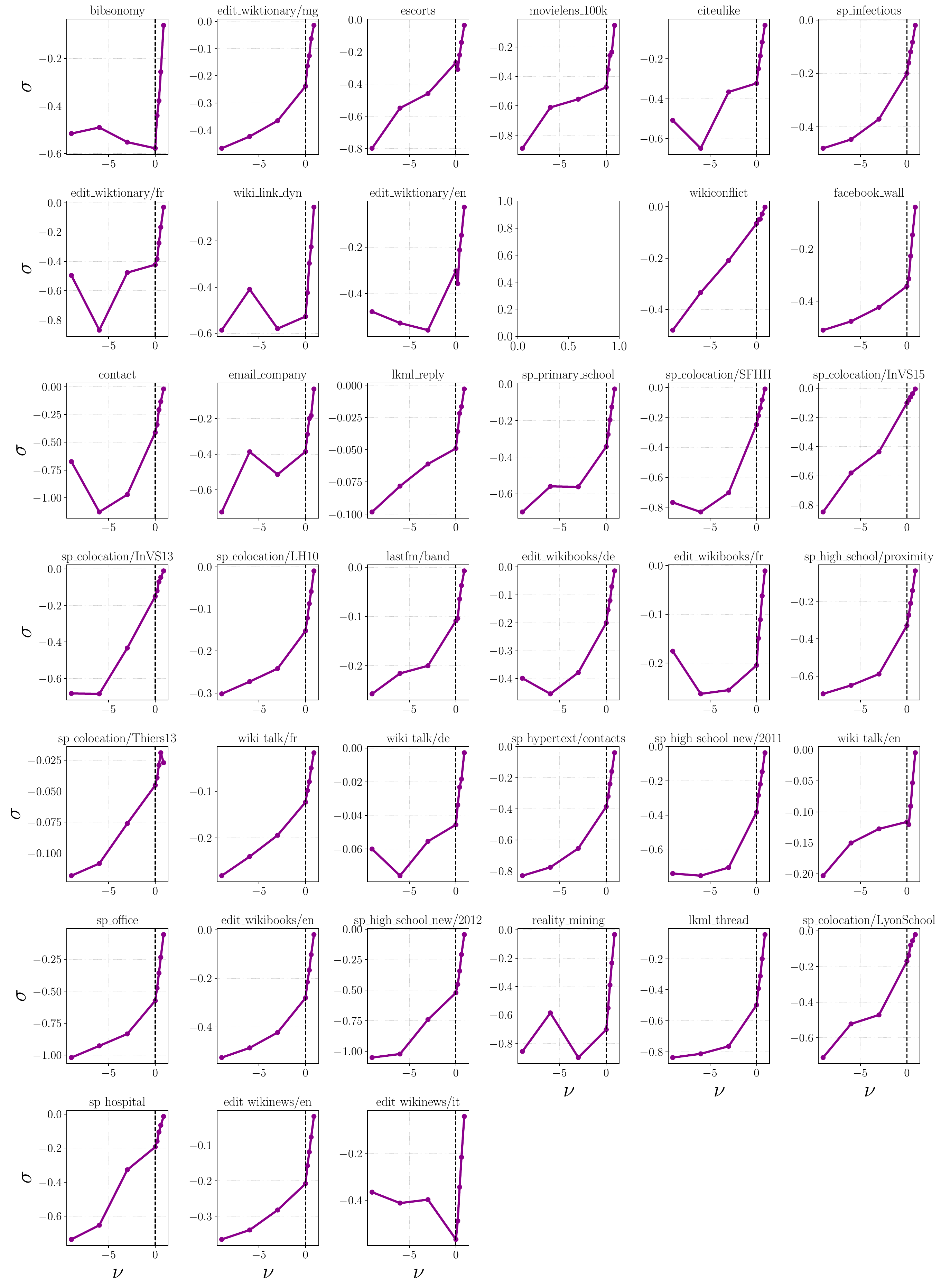} 
    \caption{Fitted linear slope in the ($\Delta k_n$,$k_n$) plane vs the fractional overlap $\nu$. From negative (data skipping) to positive overlaps, we observe a regime change: the correlation between the degree of a node in two consecutive times decays rapidly from overlaps close to 1 to overlaps close to 0 where it saturates for arbitrarily negative overlaps. The datasets are ordered by increasing system (edge) burstiness.}
    \label{fig:slopes_overlap}
\end{figure}

Here, we explore the effect of the fractional overlap $\nu$ on the fitted linear slope $\sigma$ with which degree change decays with dynamic degree.
In \frefsi{fig:slopes_overlap}, we show that $\sigma$ increases monotonically with $\nu$. 
First, we notice that the slope is always negative, indicating that we never observe a tendency for nodes to gain degree when highly connected while losing degree when poorly connected. 
Interestingly, the variation of the slope with the overlap tends to be much faster for positive overlaps than it is for negative overlaps. Since the slope $\sigma$ measures the level of correlation between the degree of a node across two consecutive windows, this implies that most correlations, but not all of them come from the overlap. When the slope becomes $-1$, we can safely assume that the two windows are no longer correlated. This however only happens for very negative overlaps (skipping data).

\paragraph{Probability of degree change.}
\label{SI:degreechange}

\begin{figure}
    \centering
    \includegraphics[width=15cm]{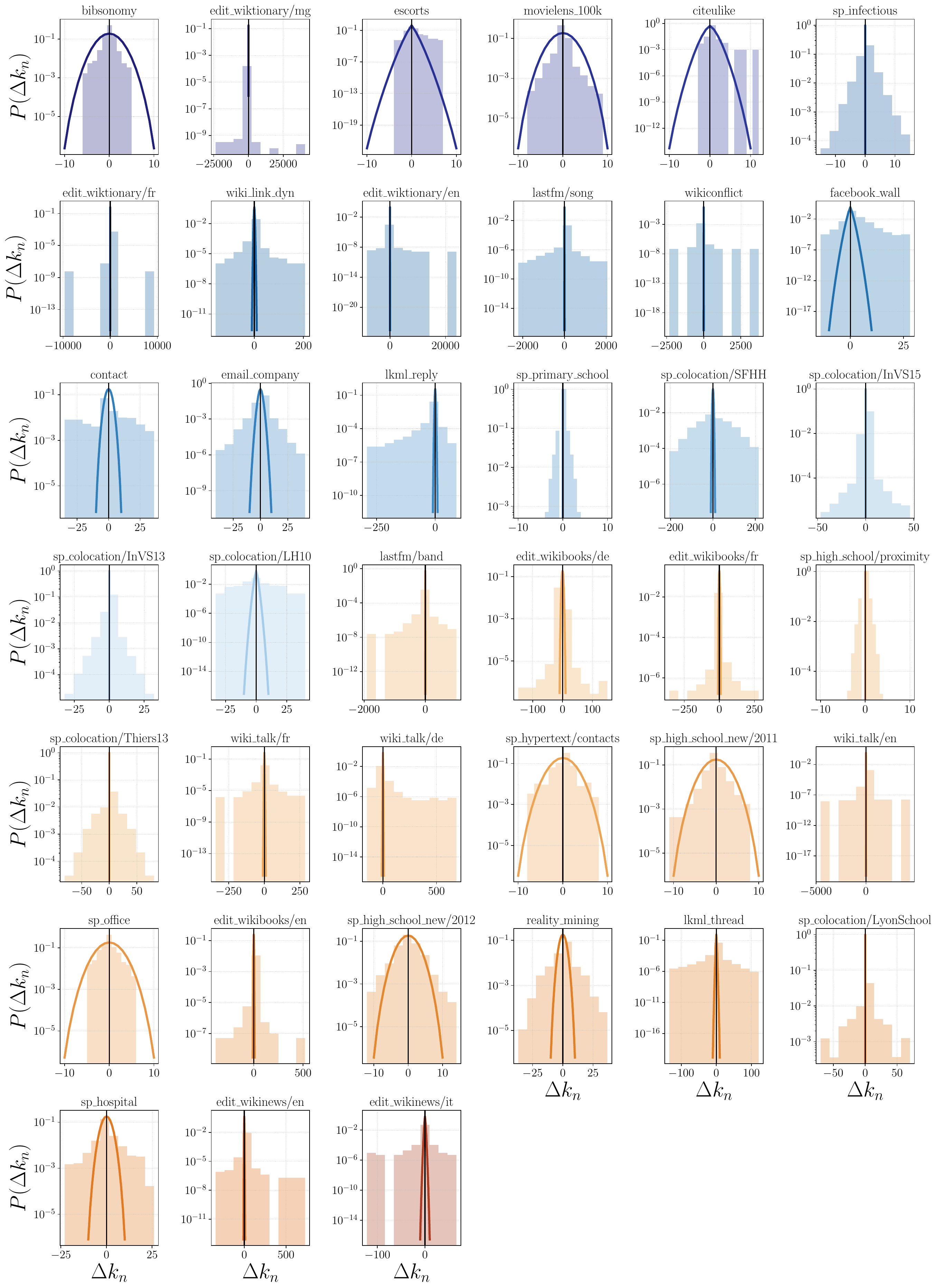} 
    \caption{Probability of degree change for all the datasets in the corpus with the analytical prediction obtained using the empirical $P(\tau)$ in \eref{eq:deg_change_binom}. The probability of a given degree change $\Delta k_n$ peaks in zero and deviations from there are exponentially suppressed. The qualititative behavior is well captured by analytical expectations, although the range of $\Delta k_n$ tends to be underestimated. Fractional overlap: $\nu = 0.2$. When the dataset exhibits an optimally dynamic aggregation scale at the node level, we consider $\eta = \eta^*_n$. The datasets are ordered by increasing system (edge) burstiness.}
    \label{fig:P_deltak_theor}
\end{figure}

We can also look at the distribution of degree changes across node-times, regardless of the dynamic degree. \frefsi{fig:P_deltak_theor} shows that a peak is observed when $\Delta k_n = 0$ in the entire corpus of datasets, and most datasets exhibit an exponentially suppressed and symmetric decay for deviations from $\Delta k_n = 0$, indicating that degree change is overall slow. This corresponds to the insets in Fig. 3c of the main.
In \frefsi{fig:P_deltak_theor}, we also compare the empirical degree change distribution with the analytical prediction obtained by plugging the empirical $P(\tau)$ in \eref{eq:deg_change_binom}.
We see that the seemingly better agreement between empirical findings and analytical expectations for `High-Burstiness' datasets in Fig. 3c of the main for a subset of only four datasets is incidental. Overall, the analytical expectations qualitatively match the observations but often fail quantitatively.

\subsection{Cycles of node degree change}
\label{SI:fig3c_supp}

\begin{figure}
    \centering
    \includegraphics[width=15cm]{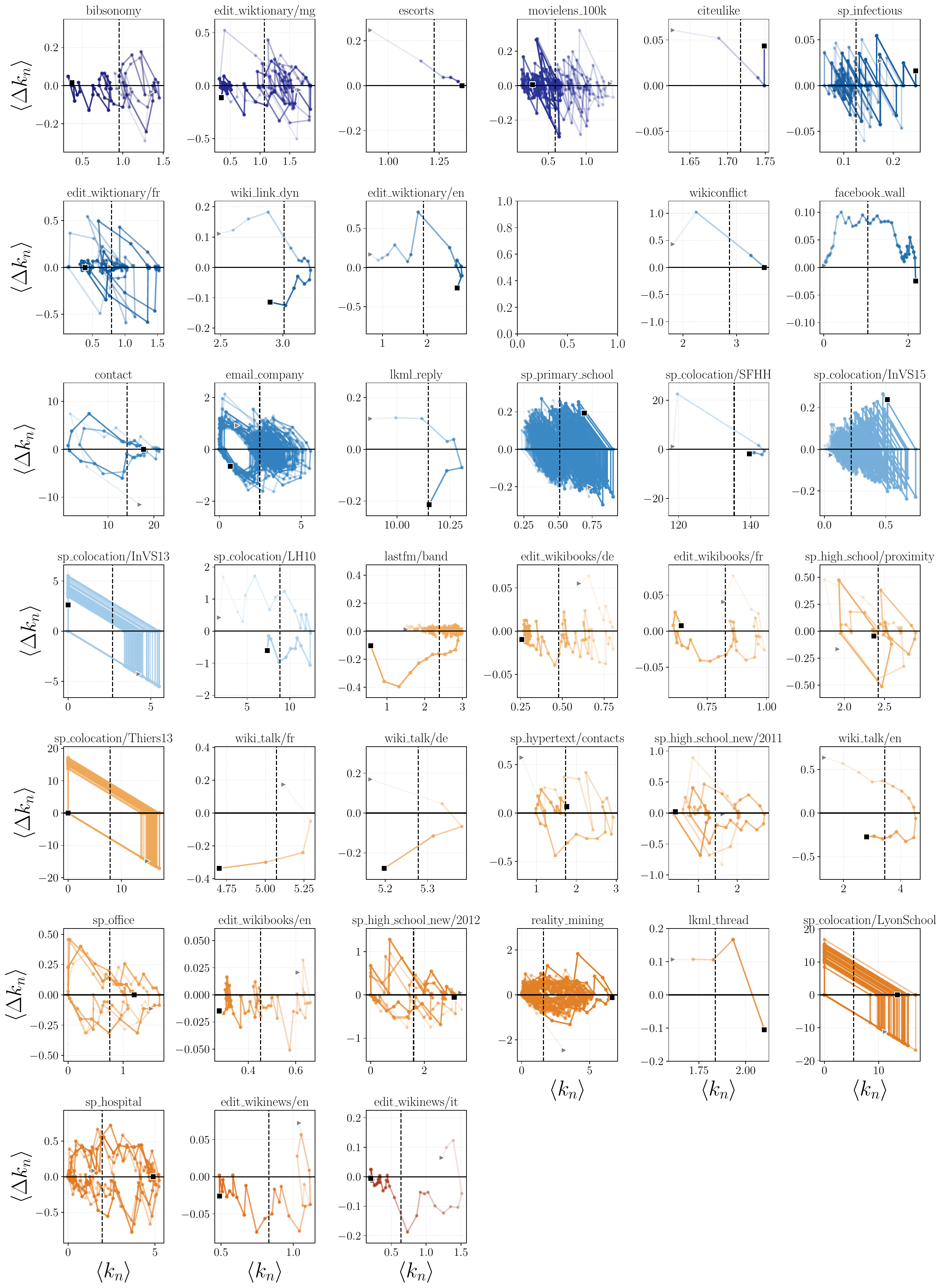} 
    \caption{Mean population trajectory in time across the ($\Delta k_n$,$k_n$) plane. Most datasets exhibit clockwise recurrent degree trajectories. Fractional overlap: $\nu = 0.9$. The datasets are ordered by increasing system (edge) burstiness.}
    \label{fig:cycles}
\end{figure}

In Fig. 3c of the main, we show the average population trajectory at each time step in the ($\Delta k_n$,$k_n$) plane, when $\eta = \eta^*_n$ and $\nu = 0.9$. A clockwise cyclic trajectory is shown for a selection of four datasets in our corpus.
In \frefsi{fig:cycles}, we show the population trajectories for all datasets. When a dataset does not exhibit an optimally dynamic aggregation scale, we consider an arbitrary intermediate value of $\eta$ for the aggregation.
We see that although some datasets have more noisy trajectories than others, the cyclic behavior is observed in most cases.

Sometimes the cycles are only partial (e.g., \texttt{wiki\textunderscore link\textunderscore dyn}, \texttt{edit\textunderscore wiktionary/en}, \texttt{wikiconflict}, \texttt{facebook\textunderscore wall}, \texttt{lkml\textunderscore reply}, \texttt{sp\textunderscore colocation/SFHH}, \texttt{sp\textunderscore colocation/LH10}), likely because the observation period is too short.
For some datasets, a small drift along the degree axis is also visible (e.g., \texttt{sp\textunderscore primary\textunderscore school}, \texttt{sp\textunderscore colocation/InVS15}), perhaps attributable to the gradual changes in the IET distribution over time for these datasets (see \frefsi{fig:IETs_period}).

\begin{figure}
    \centering
    \includegraphics[width=15cm]{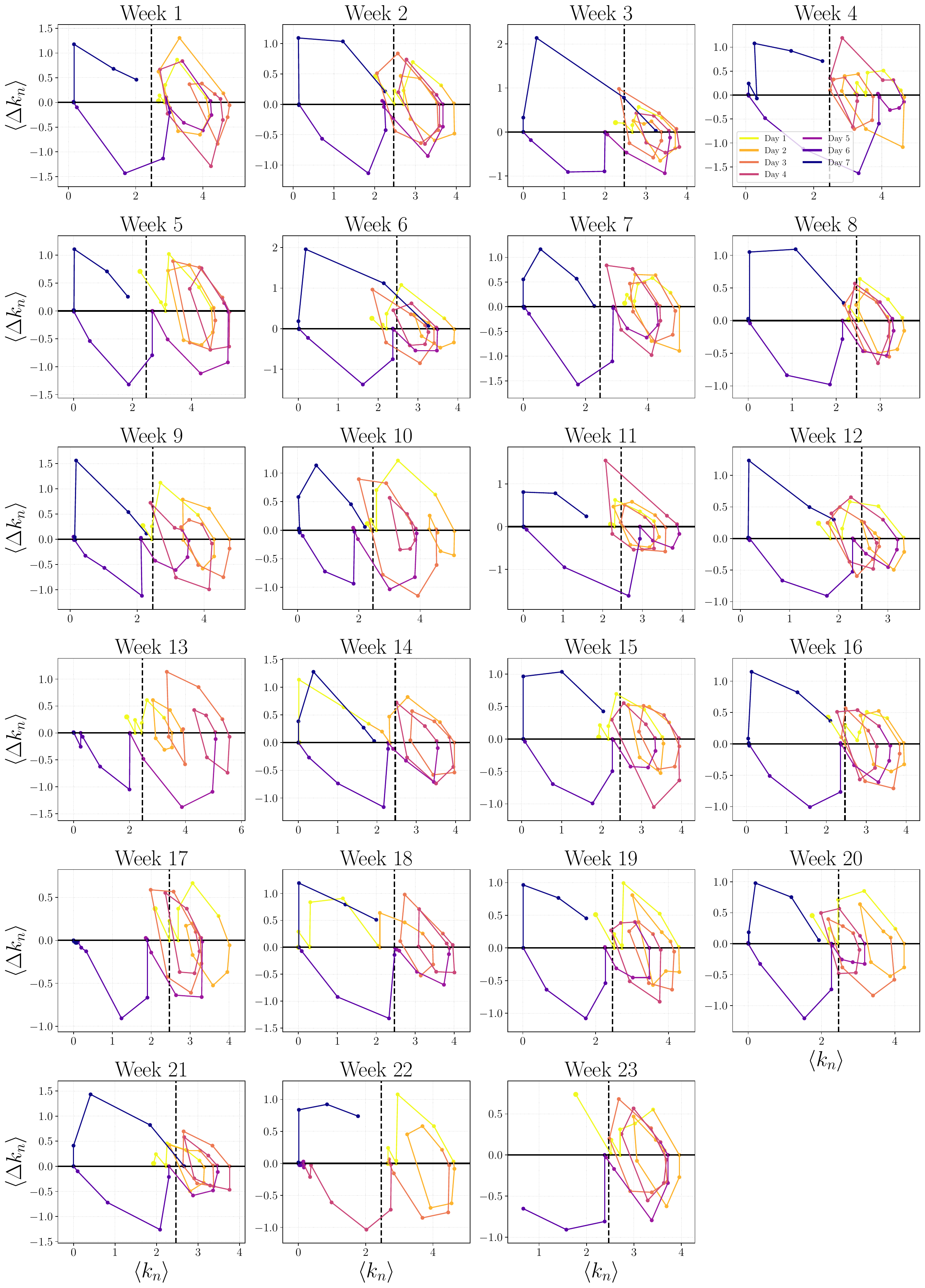} 
    \caption{Cyclic trajectory of the \texttt{email\textunderscore
    company} dataset, separated by week of data. Each color corresponds to a distinct day of data. Separating the data by day and week reveals a recurring pattern: two low-degree days corresponding to weekends, followed by five high-degree days corresponding to weekdays. The vertical dashed line is a guide for the eye indicating the average dynamical degree of the population. We considered a fractional overlap of $\nu = 0.9$.}
    \label{fig:anomalous_cycles_email}
\end{figure}

\paragraph{Anomalous cycles.}
Interestingly, one dataset, the \texttt{email\textunderscore company} dataset, exhibits a double cycle trajectory. 
One cycle takes place at higher values of degree while the other one at lower values of degree. Coloring and separating the trajectories by day and by week, respectively, we demonstrate in \frefsi{fig:anomalous_cycles_email} that the cyclic behavior occurring at higher values of the degree is performed on five consecutive days of the week, probably corresponding to weekdays, while the cycle appearing at lower values of the degree takes place in the remaining two days of the week, likely corresponding to the weekends.

\section{Temporal network model based on renewal processes}

\subsection{Analytical considerations}
\label{SI:analytical}

\subsubsection{Temporal network model}

We consider an undirected, unweighted static network of $N$ nodes as the underlying structure over which temporal interactions take place. The degree of a node (its number of neighbors in the static network) has discrete values $k = 0, 1, \ldots, N-1$ from a given degree distribution $P(k)$. The static network might be the (undirected and unweighted) aggregate of an empirical temporal network over a long observation time $T$, or a synthetic configuration-model (CM) network \cite{newman2018networks} (i.e. a network with uncorrelated degrees following a specified degree sequence) with $P(k)$ measured from the data or having a parametrized functional form.

Pairwise temporal interactions, or events, occur independently at random over each static edge via a renewal process \cite{whitt1982approximating,unicomb21} with interevent time (IET) distribution $P(\tau)$. Time $t$ is continuous, events are instantaneous, and IETs $\tau > 0$ are uncorrelated, both along a single edge and across the $k$ edges surrounding a given node. The mean IET is $\mu = \langle \tau \rangle = \int_{0}^{\infty} t P(t) dt$ and its variance is $\sigma_{\tau}^2 = \int_{0}^{\infty} t^2 P(t) dt - \langle \tau \rangle^2$. We assume that the renewal process in each link is in its steady or stationary state over the entire observation period $T$, meaning that the first ($\tau_R$) and last ($\tau_{S}$) IETs are distributed according to the complementary cumulative (e.g. residual or survival) distribution $P_{>}(\tau) = \int_{\tau}^{\infty} P(t)dt$ (i.e. the probability that an IET is at least $\tau$ or larger), while every other IET $\tau_j$ with $j = 1, 2, \ldots$ follows $P(\tau)$.

In an aggregation time window $t \in (0, \eta]$ with length parameter $\eta > 0$, the probability of observing $j-1$ IETs bounded by first and last IETs $\tau_R$ and $\tau_S$ is $P_{>}(\tau_R) P(\tau_1) \ldots P(\tau_{j-1}) P_{>}(\tau_S)$. This is equal to the probability of observing $j > 0$ events at times $t_1 < t_2 < \ldots < t_j$ (with $0 < t_1$ and $t_j \leq \eta$), given by $P_{>}(t_1) P(t_2 - t_1) \ldots P(t_j - t_{j - 1}) P_{>}(\eta - t_j)$. Thus, the probability $E_j(\eta)$ that a randomly selected edge has $j$ events in a time window of size $\eta$ (i.e. the fraction of $j$-type edges in the network) is given by the convolution
\begin{align}
\label{eq:frac_j_edges}
    E_j(\eta) &= \frac{1}{\mu} \int_{0}^{\eta} dt_1 \int_{0}^{\eta - t_1} dt_2 \ldots \int_{0}^{\eta - t_{j-2}} dt_{j-1} \int_{0}^{\eta - t_{j-1}} dt_j P_{>}(t_1) P(t_2 - t_1) \ldots P(t_j - t_{j - 1}) P_{>}(\eta - t_j) \nonumber\\
    &= \frac{1}{\mu} P_{>} \ast P^{\ast(j-1)} \ast P_{>}
\end{align}
for $j > 0$, where $P^{\ast j} = P \ast ... \ast P$ is the $j$-th convolution power of the IET distribution $P(\tau)$, and the normalization constant is the integral of the survival distribution, $\int_{0}^{\infty} P_{>}(t)dt = \mu$. For $j = 1$, the zeroth-order convolution is a Dirac delta function, $P^{\ast 0} = \delta_0$, meaning $E_1 = \mu^{-1} P_{>} \ast \delta_0 \ast P_{>}$. We also note that, because IETs $\tau$ are always positive, the convolution of two IET distributions $f$ and $g$ reduces to $(f \ast g) (t) = \int_0^{t} f(\tau) g(t - \tau) d\tau$. Since we identify active edges in the dynamical network as those having one or more events in a time window ($j \geq 1$), we can mainly focus on the fraction $E_0$ of edges with no events (i.e. in state $j = 0$). This is equal to the probability that the first IET $\tau_R$ is larger than the window of size $\eta$,
\begin{equation}
\label{eq:frac_0_edges}
E_0(\eta) = \frac{1}{\langle \tau \rangle} \int_{\eta}^{\infty} P_{>}(\tau) d\tau,
\end{equation}
where the probability of an edge having one or more events is simply $E_+ = \sum_{j = 1}^{\infty} E_j = 1 - E_0$.

If $P(\tau)$ is locally integrable, \esref{eq:frac_j_edges}{eq:frac_0_edges} can be treated analytically via the Laplace transform $\mathcal{L}[P(\tau)] = \hat{P}(s) = \int_{0}^{\infty}P(\tau) e^{-s\tau} d\tau$, where $s$ is a complex variable in the frequency domain. By calculating the integral of $P(\tau)$ as a convolution with the Heaviside step function, the Laplace transform of the residual distribution takes the form $\hat{P_{>}}(s) = [1 - \hat{P}(s)] / s$. From here, thanks to the convolution theorem, the Laplace transforms of the fraction of $j$-type edges and inactive edges ($j = 0$) are given by
\begin{equation}
\label{eq:frac_j_edges_trans}
    \hat{E}_j = \hat{E}_+ (1 - \hat{P}) \hat{P}^{j-1}
    \quad\text{and}\quad
    \hat{E}_0 = \frac{1}{s} - \hat{E}_+,
\end{equation}
with
\begin{equation}
\label{eq:frac_p_edges_trans}
    \hat{E}_+ = \frac{1 - \hat{P}}{\mu s^2},
\end{equation}
from which we can (in principle) calculate inverse Laplace transforms and obtain $E_j$ and $E_0$ explicitly as functions of $\eta$ and the parameters of the IET distribution $P(\tau)$. However, since it is unlikely that \esref{eq:frac_j_edges_trans}{eq:frac_p_edges_trans} lead to combinations of known Laplace transforms for functional forms of $P(\tau)$ of empirical interest to us (and we only need $E_0$ for the calculations below), it is more straightforward to calculate the integral in \eref{eq:frac_0_edges} directly.

\subsubsection{Edge activation probability}

Now, consider two consecutive aggregation windows of size $\eta$ and fractional overlap $\nu \in [0, 1]$, such that the part of the second window not contained in the first has length $\omega = \eta (1 - \nu)$. As seen above, the probability of having no events in the first window (including the overlap) is $E_0(\eta)$, while the probability of having one or more events in the second window (without the overlap) is $1 - E_0(\omega)$. Thus, we can approximate the probability $P^+$ that an edge gets activated (i.e. goes from having no events to having at least one, across consecutive windows) as
\begin{equation}
\label{eq:act_prob}
    P^{\pm}(\eta) = E_0(\eta) [1 - E_0(\eta [1 - \nu])].
\end{equation}
Due to time-reversal symmetry in the underlying renewal process, \eref{eq:act_prob} also gives the probability $P^-$ that an edge gets deactivated, hence the notation $\pm$. For no overlap ($\nu = 0$), \eref{eq:act_prob} is a logistic function that becomes maximal at $E_0(\eta^*) = 1/2$. To find the optimal aggregation time scale $\eta^*$ at which the activation probability $P^{\pm}$ gets maximized in general, we take the derivative of \eref{eq:act_prob} with respect to $\eta$ (for constant but arbitrary $\nu$), evaluate at $\eta^*$ and equate to zero. Using \eref{eq:frac_0_edges} we find that the optimal window length is given implicitly by
\begin{equation}
\label{eq:opt_scale}
\frac{P_{>}(\eta^*)}{E_0(\eta^*)} = \frac{(1 - \nu) P_{>}(\eta^* [1 - \nu])}{1 - E_0(\eta^* [1 - \nu])},
\end{equation}
from which we might be able to explicitly derive $\eta^*$ for particular functional forms of the IET distribution $P(\tau)$. From \eref{eq:opt_scale}, the maximum activation probability $P^{\pm}_m = P^{\pm}(\eta^*)$ is
\begin{equation}
\label{eq:act_prob_max}
    P^{\pm}_m = (1 - \nu) \frac{P_{>}(\eta^* [1 - \nu])}{P_{>}(\eta^*)} E_0^2(\eta^*),
\end{equation}
which reduces to $P^{\pm}_m = 1/4$ for $\nu = 0$, and has a limit $P^{\pm}_m \to 0$ for $\nu \to 1$, as expected.

\subsubsection{Degree change across windows}

In an egocentric network of size $k$ (i.e. the set of $k$ static edges around a given node, each with its own independent renewal process driving events), we can also calculate the probability distribution $P(\Delta k_n)$ of a randomly selected node having degree change $\Delta k = \ldots -1, 0, 1, \ldots$ between two consecutive windows of size $\eta$ and fractional overlap $\nu$. Denoting by $p_{k'}$ the probability that the ego network has degree $k'$ in a window excluding the overlap (i.e. of size $\omega = \eta [1 - \nu]$), having degree change $\Delta k_n = 0$ implies that the degrees in each window are the same, so $P(\Delta k_n = 0) = p^2_0 + p^2_1 + \ldots = \sum_{k' = 0}^{\infty} p^2_{k'}$. Similarly, we get $\Delta k_n = 1$ when the second window has one more active link than the first one, $P(\Delta k_n = 1) = \sum_{k' = 0}^{k - 1} p_{k'} p_{k' + 1}$, and $\Delta k_n = -1$ in the opposite case, $P(\Delta k_n = -1) = \sum_{k' = 1}^{k} p_{k'} p_{k' - 1}$. By induction we have
\begin{equation}
\label{eq:deg_change_prev}
    P(\Delta k_n) =
    \begin{cases}
        \sum_{k' = 0}^{k - \Delta k_n} p_{k'} p_{k' + \Delta k_n}, \quad &\Delta k_n \geq 0 \\
        \sum_{k' = \Delta k_n}^{k} p_{k'} p_{k' - \Delta k_n}, \quad &\Delta k_n \leq 0
    \end{cases},
\end{equation}
a symmetric function for all $\Delta k_n$. In a window excluding overlap, each edge is active with probability $E_+(w) = 1 - E_0(\omega)$. Since renewal processes are uncorrelated, the probability of having dynamical degree $k'$ is then binomially distributed, $p_{k'} = \binom{k}{k'} E_+^{k'} (1 - E_+)^{k - k'} = B_{k, k'}[E_+]$. Inserting into \eref{eq:deg_change_prev} we finally obtain
\begin{equation}
\label{eq:deg_change_binom}
P(\Delta k_n) = \sum_{k' = 0}^{k - \Delta k_n} B_{k, k'}[E_+(\eta[1 - \nu])] B_{k, k' + \Delta k_n}[E_+(\eta[1 - \nu])]
\end{equation}
for $\Delta k_n \geq 0$, with $P(-\Delta k_n) = P(\Delta k_n)$.

\subsubsection{Temporal network phase transition}

For a static configuration-model network with degree distribution $P(k)$ where each edge is independently and temporally active with probability $E_+(\eta)$ within an aggregation window of length $\eta$, we expect an edge percolation phase transition at a critical window size $\eta_c$, characterized by the emergence of a largest connected component (LCC) of active links \cite{unicomb21}. For $\eta < \eta_c$, active edges form dynamic but finite components with exponentially distributed sizes far smaller than network size as $N \to \infty$ (even when the static network itself has a giant component). For increasing $\eta > \eta_c$, we see instead a growing and dynamic active LCC that scales with network size, with a varying composition of nodes and edges percolating throughout the network. The critical point $\eta_c$ is given implicitly by the percolation condition \cite{newman2018networks}
\begin{equation}
\label{eq:perc_cond}
\sum_{k_n=0}^{\infty} k_n (k_n-2) P(k_n; \eta_c) = 0,
\end{equation}
where the dynamic degree distribution $P(k_n; \eta)$ within a window of size $\eta$ is obtained by randomly removing a fraction $E_0(\eta) = 1 - E_+(\eta)$ of edges from the underlying static network, that is,
\begin{equation}
\label{eq:dyn_deg_dist}
P(k_n; \eta) = \sum_{l \geq k_n}^{\infty} P(l) B_{l, k_n}[E_+(\eta)].
\end{equation}

We can find the critical window length $\eta_c$ and the size $S_{\eta}$ of the active LCC via generating functions \cite{newman2002spread,newman2018networks}. Consider the generating function of the static degree distribution $P(k)$, $g_0(x) = \sum_{k = 0}^{\infty} P(k) x^{k}$ [with normalization condition $g_0(1) = 1$], and the generating function of the static excess degree distribution, $g_1(x) = g_0'(x) / \langle k \rangle$, with $g_0'$ a derivative with respect to the dummy variable $x$, such that $g_0'(1) = \langle k \rangle$ and $g_1(1) = 1$. Since active edges are distributed binomially around any given node [according to \eref{eq:dyn_deg_dist}], the generating function of the dynamic degree distribution $P(k_n; \eta)$ is $g_0(x; \eta) = g_0[1 + (x - 1)E_+(\eta)]$. Similarly, the generating function of the dynamic excess degree distribution is $g_1(x; \eta) = g_1[1 + (x - 1)E_+(\eta)]$. Following \cite{newman2002spread}, the average size $\langle s \rangle$ of an active connected component (excluding the active LCC) can be written as
\begin{equation}
\label{eq:av_size}
\langle s \rangle = 1 + \frac{g_0'(1; \eta)}{1 - g_1'(1; \eta)} = 1 + \frac{ E_+(\eta) g_0'(1)}{1 - E_+(\eta) g_1'(1)}.
\end{equation}
Since \eref{eq:av_size} diverges for $\eta_c$ such that $E_+(\eta_c) g_1'(1) = 1$, the edge percolation critical point is given implicitly by
\begin{equation}
\label{eq:perc_scale}
E_0(\eta_c) = 1 - \frac{\langle k \rangle}{\langle k^2 \rangle - \langle k \rangle},
\end{equation}
with $\langle k^2 \rangle = \sum_{k = 0}^{\infty} k^2 P(k)$ the second moment of the static degree distribution. Using \esref{eq:frac_0_edges}{eq:perc_scale}, we may also write
\begin{equation}
\label{eq:perc_scale_exp}
\frac{1}{\langle \tau \rangle} \int_0^{\eta_c} P_{>}(\tau) d\tau = \frac{\langle k \rangle}{\langle k^2 \rangle - \langle k \rangle},
\end{equation}
which relates properties of inter-event times on the left with properties of the aggregated network on the right. As static degree heterogeneity increases ($\langle k^2 \rangle \to \infty$), $\eta_c$ vanishes and we expect an active LCC for any window size. Putting together \esref{eq:opt_scale}{eq:perc_scale_exp}, if the IET and degree distributions $P(\tau)$ and $P(k)$ are such that $\eta_c < \eta^*$, then there is an aggregation scale at which the network is both maximally dynamical and globally connected.

From \cite{newman2002spread,newman2018networks} we also know that the size $S_{\eta}$ of the active LCC follows the set of equations
\begin{subequations}
\label{eq:act_lcc}
\begin{align}
    u &= E_0(\eta) + [1 - E_0(\eta)] g_1(u), \\
    S_{\eta} &= 1 - g_0(u),
\end{align}
\end{subequations}
with $u$ a dummy variable. For given IET and degree distributions $P(\tau)$ and $P(k)$ (with tractable generating functions), \eref{eq:act_lcc} can be solved numerically to obtain $S_{\eta}$.

\subsubsection{Homogeneous IETs}

To shed more light into the dynamics of our temporal network model under aggregation, in what follows we consider archetypal examples of homogeneous and heterogeneous IET distributions.

In the case of an exponential IET distribution $P(\tau) = \mu^{-1} e^{-\tau / \mu}$ with parameter $\mu = \langle \tau \rangle$, the residual distribution is exponential as well, $P_{>}(\tau) = e^{-\tau / \langle \tau \rangle}$. Solving the integral in \eref{eq:frac_0_edges} gives
\begin{equation}
\label{eq:frac_0_edges_exp}
E_0(\eta) = e^{-\eta / \langle \tau \rangle} = P_{>}(\eta),
\end{equation}
meaning that the fraction of inactive edges is given by the residual distribution and thus decays exponentially with the window size $\eta$. The exponential case involves the relatively simple Laplace transform $\hat{E}_j = \langle \tau \rangle / (1 + \mu s)^{j+1}$ for $j \geq 0$, which implies that the fraction of $j$-type edges is Poisson-distributed,
\begin{equation}
\label{eq:frac_j_edges_exp}
E_j(\eta) = \frac{e^{-\eta / \mu}}{j!} \left( \frac{\eta}{\mu} \right)^j.
\end{equation}
\esref{eq:frac_0_edges_exp}{eq:frac_j_edges_exp} also suggest that $\eta / \mu$ is the appropriate scaled variable for window size in the temporal network aggregation process. Solving for $\eta^*$ in \eref{eq:opt_scale} leads to the optimal window length
\begin{equation}
\label{eq:opt_scale_exp}
\frac{\eta^*}{\mu} = \frac{\ln (2 - \nu)}{1 - \nu},
\end{equation}
with limiting values $\eta^* / \mu = \ln 2$ for $\nu = 0$ and $\eta^* / \mu \to 1$ for $\nu \to 1$ (i.e. non- and fully overlapping windows, respectively). Thus, when the temporal network is homogeneous and IETs are exponentially distributed, the aggregation window $\eta^*$ that maximizes the edge activation probability $P^{\pm}$ is roughly equal to the mean IET $\mu$ for any value of overlap.

\subsubsection{Heterogeneous IETs}

As an example of a heterogeneous temporal network, we consider the gamma IET distribution
\begin{equation}
\label{eq:iet_dist_gamma}
P(\tau) = \frac{1}{\Gamma (\alpha) \beta^{\alpha}} \tau^{\alpha - 1} e^{-\tau / \beta}
\end{equation}
with parameters $\alpha, \beta > 0$, where $\Gamma (\alpha) = \int_{0}^{\infty} t^{\alpha - 1} e^{-t} dt$ is the gamma function. \eref{eq:iet_dist_gamma} has mean IET $\mu = \alpha \beta$, variance $\sigma^2 = \alpha \beta^2$, and dispersion index $D = \sigma^2 / \mu = \beta$. The residual distribution is given by $P_{>}(\tau) = \Gamma(\alpha, \tau/\beta) / \Gamma(\alpha)$, with $\Gamma(\alpha, x) = \int_{x}^{\infty} t^{\alpha - 1} e^{-t}$ the upper incomplete gamma function. After solving the integral in \eref{eq:frac_0_edges} and some algebra, we obtain
\begin{equation}
\label{eq:frac_0_edges_gamma}
E_0(\eta) = \left( 1 - \frac{\eta}{\mu} \right) P_{>}(\eta) + \frac{\eta}{\mu} D P(\eta).
\end{equation}
As heterogeneity increases in the temporal network (regulated by a growing dispersion $D$), the fraction of inactive edges in \eref{eq:frac_0_edges_gamma} can remain relatively large for increasing $\eta$, leading to an optimal window length $\eta^*$ larger than the homogeneous baseline at mean IET $\mu$.

\subsubsection{Homogeneous degrees}

We also explore a couple of archetypal examples of static degree distributions, to highlight the interplay between IETs and aggregated degrees.

A random network with average static degree $z = \langle k \rangle$ and Poisson degree distribution $P(k) = z^{k} e^{-z} / k!$ has the generating function $g_0(x) = e^{z(x - 1)} = g_1(x)$. With this, \eref{eq:act_lcc} gets simplified into the transcendental equation
\begin{equation}
\label{eq:act_lcc_poisson}
S_{\eta} = 1 - e^{-z E_+(\eta) S_{\eta}},
\end{equation}
which might be solved numerically to obtain the active LCC size $S$ as a function of window length $\eta$ for a given fraction of active links $E_+(\eta) = 1 - E_0(\eta)$.

\subsubsection{Heterogeneous degrees}

As a standard example of heterogeneous static degrees we take the Zipf (or zeta) distribution
\begin{equation}
\label{eq:deg_dist_zipf}
P(k) = \frac{k^{-\gamma}}{\zeta(\gamma)}
\end{equation}
for $k \geq 1$ and $P(0) = 0$, which has a power-law decay regulated by the exponent $\gamma > 1$. The normalization constant is the Riemann zeta function $\zeta(\gamma) = \sum_{k = 1}^{\infty} k^{-\gamma}$. The generating function of $P(k)$ is $g_0(x) = \text{Li}_{\gamma}(x) / \zeta(\gamma)$ with $\text{Li}_{\gamma}(x) = \sum_{k = 1}^{\infty} x^k / k^{\gamma}$ the polylogarithm of order $\gamma$, leading to the average static degree $z = \langle k \rangle = \zeta(\gamma - 1) / \zeta(\gamma)$. Taking a derivative with respect to $x$ and dividing by $z$, the generating function of the excess degree distribution is $g_1(x) = \text{Li}_{\gamma - 1}(x) / [ x \zeta(\gamma - 1) ]$. With these two generating functions, \eref{eq:act_lcc} takes the form
\begin{subequations}
\label{eq:act_lcc_zipf}
\begin{align}
0 &= u[u - E_0(\eta)]\zeta(\gamma - 1) - [1 - E_0(\eta)]\text{Li}_{\gamma - 1}(u), \\
S_{\eta} &= 1 - \text{Li}_{\gamma}(u) / \zeta(\gamma).
\end{align}
\end{subequations}
The first equation can be solved numerically to obtain $u$ and insert it into the second equation, leading once more to the size $S$ of the active LCC as a function of window size $\eta$.

\subsection{Empirical IET distribution}
\label{SI:Gammafit}

\begin{figure}
    \centering
    \includegraphics[width=15cm]{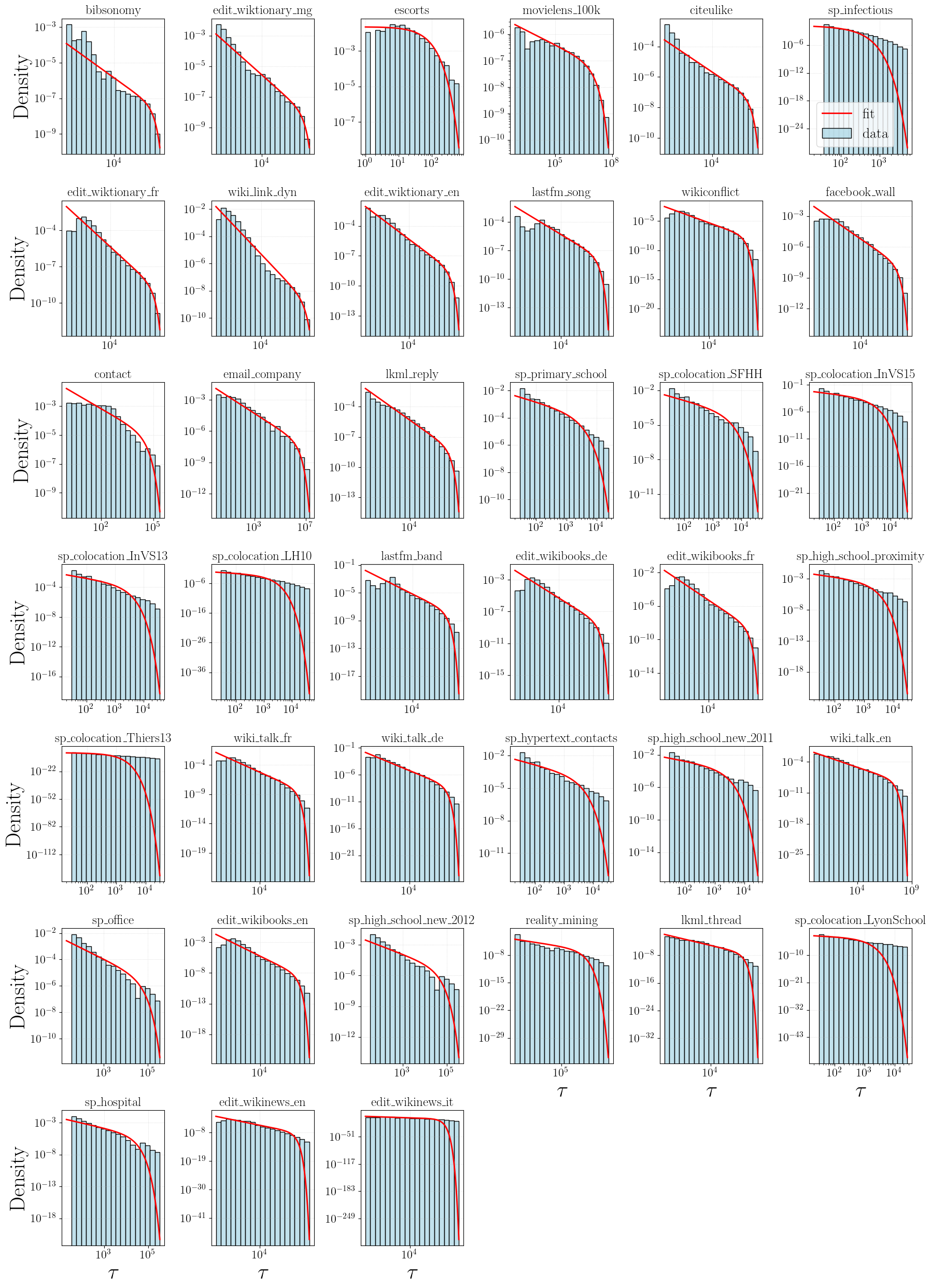} 
    \caption{IET distribution (histogram) and Gamma fit (red line), for all 39 datasets. The Gamma fits tend to fail along the tails of the distributions. However, the Wikipedia datasets -and, more generally, datasets that do not represent offline interactions- appear to be relatively well described by Gamma distributions. The datasets are ordered by increasing system (edge) burstiness.}
    \label{fig:gamma_iet_dist}
\end{figure}

To support the assumption of Gamma-distributed inter-event times when analyzing the effect of burstiness on the analytical expectation for $\eta^*$ in Fig. 2e of the main, we show in \frefsi{fig:gamma_iet_dist} how a Gamma fit compares to the fitted empirical IET distribution. As expected, the Gamma fits tend to fail along the tails of the distributions. However, the Wikipedia datasets -and, more generally, datasets that do not represent offline interactions- appear to be relatively well described by Gamma distributions.

\subsection{Numerical simulations}
\label{SI_sec:functions}

The model is fully parameterized by the data.
The dataset properties required for parameterization are the system's IET distribution $P(\tau)$, its static degree distribution $P(k)$, its number $N$ of nodes and its total observation period $T$. 
To build the network, we consider a CM network with $N$ nodes and degree distribution $P(k)$. We run a  renewal process on each of the edges, by sampling the time of first event from the residual distribution $P_{>}(\tau)$. To assign the times of the following events we sample the distance between consecutive events from $P(\tau)$. We stop sampling when the time of the next event is larger than the total observation period $T$. Refer to Fig. 1a of the main for a schematic representation of the construction of the model.
Note that analytical considerations describe the limit of $T \rightarrow \infty$. 

In Fig. 2b-c of the main, we consider $20000$ IETs and an initial time offset of $0.999$. There, we also consider two shapes for the distribution of IETs $P(\tau)$, with the same average value but different levels of heterogeneity. 
For the homogeneous IET distribution, we use the exponential parametrized by its average: $Exp(\mu) = \mu^{-1} e^{-\tau / \mu}$. 
For the heterogeneous distribution, we use the Gamma distribution parametrized by its shape and scale:
\begin{equation}
    Gamma(\alpha, \beta) = \frac{1}{\Gamma (\alpha) \beta^{\alpha}} \tau^{\alpha - 1} e^{-\tau / \beta}.
\end{equation}

We also consider two shapes for the degree distribution $P(k)$, with the same average value but different levels of heterogeneity. 
In the homogeneous case, we use the Poisson distribution parametrized by its rate: $Poisson(z) = z^k e^{-z} / k!$. 
In the heterogeneous case, we use the generalized Zipf distribution parametrized by its exponent:
\begin{equation}
    Zipf(\gamma) = \frac{k^{-\gamma}}{\zeta(\gamma)},
\end{equation}
where $\zeta(\gamma)$ is the Riemann zeta function $\zeta(\gamma) = \sum_{k = 1}^{\infty} k^{-\gamma}$.

\subsection{Independence of consecutive windows}

\label{SI_sec:independence}

In the model, we estimate the activation probability $P^+$ assuming independence between the absence of events in a window and the occurrence of at least one event in the non-overlapping portion of the following window — an assumption that is exact only for memoryless (Poissonian, $B=0$) processes. 
Since the corpus of datasets considered exhibits instead highly bursty signals ($B \in [0.08, 0.93]$), here we evaluate the extent to which such assumption of independence fails. 

\begin{figure}
    \centering
    \includegraphics[width=9cm]{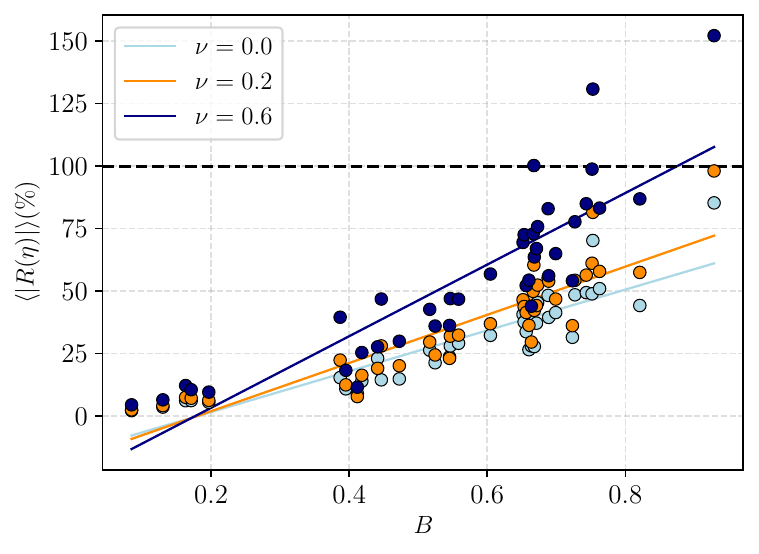} 
    \caption{Average relative bias $\langle |R(\eta)| \rangle$ against burstiness $B$, for all datasets and for increasing fractional overlap $\nu = 0, 0.2, 0.6$. The bias introduced by the assumption of independence grows with both the burstiness of the time series and the overlap between successive windows.}
    \label{fig:relative_bias}
\end{figure}

We define the relative bias on the activation probability $P^+$ as:
\begin{equation}
    R(\eta, \nu) = \frac{P^+_{\text{approx}}(\eta, \nu) - P^+_{\text{exact}}(\eta, \nu)}{P^+_{\text{exact}}(\eta, \nu)},
\end{equation}
where $P^+_{\text{approx}}(\eta, \nu) = E_0(\eta) (1-E_0(\eta(1-\nu)))$ and $P^+_{\text{exact}}(\eta, \nu) = E_0(\eta) - E_0(\eta + \eta(1-\nu))$. 
For each dataset and for a given value of the overlap $\nu$, we compute $R(\eta; \nu)$ for twenty values of $\eta$ centered around the average inter-event time [same selection used to compute the $(P^+,\eta)$ curves]. For all values of $\eta$ and for all datasets, the independence assumption overestimates the activation probability $P^+$, thus manifesting the hazard rate deteriorated by burstiness.

\frefsi{fig:relative_bias} shows that, across datasets, the average relative bias (average over values of $\eta$) $\langle |R(\eta)| \rangle$ monotonically increases with burstiness, consistently for three choices of the fractional overlap $\nu$.
The monotonic increase of the bias with burstiness demonstrates the growing correlation between successive windows resulting from burstiness. Additionally, increasing overlaps lead to smaller non-overlapping parts of the windows and eventually to dropping chances of forgetting, resulting in visible larger biases. 

While the model successfully demonstrates the emergence of a maximally dynamic scale from a simple renewal process running on a static network and its dependence on burstiness, its quantitative accuracy in predicting the relevant timescales depends on the level of openness of the system (shown in Figure 2 of the main), but also on its burstiness and the considered overlap between rolling time windows.